\documentclass[11pt,a4paper,english,twoside]{article}

\usepackage{a4wide}
\usepackage{amssymb, amsmath, amsthm}
\usepackage{graphicx}
\usepackage{subcaption}
\usepackage[all]{xy}
\usepackage{enumerate}
\usepackage[pdftex,hyperref,svgnames]{xcolor}
\usepackage[pdftex,colorlinks=true,
pdfstartview=FitV,
pdfnewwindow=true,
linktoc = page,
linkcolor= Red,
citecolor= blue,
urlcolor= blue,
hyperindex=true,
hyperfigures=false]{hyperref}
\hypersetup{linktocpage}
\usepackage{dsfont}
\usepackage{empheq}
\usepackage{cite}
\usepackage{float}
\usepackage{cancel}
\usepackage{relsize}
\usepackage{soul}
\usepackage{enumitem}
\usepackage{hhline}

\newcommand{\beq}{\begin{equation}}
\newcommand{\eeq}{\end{equation}}
\def\bea#1\eea{\begin{align}#1\end{align}}
\def\beal#1\eeal{\begin{subequations}\begin{align}#1\end{align}\end{subequations}}
\newcommand{\nn}{\nonumber}

\newcommand{\R}{\mathcal{R}}

\def\del {\partial}
\def\d {{\rm d}}
\def\mmm {\mathcal{M}}

\newtheorem*{lemma}{Lemma}

\begin{document}
\numberwithin{equation}{section}

\begin{titlepage}

\begin{center}

\phantom{DRAFT}

\vspace{2.4cm}

{\LARGE \bf{Tachyonic de Sitter solutions \vspace{0.3cm}\\ of 10d type II supergravities}}\\

\vspace{2.2 cm} {\Large David Andriot}\\
\vspace{0.9 cm} {\small\slshape Institute for Theoretical Physics, TU Wien\\
Wiedner Hauptstrasse 8-10/136, A-1040 Vienna, Austria}\\
\vspace{0.5cm} {\upshape\ttfamily david.andriot@tuwien.ac.at}\\

\vspace{2.8cm}

{\bf Abstract}
\vspace{0.1cm}
\end{center}

\begin{quotation}
\noindent Cosmological models of the early or late universe exhibit (quasi) de Sitter space-times with different stability properties. Considering models derived from string theory, the swampland program does not provide for now a definite characterisation of this stability. In this work we focus on de Sitter solutions of 10d type II supergravities, candidates for classical de Sitter string backgrounds: surprisingly, all known examples are unstable with $\eta_V < -1$. We aim at proving the existence of such a systematic tachyon, and getting formally a bound on the value of $\eta_V$. To that end, we develop three methods, giving us various sufficient conditions for having a tachyon upon assumptions, in analogy with de Sitter no-go theorems. Our analysis eventually indicates the existence of variety of different tachyons, and related bounds on $\eta_V$. We use this knowledge to find 10 new de Sitter solutions of type IIB supergravity, that have tachyons of a different kind, higher $\eta_V$ values and new 6d geometries. One solution even appears to be stable, with however non-compact extra dimensions.
\end{quotation}

\end{titlepage}

\newpage

\tableofcontents

\section{Introduction}

Dark energy, the largest energy density in our universe today, responsible for the observed accelerated expansion, is nowadays one of the greatest puzzles in physics. Observations are consistent with it being a cosmological constant to a good accuracy, even though the coming Euclid mission, or the Hubble tension problem, may give room in this interpretation (see \cite{Cao:2020evz, DAmico:2020tty, Barrau:2021vap, Park:2021jmi} for recent discussions of observational constraints). Understanding the nature of a cosmological constant remains in any case a theoretical challenge. A common description, also used to describe early universe scenarios with acceleration (inflation) or contraction (bouncing), goes through a four-dimensional (4d) theory of scalar fields minimally coupled to gravity, as given by
\begin{equation}
{\cal S} = \int \d^4 x \sqrt{|g_4|} \left(\frac{M_p^2}{2} \R_4 - \frac{1}{2} g_{ij} \del_{\mu}\phi^i \del^{\mu}\phi^j - V \right) \ ,\label{S4dgen}
\end{equation}
with the field space metric $g_{ij}$, a scalar potential $V$ depending on the scalar fields $\phi^i$, and the 4d reduced Planck mass $M_p$. In that setting, critical points or extrema of the potential ($\del_{\phi^i} V =0$) can correspond to solutions with a de Sitter space-time, whose positive cosmological constant is proportional to $V|_0 > 0$, the value of the potential at this extremum. Dark energy and the cosmological constant are then understood as the vacuum energy. Quasi de Sitter space-times can be obtained close to an extremum, allowing some rolling of the scalar fields. In such a framework, a crucial question is then that of the stability of the positive extremum or de Sitter solution. Describing our future universe, as predicted by $\Lambda$CDM, namely a pure de Sitter space-time to which are attracted to, requires a (meta)stable solution, i.e.~here a (local) minimum of $V$. Slightly unstable potentials, e.g.~away from almost flat maxima, can be interesting for plateau models, currently favored to realise single-field slow-roll inflation, or even for quintessence. Sharper maxima are also allowed by some multi-field inflation models. The stability is governed by the second derivative of $V$, and we introduce the standard parameters for $V>0$
\beq
\epsilon_V = \frac{{M_p}^2}{2} \left(\frac{|\nabla V|}{V}\right)^2 \ ,\quad \quad \eta_V= {M_p}^2\, \frac{{\rm min}\ \nabla \del V}{V} \ ,\label{par}
\eeq
where
\beq
|\nabla V| = \sqrt{g^{ij} \del_{\phi^i} V \del_{\phi^j} V} \ ,\quad {\rm min}\ \nabla \del V =\mbox{minimal eigenvalue of}\ \left( M^i{}_j = g^{ik} \nabla_{\!\phi^k} \del_{\phi^j} V \right)\ ,
\eeq
and $\nabla_{\!\phi^i}$ is the covariant derivative in field space, $M$ is the mass matrix. Having $\epsilon_V$ close to 0 indicates whether we are on a quasi-de Sitter space-time, while the sign and value of $\eta_V$ characterises the stability. Given the many possibilities for a theory \eqref{S4dgen} and their cosmological implications, we are interested in this work in those 4d models derived from string theory, and their (stability) properties. More precisely, focusing on a classical and perturbative regime of string theory, we will study a surprising phenomenon, namely a seemingly systematic instability, or tachyon, in de Sitter solutions.

What models or effective field theories can be obtained from quantum gravity, in particular string theory, is a crucial question, especially for the connection to phenomenology. It is at the heart of the swampland program \cite{Vafa:2005ui, Brennan:2017rbf, Palti:2019pca}, which aims at characterising such models, elements of the ``landscape'', distinguishing them from those which do not couple consistently to quantum gravity and are part of the ``swampland''. It is common lore to consider models like \eqref{S4dgen}, obtained from a 10d string theory by a compactification on extra dimensions gathered as a 6d compact manifold $\mmm$. One may then ask whether de Sitter solutions can be found in string theory \cite{Danielsson:2018ztv}, i.e.~a 10d string background made up of a 4d de Sitter space-time times a 6d $\mmm$, or equivalently a corresponding 4d model \eqref{S4dgen} with a de Sitter extremum. Such a realisation could match standard cosmological models described above, but alternative stringy approaches to cosmology have also been proposed (see e.g.~\cite{Brahma:2020htg, Brahma:2020tak, Gunaydin:2020ric, Bedroya:2020rac, Banerjee:2020wov, Bernardo:2020lar, Agrawal:2020xek, Faraggi:2020hpy, Cespedes:2020xpn, Cadoni:2020jxe, Dvali:2020etd, Hertog:2021jyd} for a recent sample). Finding de Sitter solutions from string theory has been attempted for a long time, and the current situation is that it is difficult to build any well-controlled example. Well-known approaches include those based on \cite{Kachru:2003aw, Balasubramanian:2005zx}, that allow various corrections or non-perturbative contributions, and recent works studying those possibilities include \cite{Farakos:2020wfc, Gao:2020xqh, Bena:2020xrh, Crino:2020qwk, Carta:2021lqg}. Further recent works on de Sitter in string compactifications can be found in \cite{CaboBizet:2020cse, Marchesano:2020uqz, Basile:2020mpt, Basiouris:2020jgp, Dine:2020vmr, Bansal:2020krz, Plauschinn:2020ram}. In this paper, we rather focus on ``classical de Sitter solutions'' \cite{Andriot:2019wrs}, where we restrict ourselves to working in a classical and perturbative regime of string theory. While being more limited in terms of possible content or ingredients, this approach, if successful, would allow a simpler control of required approximations. In practice, the focus has been on 10d type IIA/B supergravities with $D_p$-branes and orientifold $O_p$-planes \cite{Maldacena:2000mw, Hertzberg:2007wc, Silverstein:2007ac, Covi:2008ea, Haque:2008jz, Caviezel:2008tf, Flauger:2008ad, Danielsson:2009ff, deCarlos:2009fq, Caviezel:2009tu, Dibitetto:2010rg, Wrase:2010ew, Danielsson:2010bc, Andriot:2010ju, Blaback:2010sj, Danielsson:2011au, Shiu:2011zt, Burgess:2011rv, VanRiet:2011yc, Danielsson:2012by, Danielsson:2012et, Gautason:2013zw, Kallosh:2014oja, Junghans:2016uvg, Andriot:2016xvq, Junghans:2016abx, Andriot:2017jhf, Roupec:2018mbn, Andriot:2018ept, Junghans:2018gdb, Banlaki:2018ayh, Cordova:2018dbb, Cribiori:2019clo, Andriot:2019wrs, Das:2019vnx, Grimm:2019ixq, Cordova:2019cvf, Andriot:2020lea, Kim:2020ysx, Andriot:2020wpp, Andriot:2020vlg, Farakos:2020idt, Bena:2020qpa}, assumed to be in the appropriate classical and perturbative regime (small $g_s$, large volume, etc.) to be a low energy effective theory of string theory; that assumption has been further studied in \cite{Roupec:2018mbn, Junghans:2018gdb, Banlaki:2018ayh, Andriot:2019wrs, Grimm:2019ixq, Andriot:2020vlg}. In that context, despite many constraining no-go theorems (see e.g.~\cite{Andriot:2020lea}), few de Sitter solutions have been found with $\mmm$ being a group manifold \cite{Caviezel:2008tf, Flauger:2008ad, Caviezel:2009tu, Danielsson:2010bc, Danielsson:2011au, Roupec:2018mbn, Andriot:2020wpp}.\footnote{Other solutions were recently obtained and discussed in \cite{Cordova:2018dbb, Cribiori:2019clo, Cordova:2019cvf, Kim:2020ysx, Bena:2020qpa}, with a different ansatz than the one considered in this paper; we do not consider those here.} Contrary to other approaches, the supersymmetry breaking is then spontaneous and due to the geometry or the fluxes, not to the $D_p/O_p$. The ansatz used for those solutions (see \cite{Andriot:2019wrs}) allows an alternative 4d description of the type \eqref{S4dgen}, as a (gauged) supergravity. It has then been noticed that {\it all these known de Sitter solutions of type II supergravities are unstable}, with $\eta_V|_0 < - 1$. This means that the potential is at a (sharp) maximum in at least one field direction, the tachyon. Having a tachyonic direction is certainly favored statistically (see e.g.~\cite{Marsh:2011aa}), but as we will see, it remains unclear whether this is really a general feature of these solutions, i.e.~whether classical de Sitter solutions admit a systematic tachyon. If they do (conjecture 2 of \cite{Andriot:2019wrs}), one may wonder about the tachyon origin, and whether $\left|\eta_V|_0 \right|$ should always be so large, an important question for phenomenology.

These questions were naturally revived recently in the swampland program, that focuses on the outcomes of string theory. More precisely, for effective theories of quantum gravity like \eqref{S4dgen}, the shape of the potential $V$ was discussed, and characterisations were proposed, under the name of de Sitter swampland conjectures (see e.g.~\cite{Dvali:2014gua, Dvali:2017eba} for earlier arguments). The initial version \cite{Obied:2018sgi} proposed $\sqrt{2 \epsilon_V} \geq c \sim O(1)$, forbidding completely the existence of de Sitter solutions. Various refinements have then been proposed \cite{Andriot:2018wzk, Garg:2018reu, Ooguri:2018wrx, Andriot:2018mav, Rudelius:2019cfh, Bedroya:2019snp}. Some have to do with the existence of solutions, now rather forbidden in asymptotics of field space only, which may still correspond to a classical regime. In particular the TCC proposal \cite{Bedroya:2019snp} on this matter has been quantitatively very well verified \cite{Andriot:2020lea}. Other refinements were about including stability, but the statement there is less settled: among others, it was proposed in \cite{Garg:2018reu, Ooguri:2018wrx} to have $\eta_V \leq - c' \sim -O(1)$ (in agreement with above examples), while the TCC \cite{Bedroya:2019snp} allows for some metastable solutions. The constrained lifetime of the latter has been connected to the scrambling time \cite{Aalsma:2020aib, Bhattacharyya:2020kgu, Bedroya:2020rmd}; further recent works on these conjectures include \cite{Burgess:2020nec, Blumenhagen:2020doa}. This situation motivates to test further the de Sitter swampland conjectures, and make them more precise regarding stability. This is one purpose of the present analytic study of tachyons, in the well-controlled setting of 10d type II supergravities.

Stability of these de Sitter solutions has been studied either formally or through concrete examples \cite{Covi:2008ea, Danielsson:2011au, Shiu:2011zt, Danielsson:2012by, Danielsson:2012et, Kallosh:2014oja, Junghans:2016uvg, Junghans:2016abx, Andriot:2018ept, Roupec:2018mbn, Andriot:2019wrs, Andriot:2020wpp}. A first set of work \cite{Covi:2008ea, Kallosh:2014oja, Junghans:2016uvg, Junghans:2016abx} focused on 4d ${\cal N} =1$ supergravity with a de Sitter solution close to a no-scale Minkowski one. In that case, a systematic tachyon would appear, corresponding to the sgoldstino in the Minkowski limit. Beyond this situation though, it is unclear whether a tachyon would always appear for any (classical) de Sitter solution, and what would be its 10d interpretation. A surprising proposal was however made in \cite{Danielsson:2012et}: it suggested that a systematic tachyon lies among a handful of universal 4d scalar fields, namely $(\rho, \tau, \sigma_I)$. The well-known 4d dilaton $\tau$ and the 6d volume $\rho$ were introduced in \cite{Hertzberg:2007wc}, while $\sigma_I$, related to internal volumes of subspaces wrapped by the $D_p/O_p$, were introduced in \cite{Danielsson:2012et}. The general scalar potential obtained from 10d type II supergravities, restricted to these scalars only, $V(\rho, \tau, \sigma_I)$, has been derived in \cite{Andriot:2018ept, Andriot:2019wrs, Andriot:2020lea}. The proposal of \cite{Danielsson:2012et} then amounts to say that one always has $\eta_V|_0 < 0$, when considering only these fields. This has been successfully verified on examples in \cite{Danielsson:2012et, Junghans:2016uvg, Andriot:2020wpp}. It was also noticed that some solutions already admit the tachyon among the two fields $(\rho,\tau)$,\footnote{Formal results on tachyons among $(\rho,\tau)$ were obtained in \cite{Shiu:2011zt}, that pioneered some technical aspects of such studies. That work considered compactifications with $D_p/O_p$ of single size $p$ or multiple sizes. On the former, they proved under some conditions the presence of a systematic tachyon for $O_4$ with $F_0 \neq 0$, and for $O_3$. However, as later shown using Bianchi identities (see \cite{Andriot:2019wrs} and references therein), these two situations do not admit de Sitter extrema. So no result remains on stability with single size sources. However, the techniques developed in \cite{Shiu:2011zt} were interesting and we will come back to them in section \ref{sec:method1}.} but not all of them. To formally prove the existence of a systematic tachyon in classical de Sitter solutions, a strategy then consists in proving this proposal true, as already attempted in \cite{Andriot:2018ept, Andriot:2019wrs}. Given the little number of fields, and the fact the statement was always true when checked on examples, an analytical proof seems at first sight achievable.

An ideal objective would be to prove, using the potential, its first and second derivatives, as well as 10d equations (in particular Bianchi identities) that one always gets $\eta_V|_0 < 0$. Furthermore, we would like to get a bound $\eta_V|_0$, useful for phenomenology but also to compare to swampland conjectures. To that end, we focus in this paper on 10d type IIB supergravity with $D_5/O_5$ sources, the framework of \cite{Andriot:2020wpp, Andriot:2020vlg} where 17 new de Sitter solutions were recently found. This allows us to have a concrete setting, described in section \ref{sec:setting} and appendix \ref{ap:M}, with four fields $(\rho, \tau, \sigma_1, \sigma_2)$ and their explicit scalar potential $V$ (analytic in terms of solutions data), as well as examples of solutions for checks and inspiration. We believe however that the results obtained here can be reproduced in other analogous settings, as type IIA with $D_6/O_6$, where many de Sitter solutions have also been found (e.g.~\cite{Danielsson:2011au}). Given this framework, we do not fully achieve the objective. We rather get a set of 13 conditions ($C1$-$C13$) on solution data that are sufficient to ensure the existence of a tachyon; they are summarized in Table \ref{tab:sum}. All along the paper, any sufficient condition for a tachyon is boxed. These conditions can be viewed as assumptions, in analogy to those of no-go theorems on the existence of solutions.

Despite having only four fields, proving the existence of a tachyon with $V(\rho, \tau, \sigma_1, \sigma_2)$ remains technically challenging, and we develop for this three methods in section \ref{sec:methods} and appendix \ref{ap:form}, first introduced in section \ref{sec:summethods}. Those are based on two related lemmas on signs of eigenvalues of real symmetric matrices. In short, the first method consists in studying eigenvalues of $2\times 2$ blocks in the mass matrix. The second method consists in finding fixed coefficients $b_{\phi^i}$ and $c_{\phi^i}$ giving an interesting inequality (also writable with canonical fields)
\begin{empheq}[innerbox=\fbox]{align}
& V + \left( b_{\rho} \rho \del_{\rho} + b_{\tau} \tau \del_{\tau} +  b_{\sigma_1} \sigma_1 \del_{\sigma_1}  + b_{\sigma_2} \sigma_2 \del_{\sigma_2} \right) V \nn\\
& \ \ + \left( c_{\rho} \rho \del_{\rho} + c_{\tau} \tau \del_{\tau} +  c_{\sigma_1} \sigma_1 \del_{\sigma_1}  + c_{\sigma_2} \sigma_2 \del_{\sigma_2} \right)^2 V \leq 0 \label{genineqintro}
\end{empheq}
Such an inequality is sufficient to have a tachyon at a de Sitter extremum, and we prove furthermore that it implies the inequality
\begin{empheq}[innerbox=\fbox]{align}
\Rightarrow \quad - \sqrt{\hat{b}_{\hat{\rho}}^2 + \hat{b}_{\hat{\tau}}^2 + \hat{b}_{\hat{\sigma}_1}^2 + \hat{b}_{\hat{\sigma}_2}^2} \ \sqrt{2 \epsilon_V} + \left( \hat{c}_{\hat{\rho}}^2 + \hat{c}_{\hat{\tau}}^2 + \hat{c}_{\hat{\sigma}_1}^2 + \hat{c}_{\hat{\sigma}_2}^2 \right) \eta_V \leq -1 \label{genineq2intro}
\end{empheq}
where $\hat{b}_{\hat{\phi}^i}$ and $\hat{c}_{\hat{\phi}^i}$ are related to $b_{\phi^i}$ and $c_{\phi^i}$ as in \eqref{constants}. That inequality gives a bound $\eta_V|_0$, and we use it for each condition $C3$-$C10$. For $c_{\phi^i} = \hat{c}_{\hat{\phi}^i} = 0$, we recover a no-go theorem on de Sitter extrema, as well as the deduced bound on $\epsilon_V$, that one can compare to the constant $c$ of swampland conjectures, as done in \cite{Andriot:2019wrs, Andriot:2020lea}. The third method uses \eqref{genineqintro} without the term $V$: this is also sufficient for a tachyon, but does not allow to get a (non-zero) bound on $\eta_V|_0$. That method leaves some $c_{\phi^i}$ coefficients free, allowing to capture a larger class of tachyons. We obtain this way the most general sufficient condition for a tachyon, $C11$ in \eqref{C11}: it captures in one class all solutions of \cite{Andriot:2020wpp} but one.

Last but not least, we use these results to find {\it new de Sitter solutions} of 10d type IIB supergravity with $O_5/D_5$, having new physics. They are summarized in Table \ref{tab:sol}, and illustrated in Figure \ref{fig:pot}. We run a new search of de Sitter solutions, asking them to violate $C11$ \eqref{C11}, i.e.~forbidding such a tachyon. We present, in section \ref{sec:sol} and appendix \ref{ap:sol}, 10 new solutions obtained this way. Interestingly, they exhibit new physics w.r.t.~\cite{Andriot:2020wpp}, such as new 6d group manifolds $\mmm$ and higher values of $\eta_V$, one having in particular $\eta_V>0$. We study the new tachyons that appear in these solutions, without being able to characterise them all. Overall, it appears that if there is a systematic tachyon, it is not universal but splits into several different complementary classes, dependent on solution data, and it seems difficult to capture them all analytically. We summarize and illustrate (see Figure \ref{fig:arrows}) our results in section \ref{sec:ccl}, and relate them to further topics.

\section{Setting and conventions}\label{sec:setting}

In this paper, we consider a 4d theory of scalars fields minimally coupled to gravity given in \eqref{S4dgen}. Then one can define, for a scalar potential $V>0$, the parameters $\epsilon_V$ and $\eta_V$ given in \eqref{par}. Whenever possible, one introduces canonical fields, denoted $\hat{\phi}^i$: in this canonical field basis, the metric is $\delta_{ij}$, meaning
\beq
g_{ij} \del_{\mu}\phi^i \del^{\mu}\phi^j = \delta_{ij} \del_{\mu}\hat{\phi}^i \del^{\mu}\hat{\phi}^j \ .
\eeq
In our setting, this basis will exist. It is related by field space diffeomorphisms $\left(\tfrac{\del \hat{\phi}}{ \del \phi} \right)$ to the standard basis. One further defines a mass matrix $\hat{M}$ in the canonical basis, of coefficients $\hat{M}^i{}_j = \delta^{ik} \del_{\hat{\phi}^k} \del_{\hat{\phi}^j} V$. The following matrix relations hold between the two basis
\beq
\left(  \del_{\hat{\phi}} \right) = \left(\frac{\del \hat{\phi}}{ \del \phi} \right)^{-T} \left(  \del_{\phi} \right) \ ,\ \ \left(\frac{\del \hat{\phi}}{ \del \phi} \right)^T \delta \left(\frac{\del \hat{\phi}}{ \del \phi} \right) = g \ ,\ \ M =  \left(\frac{\del \hat{\phi}}{ \del \phi} \right)^{-1} \hat{M} \left(\frac{\del \hat{\phi}}{ \del \phi} \right) \ ,\label{canstandchangebasis}
\eeq
where $\left(  \del_{\phi} \right)$ denotes the column matrix of derivatives, and one should pay attention to the transpose, understandable when comparing the standard formula of diffeomorphisms with the matrix product. Eigenvalues being independent of the basis choice, $\eta_V$ can as well be expressed using $\hat{M}$, and $|\nabla V|$ can equivalently be expressed in the canonical basis. Mass matrices and relations of the kind of \eqref{canstandchangebasis} are written explicitly in appendix \ref{ap:M}.\\

We focus in this work on de Sitter solutions of 10d type II supergravities. For those we use notations and conventions of \cite{Andriot:2016xvq, Andriot:2017jhf}, reviewed in \cite{Andriot:2019wrs}. In short, we consider solutions where the 10d space-time is a product of a 4d de Sitter space-time, and a 6d compact manifold $\mmm$. The latter is taken in our ansatz to be a group manifold, determined by an underlying Lie algebra, with structure constants $f^a{}_{bc}$. Those also appear as components of the spin connection in a ``flat'' basis where the metric of $\mmm$ is $\delta_{ab}$, so these structure constants capture the 6d geometry. These $f^a{}_{bc}$ enter the Ricci tensor and scalar, and are part of the solution data. The rest of the data is made of fluxes $H$ and $F_q$, given in terms of their components in this flat basis, and the sources contributions $T_{10}^I$. All these data are constants, due to our ansatz for solutions (see \cite{Andriot:2019wrs}). The sources are $D_p$-branes and orientifold $O_p$-planes, gathered in sets $I=1 \dots N$, where in each set these extended objects are along the same dimensions and said to be parallel. The $T_{10}^I$ are proportional to a difference between the number of $O_p$ and that of $D_p$, they are therefore positive if $O_p$ contribute more, or negative if e.g.~there are only $D_p$ in the given set. More specifically, we will focus on the setting described in \cite{Andriot:2020wpp}: 10d type IIB supergravity with $O_5/D_5$ sources, gathered in $N=3$ specific sets. We refer to that work for more details.

As reviewed in section 2.2.1 of \cite{Andriot:2020lea}, one obtains from the 10d type II action a 4d effective theory of the form \eqref{S4dgen} by considering only few scalar fluctuations; here those will be $(\rho, \tau, \sigma_1, \sigma_2)$. The first two scalar fields, the volume $\rho$ and 4d dilaton $\tau$, were first considered in \cite{Hertzberg:2007wc, Silverstein:2007ac}. A third scalar $\sigma$ was introduced  in \cite{Danielsson:2012et} to distinguish parallel or transverse directions to a source, or in other words, to capture a wrapped internal volume. Their generic scalar potential, using the above ansatz, was derived in \cite{Andriot:2018ept}, and a generalization to several sets $I$ of intersecting sources with scalar fields $\sigma_I$ was given as $V(\rho, \tau, \sigma_I)$ in \cite{Andriot:2019wrs}. A correct derivation of the subtle $F_5, F_6$ terms was finally provided in \cite{Andriot:2020lea}. As recalled in the Introduction, a systematic tachyon was proposed in \cite{Danielsson:2012et} to lie among these few scalar fields, and this hypothesis was verified in examples in \cite{Danielsson:2012et, Junghans:2016uvg, Andriot:2020wpp}. This motivates the use of these fields and potential for our purposes.

This formalism was applied in \cite{Andriot:2020wpp} to the case of interest with $O_5/D_5$ sources. We thus take from \cite{Andriot:2020wpp} the following kinetic terms and scalar potential for our four fields $(\rho, \tau, \sigma_1, \sigma_2)$. The 4d action \eqref{S4dgen} is given by
\bea
{\cal S} = \int \d^4 x \sqrt{|g_4|} &\, \Bigg( \frac{M_p^2}{2} \R_4 - V(\rho, \tau, \sigma_1, \sigma_2) \label{S4dkin1}\\
&\, - \frac{M_p^2}{2} \bigg(\frac{3}{2\rho^2} (\del \rho)^2 + \frac{2}{\tau^2}(\del \tau)^2  + 12 \Big(\frac{1}{\sigma_1^2}(\del \sigma_1)^2  + \frac{1}{\sigma_2^2}(\del \sigma_2)^2 - \frac{1}{\sigma_1\sigma_2}\del_{\mu}\sigma_1 \del^{\mu}\sigma_2  \Big)\bigg) \Bigg) \nn\\
 = \int \d^4 x \sqrt{|g_4|}  &\, \Bigg( \frac{M_p^2}{2} \R_4 - V(\rho, \tau, \sigma_1, \sigma_2) \label{S4dkin2}\\
&\, - \frac{M_p^2}{2} \Big(\frac{3}{2} (\del \ln \rho)^2 + 2 (\del \ln \tau)^2  + 9 \big(\del \ln \frac{\sigma_1}{\sigma_2}\big)^2  + 3 \big(\del \ln (\sigma_1 \sigma_2 )\big)^2 \Big)  \Bigg) \ ,\nn
\eea
with $M_p^2 = \frac{1}{\kappa_{10}^2} \int \d^6 y \sqrt{|g_6|}\ g_s^{-2} $. The field space metric, read from \eqref{S4dkin1}, is given by
\begin{equation}
g_{ij} = M_p^2\,
\begin{pmatrix}
\frac{3}{2 \rho^2} & 0 & 0 &0\\[10pt]
0 &\frac{2}{\tau^2} & 0 &0 \\[10pt]
0 & 0 & \frac{12}{\sigma_1^2} & -\frac{6}{\sigma_1 \sigma_2} \\[10pt]
0 & 0 & -\frac{6}{\sigma_1 \sigma_2} & \frac{12}{\sigma_2^2}
\end{pmatrix} \ . \label{gij}
\end{equation}
From this metric or from \eqref{S4dkin2}, we determine the canonical fields
\beq
\hat{\tau}= \sqrt{2}\, M_p\, \ln \tau \, ,\ \hat{\rho}= \sqrt{\frac{3}{2}}\, M_p\, \ln \rho \, ,\ \hat{\sigma}_1= \sqrt{3}\, M_p\, \ln (\sigma_1 \sigma_2 )  \, ,\ \hat{\sigma}_2= 3\, M_p\, \ln \frac{\sigma_1}{\sigma_2} \ . \label{canfield}
\eeq
While the relation between standard fields and canonical ones is straightforward for the first two, one should be more careful with the last two because of the mixing. Let us give for $\sigma_{1,2}$ the diffeomorphism from canonical fields to standard ones, denoted above with the matrix $\left(\frac{\del \hat{\phi}}{ \del \phi} \right)$. Its elements $(i,j)$ are here given by $\frac{\del \hat{\sigma}_i}{ \del \sigma_j}$, as follows
\beq
\left(\frac{\del \hat{\phi}}{ \del \phi} \right) = M_p \sqrt{3} \left( \begin{array}{cc} \sigma_1^{-1} & \sigma_2^{-1} \\ \sqrt{3} \sigma_1^{-1} & - \sqrt{3} \sigma_2^{-1} \end{array}\right) \ . \label{diffeosigma}
\eeq
From \eqref{canstandchangebasis}, one obtains the crucial relation between derivatives
\beq
\left( \begin{array}{c} \del_{\hat{\sigma}_1} \\ \del_{\hat{\sigma}_2} \end{array}\right) = \left(\frac{\del \hat{\phi}}{ \del \phi} \right)^{-T} \left( \begin{array}{c} \del_{\sigma_1} \\ \del_{\sigma_2} \end{array}\right) = \frac{1}{M_p} \frac{1}{2 \sqrt{3}} \left( \begin{array}{c} \sigma_1 \del_{\sigma_1} + \sigma_2 \del_{\sigma_2}  \\ \frac{1}{\sqrt{3}} (\sigma_1 \del_{\sigma_1} - \sigma_2 \del_{\sigma_2}) \end{array}\right) \ . \label{reldersigma}
\eeq
For completeness, we give the same information for $(\rho,\tau)$
\beq
\left(\frac{\del \hat{\phi}}{ \del \phi} \right) = M_p  \left( \begin{array}{cc} \sqrt{\frac{3}{2}} \rho^{-1} & 0 \\ 0 &  \sqrt{2} \tau^{-1} \end{array}\right) \ ,\quad \left( \begin{array}{c} \del_{\hat{\rho}} \\ \del_{\hat{\tau}} \end{array}\right) =  \frac{1}{M_p} \left( \begin{array}{c} \sqrt{\frac{2}{3}} \rho \del_{\rho}   \\ \frac{1}{\sqrt{2}} \tau \del_{\tau} \end{array}\right) \ . \label{relderrhotau}
\eeq

The four field potential $V$ entering the above 4d action is given by\footnote{With respect to \cite{Andriot:2020wpp}, we add here the $F_5$ term building on appendix A of \cite{Andriot:2020lea}, as well as \cite{Andriot:2018ept, Andriot:2019wrs}.}
\bea
\frac{2}{M_p^{2}} \, V(\rho, \tau, \sigma_1, \sigma_2) = &-\tau^{-2}  \rho^{-1} {\cal R}_6(\sigma_1, \sigma_2) \label{potential}\\
& +\frac{1}{2}\, \tau^{-2} \rho^{-3} \left( \sigma_2^{-2A-B} \sigma_1^{-3B} \, |H^{(0)_1}|^2 + \sigma_1^{-2A-B} \sigma_2^{-3B} \, |H^{(2)_1}|^2 \right) \nn\\
&- \, g_s \, \tau^{-3} \, \rho^{-\frac{1}{2}} \, \left(\sigma_1^A  \sigma_2^B \, \frac{T_{10}^1}{6}+\sigma_1^B \sigma_2^A \, \frac{T_{10}^2}{6}+ \sigma_1^B  \sigma_2^B \, \frac{T_{10}^3}{6}\right) \nn\\
&+ \frac{1}{2}g_s^2\, \tau^{-4} \left(\rho^{2} (\sigma_1 \sigma_2)^{-B} |F_1|^2 + (\sigma_1 \sigma_2)^{-A-2B} |F_3|^2 + \rho^{-2} (\sigma_1 \sigma_2)^{-2A-3B} |F_5|^2 \right) \ ,\nn
\eea
with $A=-4,B=2$ and
\beq
{\cal R}_6 (\sigma_1,\sigma_2)=  R_1 \, \sigma_1^{-8} \sigma_2^{4} +
R_2 \, \sigma_1^{4} \sigma_2^{-8} +
R_3 \, \sigma_1^{4} \sigma_2^{4} +
R_4 \, \sigma_1^{-2} \sigma_2^{-2} +
R_5 \, \sigma_1^{4} \sigma_2^{-2} +
R_6 \, \sigma_1^{-2} \sigma_2^{4} \ .
\eeq
This potential is given in terms of the 10d solution data mentioned above, where the $R_{i=1\dots 6}$ curvature terms are given as follows in terms of the structure constants
\begin{align}
-2 R_1 &= \left(f^{1}{}_{35} \right)^2 + \left(f^{1}{}_{36} \right)^2 + \left(f^{1}{}_{45} \right)^2 + \left(f^{1}{}_{46} \right)^2 + \left(f^{2}{}_{35} \right)^2 + \left(f^{2}{}_{36} \right)^2 + \left(f^{2}{}_{45} \right)^2 + \left(f^{2}{}_{46} \right)^2 \,, \nn \\[6pt]
-2 R_2 &= \left(f^{3}{}_{15} \right)^2 + \left(f^{3}{}_{16} \right)^2 + \left(f^{3}{}_{25} \right)^2 + \left(f^{3}{}_{26} \right)^2 + \left(f^{4}{}_{15} \right)^2 + \left(f^{4}{}_{16} \right)^2 + \left(f^{4}{}_{25} \right)^2 + \left(f^{4}{}_{26} \right)^2 \,, \nn \\[6pt]
-2 R_3 &= \left(f^{5}{}_{13} \right)^2 + \left(f^{5}{}_{14} \right)^2 + \left(f^{5}{}_{23} \right)^2 + \left(f^{5}{}_{24} \right)^2 + \left(f^{6}{}_{13} \right)^2 + \left(f^{6}{}_{14} \right)^2 + \left(f^{6}{}_{23} \right)^2 + \left(f^{6}{}_{24} \right)^2 \,, \nn \\[6pt]
-R_4 &= f^{1}{}_{35} \, f^{3}{}_{15} + f^{1}{}_{36} \, f^{3}{}_{16} + f^{2}{}_{35} \, f^{3}{}_{25} + f^{2}{}_{36}  \, f^{3}{}_{26} \nn \\
&+f^{1}{}_{45} \, f^{4}{}_{15} + f^{1}{}_{46} \, f^{4}{}_{16} + f^{2}{}_{45} \, f^{4}{}_{25} + f^{2}{}_{46} \, f^{4}{}_{26} \,, \label{coeff} \\[6pt]
-R_5 &= f^{3}{}_{51} \, f^{5}{}_{31} + f^{4}{}_{51} \, f^{5}{}_{41} + f^{3}{}_{52} \, f^{5}{}_{32} + f^{4}{}_{52}  \, f^{5}{}_{42} \nn \\
&+f^{3}{}_{61} \, f^{6}{}_{31} + f^{4}{}_{61} \, f^{6}{}_{41} + f^{3}{}_{62} \, f^{6}{}_{32} + f^{4}{}_{62} \, f^{6}{}_{42} \,, \nn \\[6pt]
-R_6 &= f^{5}{}_{13} \, f^{1}{}_{53} + f^{5}{}_{14} \, f^{1}{}_{54} + f^{5}{}_{23} \, f^{2}{}_{53} + f^{5}{}_{24}  \, f^{2}{}_{54} \nn \\
&+f^{6}{}_{13} \, f^{1}{}_{63} + f^{6}{}_{14} \, f^{1}{}_{64} + f^{6}{}_{23} \, f^{2}{}_{63} + f^{6}{}_{24} \, f^{2}{}_{64} \,. \nn
\end{align}
To match notations used in previous papers, we also rewrite the above as
\bea
& |f^{{}_{||_1}}{}_{{}_{\bot_1} {}_{\bot_1}}|^2 = - 2R_1  \ ,\ |f^{{}_{||_2}}{}_{{}_{\bot_2} {}_{\bot_2}}|^2 = -2 R_2 \ , \ |f^{{}_{||_3}}{}_{{}_{\bot_3} {}_{\bot_3}}|^2 = -2R_3 \\
& - \delta^{cd}   f^{b_{\bot_1}}{}_{a_{||_1}  c_{\bot_1}} f^{a_{||_1}}{}_{ b_{\bot_1} d_{\bot_1}} = \lambda_1 |f^{{}_{||_1}}{}_{{}_{\bot_1} {}_{\bot_1}}|^2 = R_4 +R_6 \ .\label{lambda1}
\eea
We recall for completeness the following notations in terms of the solution data
\beq
|H^{(0)_1}|^2 = (H_{345})^2 + (H_{346})^2 \ ,\ |H^{(2)_1}|^2 = (H_{125})^2 + (H_{126})^2 \ , \ |F_q|^2 = \frac{1}{q!} F_{q\, a_1 \dots a_q} F_q{}^{a_1 \dots a_q} \ .\nn
\eeq

In our conventions, a de Sitter extremum or background is obtained at
\beq
\mbox{De Sitter background:}\quad \rho = \tau = \sigma_1 = \sigma_2 = 1 \ ,\ |\nabla V|_0 = 0 \ ,\ V|_0 > 0 \ .
\eeq
In particular, it was verified explicitly in \cite{Andriot:2018ept, Andriot:2019wrs, Andriot:2020lea} that the vanishing of the first derivatives corresponds to 10d equations of motion, and the 4d Einstein equation is
\beq
\frac{2}{M_p^{2}} V|_0 = \frac{1}{2} {\cal R}_4 \ . \label{VR4}
\eeq
Other 10d equations of motion (and Bianchi identities) are not captured here though, by this restricted set of fields.

Setting $\sigma_{1,2}=1$, we recover the particular case of the two fields $(\rho, \tau)$ \cite{Hertzberg:2007wc}, with
\bea
& {\cal S} = \int \d^4 x \sqrt{|g_4|} \Bigg( \frac{M_p^2}{2} \R_4 - V(\rho, \tau)  - \frac{M_p^2}{2} \bigg(\frac{3}{2\rho^2} (\del \rho)^2 + \frac{2}{\tau^2}(\del \tau)^2  \bigg) \Bigg) \ , \label{S4d2fields}\\
& \frac{2}{M_p^2} V =  - \tau^{-2} \bigg( \rho^{-1} {\cal R}_6 -\frac{1}{2} \rho^{-3} |H|^2 \bigg) - g_s \tau^{-3} \rho^{\frac{p-6}{2}}  \frac{T_{10}}{p+1} +\frac{1}{2} g_s^2 \tau^{-4} \sum_{q=0}^{6} \rho^{3-q} |F_q|^2  \label{potrhotau} \ ,
\eea
to be considered here in IIB with $p=5$. We then recall notations on the background values
\beq
T_{10}= \sum_{I=1,2,3} T_{10}^I \ ,\quad {\cal R}_6 = \sum_{i=1,\dots,6} R_i \ ,\quad |H|^2 = |H^{(0)_1}|^2 + |H^{(2)_1}|^2 \ .
\eeq

Last but not least, we recall some useful knowledge on signs of quantities. We will use
\beq
T_{10} > 0 \ ,\ |F_1|^2 > 0 \ ,\ {\cal R}_6 < 0 \ ,\ T_{10}^3 \leq 0 \ , \ T_{10}^1 > 0 \ . \label{entriessign}
\eeq
The first three signs come from well known no-go theorems (see e.g.~no-gos 1, 2 and 3 in \cite{Andriot:2020lea} and references therein) and are requirements for de Sitter. The sign of $T_{10}^3$ is imposed in the setting of \cite{Andriot:2020wpp} we use here, also as a requirement for de Sitter. Because $T_{10}>0$, it means that $T_{10}^1$ or $T_{10}^2$ has to be positive, and we choose without loss of generality $T_{10}^1>0$, leaving the sign of $T_{10}^2$ free.

\section{Identifying tachyons}\label{sec:methods}

In this section, we use the material of section \ref{sec:setting} to build sufficient conditions for having a tachyon. Any such condition is signaled with a box. To obtain these conditions, we use three different methods, that we first introduce in section \ref{sec:summethods}. Method 1 is implemented in section \ref{sec:method1}, method 2 in section \ref{sec:method2} and method 3 in section \ref{sec:method3}. Further formalism required for methods 2 and 3 is presented in appendix \ref{ap:form}, and summarized in section \ref{sec:interlude}. We typically check the conditions obtained for a tachyon on the 17 solutions obtained in \cite{Andriot:2020wpp}, to evaluate how relevant they are. We recall that $F_5=0$ in all these solutions. A summary of the conditions found is given in Table \ref{tab:sum}.

\subsection{Introduction to the three methods used}\label{sec:summethods}

As explained in the Introduction, there are hints of a systematic tachyon in de Sitter solutions of 10d type II supergravities (with the standard ansatz of e.g.~\cite{Andriot:2019wrs}). We aim here at identifying such a systematic tachyon, ideally proving it is always present. To that end, we will consider a 4d theory of the type \eqref{S4dgen}, focusing more precisely here on the one given in \eqref{S4dkin1}, depending on the four fields $(\rho, \tau, \sigma_1, \sigma_2)$: the tachyon is believed to be among those. Let us recall that these four fields, scalar potential and action, are obtained through fluctuations around a 10d de Sitter solution or background. It has been explicitly verified that such a solution corresponds to an extremum of the potential, around which we can then study the stability. One may however worry that keeping only these four fields (and freezing the others) is a too restricted truncation,\footnote{Recent discussions on consistent truncations and low energy truncations on group manifolds can be found e.g.~in \cite{Andriot:2018tmb, Andriot:2020ola}.} and that the physics of the 10d solution is not fully captured. However, as we will crucially recall below thanks to the lemmas, finding a tachyon among these few scalar fields {\it is sufficient} to prove a general instability of the solution. We thus focus from now on these four fields only.\\

The instability of a solution is captured in 4d by studying the mass matrix $M= g^{-1} \nabla \del V$, of coefficients $M^i{}_k= g^{ij} \nabla_j \del_k V= g^{ij} (\del_j \del_k V - \Gamma^l_{jk} \del_l V)$. Eigenvalues of $M$ correspond to ${\rm masses}^2$. If one of them is negative, we have a tachyon, i.e.~an instability. This is equivalent to $\eta_V < 0$. The associated eigenvector corresponds to a direction in field space, the tachyonic direction, along which the potential is at a maximum. Identifying a systematic tachyon then amounts to prove that an eigenvalue is always negative, and to find a corresponding tachyonic direction.

The problem then requires an {\it analytic study} of eigenvalues of the mass matrix. We may as well consider the mass matrix $\hat{M}$ in the canonical basis, of coefficient $\hat{M}^i{}_j = \delta^{ik} \del_{\hat{\phi}^k} \del_{\hat{\phi}^j} V$. We recall from section \ref{sec:setting} that $M$ and $\hat{M}$ are related by a field space diffeomorphism, so their eigenvalues are the same. We however face in any case a basic issue here: with the four fields $(\rho, \tau, \sigma_1, \sigma_2)$, $M$ is $4\times 4$ matrix. This makes it difficult to analytically determine the eigenvalues. As indicated in \cite{Danielsson:2011au, Andriot:2020wpp}, some solutions already admit a tachyon among the two fields $(\rho, \tau)$, but not all of them. The determination of eigenvalues for a $2\times 2$ matrix is of course possible, but it is then not enough here. This is where interesting mathematical results come to hand. Before stating them, let us recall that $\hat{M}$ is, by definition, a real square symmetric matrix, thanks to the commutation of derivatives. The first result is the following lemma, introduced and proved in \cite{Andriot:2020wpp}
\begin{lemma}
Let $\hat{M}$ be a square symmetric matrix of finite size, and $A$ an upper left square block of $\hat{M}$. Let $\mu_1$ be the minimal eigenvalue of $\hat{M}$ and $\alpha$ any eigenvalue of $A$. Then one has $\mu_1 \leq \alpha$.
\end{lemma}
\noindent Another useful lemma, that one can deduce from the above and Sylvester's criterion (see e.g.~\cite{Shiu:2011zt}), is the following
\begin{lemma}
Let $\hat{M}$ be a real square symmetric matrix of finite size, and $A$ an upper left square block of $\hat{M}$. If $A$ has a negative determinant, then $\hat{M}$ admits a negative eigenvalue.
\end{lemma}
These lemmas are crucial: they allow to prove the existence of a tachyon, i.e.~a negative eigenvalue of $\hat{M}$, by studying only eigenvalues of smaller diagonal blocks of this matrix. This has several important consequences. First, method 1 in section \ref{sec:method1} will consist in studying $2\times 2$ blocks, for which we can compute analytically eigenvalues. Secondly, the lemmas also apply to a $1\times 1$ diagonal block, i.e.~a diagonal entry! This is obvious in the eigenvector basis where a diagonal entry is directly the eigenvalue, but this will also be of interest in other basis. Last but not least, we can take the block to be made of the fields of a truncation, e.g.~the four fields considered here. If a tachyon is found among those, then the tachyon can only remain present (and get ``worse'') by considering more fields \cite{Andriot:2020wpp}. Mathematically, we mean that a negative eigenvalue can only get more negative by making the matrix bigger.\footnote{This holds provided the matrix remains symmetric, e.g.~if one can always find a canonical basis.} This explains why finding a tachyon among our four fields {\it is sufficient} to prove the existence of a systematic tachyon in the solutions.\\

The tachyon observed in the solutions is not always among the two fields $(\rho, \tau)$. Studying the $2\times 2$ blocks in section \ref{sec:method1}, we will see that it is not necessarily among pairs of fields within our four fields either. This implies that the tachyonic direction is generally a combination of all four field directions, as is a priori an eigenvector through a change of basis. We will then introduce a canonical field $\hat{t}_c$, eventually playing the role of the tachyon, that corresponds to a combination of our four fields
\bea
\frac{1}{M_p} \big( & c_{\rho} \rho \del_{\rho} + c_{\tau} \tau \del_{\tau} +  c_{\sigma_1} \sigma_1 \del_{\sigma_1}  + c_{\sigma_2} \sigma_2 \del_{\sigma_2} \big) \\
 =\ & \hat{c}_{\hat{\rho}} \del_{\hat{\rho}} + \hat{c}_{\hat{\tau}} \del_{\hat{\tau}} + \hat{c}_{\hat{\sigma}_1} \del_{\hat{\sigma}_1} + \hat{c}_{\hat{\sigma}_2} \del_{\hat{\sigma}_2} = \sqrt{\hat{c}_{\hat{\rho}}^2 + \hat{c}_{\hat{\tau}}^2 + \hat{c}_{\hat{\sigma}_1}^2 + \hat{c}_{\hat{\sigma}_2}^2}\ \del_{\hat{t}_c} \nn\ ,
\eea
where coefficients $c_{\phi^i}$ and $\hat{c}_{\hat{\phi}^i}$ are related to one another. We will show that this can be viewed, on very general grounds, as a change of basis between that of the original fields, and a basis where $\del_{\hat{t}_c}^2  V$ is a diagonal entry in the canonical mass matrix. We will then look for coefficients $c$'s such that
\beq
\del_{\hat{t}_c}^2  V < 0 \ , \label{condtc}
\eeq
or generalizations of this condition such as \eqref{genineqintro}. Thanks to the above lemmas, proving a condition \eqref{condtc} will be sufficient to prove the existence of a tachyon.

We will look at first for coefficients $c$'s that are universal, i.e.~fixed numbers, for which we prove the existence of a tachyon. In other words, a (universal) tachyon is then always present along the direction specified by four fixed values of $c$'s. We will develop tools and search for such values with method 2 in section \ref{sec:method2}. However, having in general such a universal tachyon is unlikely. The tachyon is expected to depend on the precise values of fluxes, curvatures, etc., i.e.~data of the solution. One may still hope to generically relate $c$'s to flux values, etc., but there can also be different classes of solutions, giving different origins to the tachyon, and different relations of the $c$'s to the data. We will find a way to capture such classes with method 3 in section \ref{sec:method3}, and this will be the most successful approach.

Finally, we note that the idea of having a tachyonic direction captured by coefficients along the various fields is similar to what was done in \cite{Junghans:2016uvg}, where coefficients $(t, s_i)$ were introduced. The present method 3 is then somehow related to what is described by figure 1 in that paper, put aside the discussion on the sgoldstino. While the analysis there was made on concrete solutions, the approach here is however more formal, and aims at some generality.\\

We started with the hope of proving the existence of a systematic tachyon. Giving up on a restriction to two fields, we could still hope to get a universal tachyon along some fixed direction in the four fields. But it seems that several classes of different tachyons exist among the four fields, depending on the solution. If we had all such classes at hand, we might still be able to prove the systematic existence of a tachyon, but it will not be the case here. Thanks to these three methods, we will still obtain a list of 13 sufficient conditions to have a tachyon, $C1$-$C13$. Those might be extended in the future to a complete set, allowing then to prove the existence of a systematic tachyon.

\subsection{Method 1: $2\times 2$ blocks}\label{sec:method1}

We consider here $2\times 2$ diagonal blocks of the canonical mass matrix $\hat{M}$, and study the possibility for them to admit a negative eigenvalue. As explained in section \ref{sec:summethods}, this would be sufficient to conclude on the existence of a tachyon. Some of the techniques presented here are inspired from \cite{Shiu:2011zt}, especially the simplification of ${\rm det} \hat{M}$ at a de Sitter extremum using first derivatives.

For a $2\times2$ matrix, the eigenvalues can be determined analytically. Considering only the two fields $(\rho, \tau)$, the minimal eigenvalue of $\hat{M}$ is given by
\bea
\lambda_- &= \frac{1}{2} \left({\rm Tr} \hat{M} - \sqrt{({\rm Tr} \hat{M})^2 - 4 {\rm det} \hat{M}}  \right) \\
&= \frac{1}{2} \left(\del_{\hat{\rho}}^2  V +\del_{\hat{\tau}}^2  V  - \sqrt{(\del_{\hat{\rho}}^2  V - \del_{\hat{\tau}}^2  V )^2 + 4 (\del_{\hat{\rho}} \del_{\hat{\tau}}  V)^2 }  \right) \ .\nn
\eea
One shows in general that
\beq
\lambda_- < 0\quad \Leftrightarrow \quad  {\rm Tr} \hat{M} < 0\ \ {\rm or} \ \ {\rm Tr} \hat{M} \geq 0 \ {\rm and} \ {\rm det} \hat{M} < 0 \ .\label{lambdacond}
\eeq
With the potential \eqref{potrhotau} for $p=5$, one shows at an extremum that
\beq
\frac{2}{M_p^2} \left( -\frac{5}{3} \rho \del_{\rho} V + \frac{11}{6} \tau \del_{\tau} V + \del_{\hat{\rho}}^2  V + \del_{\hat{\tau}}^2  V \right)_0 =-\frac{2}{3} {\cal R}_6 + \frac{14}{3} |H|^2 + \frac{1}{3}g_s^2 (|F_3|^2 + 10 |F_5|^2) \ ,\label{trrhotau}
\eeq
where here and in the rest of this subsection, we set $M_p=1$ in the $\del_{\hat{\phi}^i}$ derivatives for simplicity. This technique of combining quantities with first derivatives, that vanish at the extremum, will used extensively in this paper. From \eqref{trrhotau} and \eqref{entriessign}, we deduce
\beq
\mbox{De Sitter extremum:}\quad  {\rm Tr} \hat{M} =  \del_{\hat{\rho}}^2  V + \del_{\hat{\tau}}^2  V > 0 \ .
\eeq
From \eqref{lambdacond}, this implies that a tachyon in these two fields is equivalent to ${\rm det} \hat{M} < 0$. This quadratic condition looks at first sight difficult to use, but again, we can use first derivatives to simplify its entries, as done in \cite{Shiu:2011zt}. We do this here to remove $F_3$ and $H$ terms, and compute at an extremum
\bea
\frac{4}{M_p^4} {\rm det} \hat{M}|_0 =\, &\frac{4}{M_p^4} \left(2 \rho \del_{\rho} V + \del_{\hat{\rho}}^2  V\right)\left(-\frac{2}{3} \rho \del_{\rho} V + 2 \tau \del_{\tau} V + \del_{\hat{\tau}}^2  V \right) - \frac{4}{M_p^4} \left( \frac{2}{\sqrt{3}} \rho \del_{\rho} V  + \del_{\hat{\rho}} \del_{\hat{\tau}}  V \right)^2 \nn\\
=\, & \frac{2}{81} \Big(3 g_s^2 |F_5|^2 ( - 24 g_s^2 |F_5|^2 + g_s T_{10} + 72 g_s^2 |F_1|^2 )   -36 g_s^2 |F_1|^2 ( 4 g_s^2 |F_1|^2 - {\cal R}_6 ) \Big)\nn\\
 +\, & \frac{2}{81}  (6 {\cal R}_6 + g_s T_{10}) \left( 18 g_s^2 |F_1|^2 + 12 {\cal R}_6 + g_s T_{10}  \right) \ . \label{detM}
\eea
This expression can be used to check on a solution if ${\rm det} \hat{M}|_0 < 0$. But it remains a complicated one; let us try to find a simpler condition. With $F_5=0$, we deduce
\beq
\frac{4}{M_p^4} {\rm det} \hat{M}|_0 \leq \frac{2}{81}  (6 {\cal R}_6 + g_s T_{10})\ \left( 18 g_s^2 |F_1|^2 + 12 {\cal R}_6 + g_s T_{10}  \right) \ .
\eeq
One can further show
\beq
6 {\cal R}_6 + g_s T_{10} = 3 \frac{2}{M_p^2} ( \rho \del_{\rho} V + \frac{1}{2} \tau \del_{\tau} V )_0 + 3 (g_s^2 |F_3|^2 + 2 g_s^2 |F_5|^2 + 2 |H|^2) \geq 0 \ .
\eeq
We deduce our first sufficient condition for a tachyon
\beq
\boxed{C1}:\quad F_5=0 \ ,\quad 18 g_s^2 |F_1|^2 + 12 {\cal R}_6 + g_s T_{10} < 0 \ . \label{C1}
\eeq
This condition is however not verified by any of the solutions of \cite{Andriot:2020wpp}, even though several possess a tachyon in those two fields. Still, it is close to other conditions, \eqref{C4} and \eqref{C5}, verified by some solutions, that we will obtain differently.

In general, we also note the following
\bea
& \frac{2}{M_p^2} \left( -4 V -\frac{4}{3} \rho \del_{\rho} V - \tau \del_{\tau} V + \del_{\hat{\rho}}^2  V \right)_0 = 4 |H|^2 + g_s \frac{T_{10}}{36} + \frac{8}{3}g_s^2 |F_5|^2 \\
\Rightarrow\ & \mbox{De Sitter extremum:}\quad \del_{\hat{\rho}}^2  V > 0 \ .
\eea
This equation was obtained as (3.22) in  \cite{Andriot:2018ept}, up to the $F_5$ term corrected in \cite{Andriot:2020lea}. The sign of $\del_{\hat{\tau}}^2 V$ is more difficult to constrain. We obtain
\beq
\frac{2}{M_p^2} \left(3 \tau \del_{\tau} V + \del_{\hat{\tau}}^2 V \right)_0 = -2 {\cal R}_4 + g_s \frac{T_{10}}{12} \ .
\eeq
As indicated in section \ref{sec:summethods} with the lemmas, if one diagonal entry of $\hat{M}$ is negative, this is sufficient to conclude on a tachyon. We deduce a second sufficient condition for a tachyon
\beq
\boxed{C2}:\quad -2 {\cal R}_4 + g_s \frac{T_{10}}{12} < 0 \ . \label{C2}
\eeq

We now turn to other $2\times2$ diagonal blocks in the mass matrix $\hat{M}$: those involve $\del_{\hat{\sigma}_1}$ and $\del_{\hat{\sigma}_2}$ that we read from \eqref{reldersigma}, and we consider the potential \eqref{potential}. We first focus on the block along $(\hat{\tau}, \hat{\sigma}_1)$. Interestingly, one can show that $\del_{\hat{\sigma}_1}^2 V|_0 > 0$, with the following combination
\bea
& \frac{2}{M_p^2} \left( - \frac{8}{3} V  -\frac{1}{3} \rho \del_{\rho} V - \frac{5}{6} \tau \del_{\tau} V + \frac{1}{6} \sigma_1 \del_{\sigma_1} V + \frac{1}{6} \sigma_2 \del_{\sigma_2} V + \del_{\hat{\sigma}_1}^2 V \right)_0 \\
 =\, & \frac{1}{3}  \left( g_s^2 |F_1|^2 +  g_s^2 |F_3|^2 + 5 g_s^2 |F_5|^2 - 18 R_3 - g_s T_{10}^3 \right) >0  \ , \nn\\
\Rightarrow\ & \mbox{De Sitter extremum:}\quad \del_{\hat{\sigma}_1}^2  V > 0 \ ,
\eea
thanks to \eqref{entriessign}, where we recall any linear combination of $\sigma_1 \del_{\sigma_1} V$ and $\sigma_2 \del_{\sigma_2} V $ can be brought to one of $\del_{\hat{\sigma}_1} V $ and $\del_{\hat{\sigma}_2} V$. Now using the same formulas as above for $\lambda_-$, we are interested in the trace of this block: we show again that it is positive on a de Sitter extremum
\bea
& \frac{2}{M_p^2} \left( -\frac{1}{2} \rho \del_{\rho} V + \frac{19}{12} \tau \del_{\tau} V + \frac{1}{6} \sigma_1 \del_{\sigma_1} V + \frac{1}{6} \sigma_2 \del_{\sigma_2} V + \del_{\hat{\tau}}^2 V + \del_{\hat{\sigma}_1}^2 V  \right)_0\\
 =\, & \frac{1}{6}  \left( 4 g_s^2 |F_1|^2 + 5 g_s^2 |F_3|^2 + 14 g_s^2 |F_5|^2 + |H|^2 - 36 R_3 - 2 g_s T_{10}^3  \right) >0  \ .\nn
\eea
We then compute the determinant of this $2\times2$ block, and look for conditions to have it negative. We use as above combinations with first derivatives to simplify the expressions. However, we do not reach any satisfying expression in the end, and thus have difficulties concluding on this block. The same holds for the block $(\hat{\rho} , \hat{\sigma}_1)$, or a block with $\hat{\sigma}_2$. We also note that the sign of $\del_{\hat{\sigma}_2}^2 V|_0$ is hard to constrain. So we do not obtain any other useful conditions on tachyons from this first method. As explained in section \ref{sec:summethods}, we will now leave such $2\times 2$ blocks or field subspaces to consider tachyons possibly along the four fields.

\subsection{Interlude: preliminary formalism}\label{sec:interlude}

Methods 2 and 3 of sections \ref{sec:method2} and \ref{sec:method3} require some formalism, detailed in appendix \ref{ap:form}, and briefly introduced in section \ref{sec:summethods}. We summarize here the main results to be used. We consider the four fields $(\rho, \tau, \sigma_1, \sigma_2)$, with $4\times 4$ mass matrices. We prove for them in appendix \ref{ap:form} several key formulas. First, we prove the following sufficient condition for a tachyon
\begin{empheq}[innerbox=\fbox]{align}
\left( c_{\rho} \rho \del_{\rho} + c_{\tau} \tau \del_{\tau} +  c_{\sigma_1} \sigma_1 \del_{\sigma_1}  + c_{\sigma_2} \sigma_2 \del_{\sigma_2} \right)^2 V < 0  \label{conddel2}
\end{empheq}
for any real constants $(c_{\rho} , c_{\tau}, c_{\sigma_1} , c_{\sigma_2})$. Secondly, we prove the following implication
\begin{empheq}[innerbox=\fbox]{align}
& V + \left( b_{\rho} \rho \del_{\rho} + b_{\tau} \tau \del_{\tau} +  b_{\sigma_1} \sigma_1 \del_{\sigma_1}  + b_{\sigma_2} \sigma_2 \del_{\sigma_2} \right) V \nn\\
& \ \ + \left( c_{\rho} \rho \del_{\rho} + c_{\tau} \tau \del_{\tau} +  c_{\sigma_1} \sigma_1 \del_{\sigma_1}  + c_{\sigma_2} \sigma_2 \del_{\sigma_2} \right)^2 V \leq 0 \label{genineq}\\
\Rightarrow & \, - \sqrt{\hat{b}_{\hat{\rho}}^2 + \hat{b}_{\hat{\tau}}^2 + \hat{b}_{\hat{\sigma}_1}^2 + \hat{b}_{\hat{\sigma}_2}^2} \, \sqrt{2 \epsilon_V} + \left( \hat{c}_{\hat{\rho}}^2 + \hat{c}_{\hat{\tau}}^2 + \hat{c}_{\hat{\sigma}_1}^2 + \hat{c}_{\hat{\sigma}_2}^2 \right) \eta_V \leq -1 \nn
\end{empheq}
for arbitrary real constants $b_{\phi^i}$ and $c_{\phi^i}$, related to $\hat{c}_{\hat{\phi}^i}$ as
\beq
\hat{c}_{\hat{\rho}} =  \sqrt{\frac{3}{2}}\, c_{\rho}  \ ,\ \hat{c}_{\hat{\tau}} = \sqrt{2}\, c_{\tau} \ , \ \hat{c}_{\hat{\sigma}_1} = \sqrt{3}\, (c_{\sigma_1} + c_{\sigma_2}) \ , \ \hat{c}_{\hat{\sigma}_2} = 3\, (c_{\sigma_1} - c_{\sigma_2}) \ , \label{constants}
\eeq
and similarly for $\hat{b}_{\hat{\phi}^i}$. Conditions \eqref{genineq} are as well sufficient to have a tachyon (at a de Sitter extremum), and give the following upper bound on $\eta_V$
\beq
\eta_V|_0 \leq -\frac{1}{\hat{c}_{\hat{\rho}}^2 + \hat{c}_{\hat{\tau}}^2 + \hat{c}_{\hat{\sigma}_1}^2 + \hat{c}_{\hat{\sigma}_2}^2} = -\frac{1}{\frac{3}{2} c_{\rho}^2 + 2 c_{\tau}^2 + 3 (c_{\sigma_1}+c_{\sigma_2})^2 + 9 (c_{\sigma_1}-c_{\sigma_2})^2} \ . \label{boundetaV}
\eeq
Finally, we obtain the following sufficient condition for a tachyon, with $a\in [0,1]$, related to the above
\bea
& V + \left( b_{\rho} \rho \del_{\rho} + b_{\tau} \tau \del_{\tau} +  b_{\sigma_1} \sigma_1 \del_{\sigma_1}  + b_{\sigma_2} \sigma_2 \del_{\sigma_2} \right) V \label{genineqa}\\
& + \left( c_{\rho} \rho \del_{\rho} + c_{\tau} \tau \del_{\tau} +  c_{\sigma_1} \sigma_1 \del_{\sigma_1}  + c_{\sigma_2} \sigma_2 \del_{\sigma_2} \right)^2 V \leq (1-a) V \ , \nn
\eea
from which one deduces the following bound on $\eta_V$
\beq
\eta_V|_0 \leq  -\frac{a}{\frac{3}{2} c_{\rho}^2 + 2 c_{\tau}^2 + 3 (c_{\sigma_1}+c_{\sigma_2})^2 + 9 (c_{\sigma_1}-c_{\sigma_2})^2} \ . \label{boundetaVa}
\eeq

Methods 2 and 3 will require to look for constants $b$'s and $c$'s that lead to interesting conditions of the form \eqref{conddel2}, \eqref{genineq} or \eqref{genineqa}, and when possible deduce a bound on $\eta_V$. To obtain useful conditions, we analyse the linear combination $LC$ that appears in the above
\bea
LC= & \frac{2}{M_p^{2}} \Big( V + \left( b_{\rho} \rho \del_{\rho} + b_{\tau} \tau \del_{\tau} +  b_{\sigma_1} \sigma_1 \del_{\sigma_1}  + b_{\sigma_2} \sigma_2 \del_{\sigma_2} \right) V \\
&\phantom{\frac{2}{M_p^{2}} \Big(} + \left( c_{\rho} \rho \del_{\rho} + c_{\tau} \tau \del_{\tau} +  c_{\sigma_1} \sigma_1 \del_{\sigma_1}  + c_{\sigma_2} \sigma_2 \del_{\sigma_2} \right)^2 V \Big) \ , \nn
\eea
and write it down explicitly in \eqref{LC4}, using the four field potential \eqref{potential}. We explain in appendix \ref{ap:form} the strategy to obtain interesting conditions from it. Replacements of quantities with indefinite signs are in particular proposed. The restriction to two fields $(\rho, \tau)$ is also considered as
\bea
LC=\frac{2}{M_p^2} \Big( V +\, & b_{\rho} \rho \del_{\rho} V + b_{\tau} \tau \del_{\tau} V + (c_{\rho} \rho \del_{\rho} + c_{\tau} \tau \del_{\tau})^2\, V \Big)
\eea
and expressed explicitly in \eqref{LC}. All these tools will now be used in sections \ref{sec:method2} and \ref{sec:method3} for methods 2 and 3.

\subsection{Method 2: the universal tachyon}\label{sec:method2}

This method makes use of the formalism and tools presented in appendix \ref{ap:form} and summarized in section \ref{sec:interlude}. It consists in finding values of $b$'s and $c$'s coefficients that provide inequalities \eqref{conddel2}, \eqref{genineq} or \eqref{genineqa}, up to simple conditions on solutions data. This implies the existence of a tachyon, and sometimes allows to get a bound on $\eta_V$. The specificity of this method is that the $b$'s and $c$'s coefficients are fixed constants. The hope is to find a tachyonic direction which is always the same, in other words a universal tachyon. We will not manage to find such a set of constants, not even for the solutions 1-17 of \cite{Andriot:2020wpp}, but we will still obtain interesting conditions. We start with the two fields $(\rho, \tau)$ as a warm-up, and then turn to the four fields.

\subsubsection{Two fields}

For the two fields $(\rho, \tau)$, we already know from section \ref{sec:method1} an analytic necessary and sufficient condition for the existence of a tachyon at a de Sitter extremum: ${\rm det} \hat{M}|_0 <0$, with the expression given in \eqref{detM}. This condition is however complicated, and quadratic in the entries of the potential. Finding a linear sufficient condition could be useful, as for instance \eqref{C1}. We find here such conditions using the linear combination \eqref{LC}.

We first verify that there exists no set of coefficients allowing to prove that each line of \eqref{LC} is negative. This is consistent with the fact that not all solutions admit a tachyon in $(\rho,\tau)$. A first set of coefficients giving an interesting combination is the following
\bea
& b_{\rho} = \frac{65}{8} \ ,\ b_{\tau} = \frac{325}{16}\ ,\ c_{\rho} = -1 \ ,\ c_{\tau} = -\frac{5}{2}\ ,\\
\Rightarrow \  LC\, = &\ \frac{95}{4} g_s^2 |F_5|^2\, \tau^{-4}\rho^{-2} + \frac{79}{8} \Big(g_s^2 |F_3|^2\, \tau^{-4} +  \frac{94}{79}\, \R_6\, \tau^{-2}\rho^{-1} \Big) \ . \nn
\eea
We deduce with $LC|_0 \leq 0$ the following sufficient condition for a tachyon at a de Sitter extremum
\beq
\boxed{C3}:\quad -\R_6 \geq \frac{79}{94} g_s^2 |F_3|^2 + \frac{95}{47} g_s^2 |F_5|^2 \ . \label{C3}
\eeq
This is interesting given the existence constraint (see e.g.~\cite{Andriot:2016xvq})
\beq
\mbox{De Sitter extremum:}\  -\R_6 \geq \frac{1}{2} g_s^2 |F_3|^2 + g_s^2 |F_5|^2 \ .
\eeq
Indeed, in all known solutions, $F_5=0$, while one has $\frac{79}{94} \approx \frac{1}{1.18987} $. Then, avoiding the tachyon \eqref{C3} with $F_5=0$ only leaves a little window when comparing $-\R_6$ and $|F_3|^2$. All solutions of \cite{Andriot:2020wpp} actually fall in this window since none of them obey \eqref{C3}. Still one deduces from this inequality the following bound thanks to \eqref{boundetaV}
\beq
\eta_V|_0 \leq -\frac{1}{14} \approx -0.0714286 \ .
\eeq

We now rather look for coefficients giving combinations that capture many if not all solutions of \cite{Andriot:2020wpp}. More precisely, as displayed in Table 2 of \cite{Andriot:2020wpp}, solutions 1-9, 11 and 15 admit a tachyon in $(\rho,\tau)$. Solution 15 has the smallest $|\eta_V|$ value when restricting to those two fields, and solution 1 the second smallest. We find the following interesting coefficient set
\bea
& b_{\rho} = \frac{1981426373}{30408704} \ ,\ b_{\tau} = \frac{3908638849}{60817408}\ ,\ c_{\rho} = -\frac{851}{256} \ ,\ c_{\tau} = -4\ ,\\
\Rightarrow \  LC\, = &  - \frac{145590343}{7602176} g_s^2 |F_1|^2\, \tau^{-4}\rho^{2} - \frac{1120385}{30408704} g_s^2 |F_3|^2\, \tau^{-4} + \frac{962118829}{15204352} g_s^2 |F_5|^2\, \tau^{-4}\rho^{-2} \nn\\
&  + \frac{980053627}{15204352} \R_6\, \tau^{-2}\rho^{-1} + \frac{23903747}{3801088} g_s T_{10}\, \tau^{-3}\rho^{-\frac{1}{2}} \\
& \hspace{-0.7in} < \,  \frac{962118829}{15204352} g_s^2 |F_5|^2\, \tau^{-4}\rho^{-2} + \frac{23903747}{3801088} \Big(-3 g_s^2 |F_1|^2\, \tau^{-4}\rho^{2}+ 10.25\, \R_6\, \tau^{-2}\rho^{-1} + g_s T_{10}\, \tau^{-3}\rho^{-\frac{1}{2}} \Big) \nn
\eea
where we used $F_1 \neq 0$ (see \eqref{entriessign}) for the inequality. We deduce the following sufficient condition for a tachyon at a de Sitter extremum
\beq
\boxed{C4}:\quad F_5 = 0 \ ,\quad -3 g_s^2 |F_1|^2 + 10.25\, \R_6 + g_s T_{10} \leq 0  \ . \label{C4}
\eeq
Interestingly, \eqref{C4} is satisfied by all solutions but the 15, i.e.~solutions 1-9 and 11 of \cite{Andriot:2020wpp}. Note also that one can show the following existence requirement
\beq
\mbox{De Sitter extremum:}\ -3 g_s^2 |F_1|^2 + 2\, \R_6 + g_s T_{10} > 0 \ ,
\eeq
which indicates again a tight window for these solutions. Finally, we deduce from these coefficients the bound
\beq
\eta_V|_0 \leq -\frac{131072}{6366907} \approx -0.0205864 \ ,
\eeq
which is more than a factor of 10 smaller than the values of $|\eta_V|$ of the solutions.

We get a much better bound with the following set
\bea
& b_{\rho} = \frac{2893}{496} \ ,\ b_{\tau} = \frac{4217}{992}\ ,\ c_{\rho} = 1 \ ,\ c_{\tau} = 1\ ,\\
\Rightarrow \  LC\, = &  - \frac{21}{124} g_s^2 |F_1|^2\, \tau^{-4}\rho^{2} - \frac{1}{496} g_s^2 |F_3|^2\, \tau^{-4} + \frac{1033}{248} g_s^2 |F_5|^2\, \tau^{-4}\rho^{-2} \\
&  + \frac{1075}{248} \R_6\, \tau^{-2}\rho^{-1} + \frac{25}{62} g_s T_{10}\, \tau^{-3}\rho^{-\frac{1}{2}} \nn\\
< &\,  \frac{1033}{248} g_s^2 |F_5|^2\, \tau^{-4}\rho^{-2} + \frac{25}{62} \Big( 10.75\, \R_6\, \tau^{-2}\rho^{-1} + g_s T_{10}\, \tau^{-3}\rho^{-\frac{1}{2}} \Big) \ , \nn
\eea
which gives the following sufficient condition for a tachyon
\beq
\boxed{C5}:\quad F_5 = 0 \ ,\quad 10.75\, \R_6 + g_s T_{10} \leq 0  \ . \label{C5}
\eeq
This is however only verified by solutions 2-9 and 11 of \cite{Andriot:2020wpp}, i.e.~solutions 1 and 15 are missing. This set of coefficient leads to the following interesting bound
\beq
\eta_V|_0 \leq -\frac{2}{7} \approx -0.285714 \ ,
\eeq
in agreement with the values obtained for the solutions of \cite{Andriot:2020wpp}.

Solution 15 has not been captured by any of the above conditions. We find the following combination
\bea
& b_{\rho} =  b_{\tau} = 0\ ,\ c_{\rho} = -\frac{13}{4} \ ,\ c_{\tau} = -\frac{7}{4}\ ,\\
\Rightarrow \  LC\, = &  \frac{5}{8} g_s^2 |F_1|^2\, \tau^{-4}\rho^{2} + 25 g_s^2 |F_3|^2\, \tau^{-4} + \frac{733}{8} g_s^2 |F_5|^2\, \tau^{-4}\rho^{-2} +  \frac{2825}{32} |H|^2\, \tau^{-2}\rho^{-3} \\
&  - \frac{745}{16} \R_6\, \tau^{-2}\rho^{-1} - \frac{3089}{384} g_s T_{10}\, \tau^{-3}\rho^{-\frac{1}{2}} \ , \nn
\eea
giving the following sufficient condition for a tachyon
\beq
\boxed{C6}:\quad g_s^2 \left( \frac{5}{8} |F_1|^2 + 25 |F_3|^2 + \frac{733}{8}  |F_5|^2 \right) + \frac{2825}{32} |H|^2 - \frac{745}{16} \R_6 - \frac{3089}{384} g_s T_{10} \leq 0  \ . \label{C6}
\eeq
This condition is verified by solution 15, and surprisingly solution 4, but none of the others. This set gives the following bound
\beq
\eta_V|_0 \leq -\frac{32}{703} \approx -0.0455192 \ ,
\eeq
in agreement with the value for solution 15. This situation indicates already the difficulty of finding a universal tachyon, and rather hints at the possibility of different tachyonic directions. We note that in this last set of coefficients, $|c_{\rho}| > |c_{\tau}|$, which is not the case so far for any of the other conditions. This is a concrete illustration of different tachyonic directions.

\subsubsection{Four fields}

We turn to the case of four fields and use the linear combination \eqref{LC4}. We first present few interesting sets of coefficients that give conditions capturing a maximum of the 17 solutions of \cite{Andriot:2020wpp}. For simplicity, we only give the $LC$ at extrema.\footnote{In expressions given, $T_{10}^{2}$ of indefinite sign has been traded for $T_{10}-T_{10}^1 - T_{10}^3$, of definite signs (see \eqref{entriessign}).} A first interesting set is the following
\bea
& b_{\sigma_1}= b_{\sigma_2} = 27 \ ,\ b_{\rho}= 108 \ ,\ b_{\tau}= \frac{229}{8} \ , \ c_{\sigma_1}=c_{\sigma_2}=1 \ ,\ c_{\rho}=\frac{7}{2} \ ,\ c_{\tau}=\frac{9}{2} \nn\\
\Rightarrow\ & LC|_0= \frac{439}{4} g_s^2 |F_1|^2 + \frac{421}{4} g_s^2|F_3|^2 + \frac{439}{4} g_s^2|F_5|^2 - 72 R_3 - \frac{1675}{96} g_s T_{10} + \frac{3}{2} g_s T_{10}^3
\ ,
\eea
from which we get the sufficient condition for a tachyon
\beq
\boxed{C7}:\quad \frac{439}{4} g_s^2 |F_1|^2 + \frac{421}{4} g_s^2|F_3|^2 + \frac{439}{4} g_s^2|F_5|^2 - 72 R_3 - \frac{1675}{96} g_s T_{10} + \frac{3}{2} g_s T_{10}^3 \leq 0  \ . \label{C7}
\eeq
This condition is obeyed by all 17 solutions of \cite{Andriot:2020wpp} except solutions 9, 10, 14: those three actually have the smallest values of $|\eta_V|$, solution 14 having the smallest. These values remain far away from the corresponding bound given by
\beq
\eta_V \leq -\frac{8}{567} \approx -0.0141093 \ .
\eeq
A second set is the following
\bea
& b_{\sigma_1}=\frac{194}{5} \ ,\ b_{\sigma_2} = \frac{1128}{25} \ ,\ b_{\rho}= 184 \ ,\ b_{\tau}= \frac{3773}{50} \ , \ c_{\sigma_1}=1 \ ,\ c_{\sigma_2}=\frac{6}{5} \ ,\ c_{\rho}=4 \ ,\ c_{\tau}= 7 \nn\\
\Rightarrow\ & LC|_0= \frac{12367}{50} g_s^2 |F_1|^2 + \frac{12079}{50} g_s^2 |F_3|^2 + \frac{12439}{50} g_s^2 |F_5|^2 - \frac{612}{25} |H^{(0)_1}|^2 + \frac{72}{5} R_1 - \frac{432}{25} R_2 \nn\\
&\phantom{LC|_0} - \frac{432}{5} R_3 - \frac{12269}{300} g_s T_{10} + \frac{94}{25} g_s T_{10}^1 +  \frac{204}{25} g_s T_{10}^3
\ .
\eea
This complicated combination gives the sufficient condition for a tachyon
\bea
\boxed{C8}:\quad & \frac{12367}{50} g_s^2 |F_1|^2 + \frac{12079}{50} g_s^2 |F_3|^2 + \frac{12439}{50} g_s^2 |F_5|^2 - \frac{612}{25} |H^{(0)_1}|^2 + \frac{72}{5} R_1 - \frac{432}{25} R_2 \nn \\
 & - \frac{432}{5} R_3 - \frac{12269}{300} g_s T_{10} + \frac{94}{25} g_s T_{10}^1 +  \frac{204}{25} g_s T_{10}^3 \leq 0  \ , \label{C8}
\eea
obeyed by solutions 1-12 of \cite{Andriot:2020wpp}. It gives the following bound
\beq
\eta_V \leq -\frac{25}{3422} \approx -0.00730567 \ .
\eeq
The only solution not captured so far is thus solution 14, which is special on several aspects. We found for it the following combination
\bea
& b_{\sigma_1}=\frac{107}{160} \ ,\ b_{\sigma_2} = -\frac{67}{100} \ ,\ b_{\rho}= \frac{2257}{400} \ ,\ b_{\tau}= \frac{373}{160} \ , \ c_{\sigma_1}=\frac{1}{8} \ ,\ c_{\sigma_2}=-\frac{1}{10} \ ,\ c_{\rho}=\frac{1}{2} \ ,\ c_{\tau}= \frac{5}{4} \nn\\
\Rightarrow\ & LC|_0= \frac{3873}{400} g_s^2|F_1|^2 + \frac{667}{80} g_s^2|F_3|^2 + \frac{1579}{200} g_s^2|F_5|^2 - \frac{1107}{400} |H^{(2)_1}|^2 - \frac{81}{40} R_1 - \frac{81}{50} R_2 \nn\\
&\phantom{LC|_0}  + \frac{9}{10} R_3 - \frac{257}{240} g_s T_{10} - \frac{189}{400} g_s T_{10}^1 - \frac{3}{50} g_s T_{10}^3
\ ,
\eea
giving the sufficient condition for a tachyon
\bea
\boxed{C9}:\quad & \frac{3873}{400} g_s^2|F_1|^2 + \frac{667}{80} g_s^2|F_3|^2 + \frac{1579}{200} g_s^2|F_5|^2 - \frac{1107}{400} |H^{(2)_1}|^2 - \frac{81}{40} R_1 - \frac{81}{50} R_2 \nn\\
 &  + \frac{9}{10} R_3  - \frac{257}{240} g_s T_{10} - \frac{189}{400} g_s T_{10}^1 - \frac{3}{50} g_s T_{10}^3 \leq 0  \ , \label{C9}
\eea
only satisfied by solution 14 of \cite{Andriot:2020wpp}. It gives the following bound
\beq
\eta_V \leq -\frac{400}{1583} \approx -0.252685 \ .
\eeq

We see, as for two fields, that it is difficult to capture all solutions with a single universal tachyon. The different signs of $c_{\sigma_1}, c_{\sigma_2}$ in the above also underlines the different tachyonic directions. More generally, we have tried a direct search a universal tachyon by having all lines in $LC$ \eqref{LC4} negative. As explained there, one should however have only entries of definite sign. This situation can be reached with $T_{10}^2=T_{10} - T_{10}^1 - T_{10}^3$ and replacing the $R_{4,5,6}$ by $X_{4,5,6}$, as detailed below \eqref{LC4}. We then tried to find a set of coefficients providing a purely negative right-hand side. This search was computationally challenging, and we could not conclude, but it has most likely no solution.

What is more likely is that there are several different classes corresponding to a few different tachyons, possibly covering the whole parameter space of de Sitter solutions, thus leading to the systematic existence of a tachyon. The various conditions found above rather hint at such a situation. To cover completely the parameter space, we should find several complementary combinations or conditions, that e.g.~once added, are of definite sign. We looked for other combinations, close to the opposite of the three above, but could not find any relevant ones. At this stage, it appears difficult to cover all possible solutions.\\

Last but not least, we obtained the following interesting combination
\bea
& b_{\sigma_1}= b_{\sigma_2} = \frac{1}{4} \ ,\ b_{\rho}= \frac{1}{2} \ ,\ b_{\tau}= \frac{3}{4} \ , \ c_{\sigma_1}=c_{\sigma_2}=\frac{1}{6\sqrt{2}} \ ,\ c_{\rho}=\frac{1}{3\sqrt{2}} \ ,\ c_{\tau}=\frac{1}{2\sqrt{2}} \nn\\
\Rightarrow\ & LC= -R_3\ \tau^{-2}\rho^{-1} \sigma_1^{4} \sigma_2^{4} \geq 0 \ . \label{LCint0}
\eea
Even though this is not negative, this is remarkably simple. We can use this combination with the less demanding condition \eqref{genineqa}: let us consider the sufficient condition for a tachyon $LC|_0 \leq  (1-a) \frac{2}{M_p^{2}} V|_0$, i.e.~here
\beq
\boxed{C10}:\quad -R_3 \leq (1-a) \frac{1}{2} {\cal R}_4 \ , \ \ a \in [0,1] \ , \label{C10}
\eeq
we then deduce from \eqref{boundetaVa} the following bound
\beq
\eta_V \leq  - 2 a \ . \label{etaa}
\eeq
This is obeyed for all 17 solutions of \cite{Andriot:2020wpp} except solutions 2, 7, 9, 14. For these 13 solutions, we can pick $a = \frac{2}{3}$, giving the bound
\beq
\eta_V \leq - \frac{4}{3} \ .
\eeq
This value is certainly verified by the solutions, and for once of the same order of the actual values. Let us mention that the value for this bound has been obtained as well in \cite{Junghans:2016abx}, even though at first sight, we see no obvious relation between the two derivations and settings. We come back in section \ref{sec:under} to the ratio of $-R_3$ and $\frac{1}{2}{\cal R}_4$, which plays a role in further tachyon conditions. This last combination gives two ideas to be used in the following: using the less demanding condition \eqref{genineqa} or even \eqref{conddel2}, and leaving some free parameters (here the $a$) while still being sure of having a tachyon, to capture more solutions.

\subsection{Method 3: classes of non-universal tachyons}\label{sec:method3}

This method uses the tools of appendix \ref{ap:form} and section \ref{sec:interlude}. Instead of looking for a universal tachyon, i.e.~a tachyonic direction with fixed coefficients, we will leave some coefficients free, say $c_{\tau}, c_{\rho}$, and find a condition ensuring that a certain range of their values guarantees the existence of a tachyon. This freedom will allow us to capture more solutions in one ``class'' of tachyons. In addition, we will use the condition \eqref{conddel2} for a tachyon, with first derivatives to simplify the expressions, or in other words condition \eqref{genineqa} with $a=0$. This is less demanding, allowing to capture more cases, but it prevents us from getting a bound on $\eta_V$.

More concretely, we recall from \eqref{VR4} that
\beq
LC|_0( c_{\sigma_1}= c_{\sigma_2}= c_{\rho}= c_{\tau}= 0)  = \frac{1}{2} {\cal R}_4 \ .
\eeq
So removing $\frac{1}{2} {\cal R}_4$ from $LC|_0$ amounts to leave only first and second derivatives of the potential, which boils down to \eqref{genineqa} with $a=0$. The method then consists in fixing all coefficients but two, and bring $LC|_0$ to the form
\beq
LC|_0 = {\cal P}(c_{\tau}, c_{\rho}) + \frac{1}{2} {\cal R}_4\ ,\qquad {\cal P}(c_{\tau}, c_{\rho}) = c_{\tau}^2 A_0 + c_{\rho}^2 B_0  -c_{\rho} c_{\tau} C_0 \ , \label{LCP}
\eeq
with a polynomial ${\cal P}$. A sufficient condition for a tachyon is then that ${\cal P} <0$. In the following, we will have $A_0> 0$, which means that there exist $c_{\tau}, c_{\rho}$ for which ${\cal P} <0$ if and only if
\begin{empheq}[innerbox=\fbox]{align}
4A_0 B_0 - C_0^2 <0
\end{empheq}
This is a sufficient condition for a tachyon.

We apply successfully this method to the following combination
\bea
& b_{\sigma_1}=b_{\sigma_2} = \frac{1}{24}  (4 - 27 c_{\rho}^2 - 36 c_{\rho} c_{\tau} + 52 c_{\tau}^2) \ ,\ b_{\rho}= \frac{1}{3}+ \frac{1}{12} (27 c_{\rho}^2 + 36 c_{\rho} c_{\tau} - 20 c_{\tau}^2) \ ,\\
& b_{\tau}= \frac{1}{8}  (4 - 9 c_{\rho}^2 - 12 c_{\rho} c_{\tau} + 28 c_{\tau}^2) \ ,\ c_{\sigma_1}= c_{\sigma_2}= \frac{1}{4} (-c_{\rho} + 2 c_{\tau}) \nn\\
\Rightarrow\ & LC|_0 = (3 c_{\rho} - 2 c_{\tau})^2 g_s^2 |F_1|^2 -
 \frac{9}{2} (c_{\rho} - 2 c_{\tau})^2 R_3 + \left(\frac{1}{2} -\frac{1}{4} (3 c_{\rho} + 2c_{\tau})^2\right) {\cal R}_4 \label{LCint} \ ,
\eea
where we replaced $R_4$ using equation \eqref{eqR4R3}, introducing this way ${\cal R}_4$. Note that \eqref{LCint} boils down to \eqref{LCint0} for $c_{\tau}=\frac{1}{2 \sqrt{2}}$, $c_{\rho} = \frac{2}{3} c_{\tau}$. But leaving those coefficients free, we write $LC|_0$ as \eqref{LCP} with
\bea
& A_0= 4 g_s^2|F_1|^2  - {\cal R}_4 - 18 R_3 \ ,\ B_0 = \frac{9}{4}(4 g_s^2|F_1|^2  -  {\cal R}_4 - 2 R_3) \ ,\\
& C_0=3(4 g_s^2|F_1|^2 +  {\cal R}_4 - 6 R_3 ) \ , \nn
\eea
where $|H|^2 + g_s^2|F_5|^2 =g_s^2 |F_1|^2 -  {\cal R}_4 \geq 0$ (see e.g.~(3.10) of \cite{Andriot:2016xvq}) ensures that $A_0, B_0, C_0 \geq 0$. We then compute
\bea
4 A_0 B_0 - C_0^2  & = 144\, \left( -2  R_3 (g_s^2 |F_1|^2  - {\cal R}_4) - g_s^2 |F_1|^2 {\cal R}_4 \right) \label{tachyoncond} \\
& = 144\, \left( -2  R_3 (|H|^2+ g_s^2|F_5|^2) - g_s^2 |F_1|^2 {\cal R}_4 \right) \ .\nn
\eea
Despite the comparatively simple form of this quantity, it is hard to get any definite sign for it: bounding $-R_3$ from above is difficult. We deduce the following sufficient condition for a tachyon
\beq
\boxed{C11}:\quad  -2  R_3 (g_s^2 |F_1|^2  - {\cal R}_4) - g_s^2 |F_1|^2 {\cal R}_4  < 0 \ . \label{C11}
\eeq
Very interestingly, all 17 solutions of \cite{Andriot:2020wpp} except the special one 14 satisfy this condition. This is the best condition obtained so far. A particularity is also that it is not linear anymore.

Let us briefly illustrate the resulting tachyonic directions. One verifies that
\beq
{\cal P} = 0 \Leftrightarrow 2 A_0 c_{\tau} = c_{\rho} \left(C_0 \pm \sqrt{C_0^2 - 4A_0 B_0}\right)\ {\rm or} \ 2 B_0 c_{\rho} = c_{\tau} \left(C_0 \pm \sqrt{C_0^2 - 4A_0 B_0}\right) \ ,
\eeq
and $ {\cal P} < 0$ between these values. Therefore for instance, ${\cal P} < 0$ for
\beq
 2 A_0 c_{\tau} = c_{\rho} C_0 \Rightarrow  {\cal P} =c_{\rho}^2\, \frac{4 A_0 B_0 - C_0^2}{4 A_0} \quad {\rm or} \quad  2 B_0 c_{\rho} = c_{\tau} C_0 \Rightarrow  {\cal P} =c_{\tau}^2\, \frac{4 A_0 B_0 - C_0^2}{4 B_0} \ .
\eeq
The tachyonic directions are then solution dependent, as seen here through the dependence of $\frac{c_{\rho}}{c_{\tau}}$ on solution data. The tachyon is then not universal as for method 2. We rather have here a class of tachyons captured by the condition \eqref{C11}.

We tried to find complementary conditions to the one above, without success. At least we could find one for solution 14, not included above. We considered
\bea
& b_{\sigma_2} = -\frac{1}{6} - b_{\sigma_1} + 450 c_{\sigma_1}^2 - 96 c_{\sigma_1} c_{\tau} + \frac{16}{3} c_{\tau}^2 \ ,\ b_{\rho}= \frac{1}{3} + 864 c_{\sigma_1}^2 - 216 c_{\sigma_1} c_{\tau} + \frac{40}{3} c_{\tau}^2 \ ,\\
& b_{\tau}= \frac{1}{2} - 720 c_{\sigma_1}^2 + 132 c_{\sigma_1} c_{\tau} - 4 c_{\tau}^2 \ ,\ c_{\sigma_2}= -2 c_{\sigma_1} \ ,\ c_{\rho}= 2 (-8 c_{\sigma_1} + c_{\tau}) \nn
\eea
and replaced $R_{5,6}$ in terms of ${\cal R}_4$ using \eqref{eqR5}, \eqref{eqR6}. This gave us $LC|_0$ of the form \eqref{LCP}, however in terms of different coefficients
\bea
& LC|_0 = {\cal P}(c_{\tau}, c_{\sigma_1}) + \frac{1}{2} {\cal R}_4\ ,\nn\\
& {\cal P}(c_{\tau}, c_{\sigma_1}) = c_{\tau}^2 A_0 + c_{\sigma_1}^2 B_0  -c_{\sigma_1} c_{\tau} C_0 \ ,\nn\\
& A_0 = 16 (g_s^2|F_1|^2 - {\cal R}_4) \ ,\ C_0 = 24(16 g_s^2|F_1|^2 - 6 g_s^2|F_5|^2 - 6 |H^{(0)_1}|^2 - 11 {\cal R}_4) \ ,\label{A014}\\
& B_0 = 36 (64 g_s^2|F_1|^2 + 3 g_s^2|F_3|^2 - 33 g_s^2|F_5|^2 - 33 |H^{(0)_1}|^2 - 6 R_1 - 12 R_2 + 4 R_3 - 37 {\cal R}_4 - g_s T_{10}^1) \ .\nn
\eea
We may simplify $B_0$ but its sign is not clear. Still, $A_0 \geq 0$ so we get as above the following sufficient condition for a tachyon
\beq
\boxed{C12}:\quad 4A_0 B_0 - C_0^2 < 0 \ \mbox{with values of}\ \eqref{A014} \ . \label{C12}
\eeq
This condition is verified by solution 14 of \cite{Andriot:2020wpp} (and few others, see Table \ref{tab:sum}). It is however less simple than \eqref{C11}, and is not especially complementary. Furthermore, we could not get a complete set of conditions starting from those two. Analogous conditions with $T_{10}^2$ instead $T_{10}^1$ could still be obtained, though not helping more. Despite the generality of condition \eqref{C11}, we face once again the possibility that there exists several different classes of tachyons, that we do not manage to all identify. In the next section, we change strategy and try to find directly new solutions in different classes. They could carry new interesting physics.

\section{New de Sitter solutions}\label{sec:sol}

Using the same tools and strategy as in \cite{Andriot:2020wpp}, we now look for new de Sitter solutions of type IIB supergravity with intersecting $O_5/D_5$. We however ask in our search for an additional condition w.r.t.~\cite{Andriot:2020wpp}, namely that the solution violates condition $C11$ given in \eqref{C11}. The latter was so far the most general sufficient condition found to have a tachyon. Indeed, it captures all solutions of \cite{Andriot:2020wpp} but one, solution 14. Asking for a violation of this condition will give us solutions with new interesting physics: we will indeed obtain new 6d group manifolds, and higher $\eta_V$ values. All solutions found are however still tachyonic (except one, on a non-compact manifold), and we will again study this instability. We first describe the solutions obtained in section \ref{sec:descr}, and give some characterisation of their tachyons, before discussing some analytic interpretation in section \ref{sec:under}. The complete list of new solutions is given in appendix \ref{ap:sol}.

\subsection{Description}\label{sec:descr}

We present here 10 new de Sitter solutions, obtained by asking for a violation of \eqref{C11}, i.e.~the absence of such a tachyon. We label these solutions 18 to 27, such that solutions 1 to 17 refer to those of \cite{Andriot:2020wpp}. A first concise summary of the solutions properties, namely the underlying Lie algebra of the group manifold and the value of $\eta_V$ obtained for four fields, is given in Table \ref{tab:sol}. The complete data of the solutions is given in appendix \ref{ap:sol}.

\begin{table}[h]
  \begin{center}
    \begin{tabular}{|c||c|c|c|c|}
    \hline
     & & & & \\[-8pt]
\mbox{Solutions}  & 18 & 19 & 20, 21 & 22-27 \\[3pt]
    \hhline{=====}
     & & & & \\[-8pt]
Algebra of $\mmm$ & $\mathfrak{so}(2,1) \oplus {\rm Heis}_3$ & ? & $\mathfrak{so}(3) \oplus {\rm Heis}_3$ & $\mathfrak{g}_{3.4}^{-1} \oplus \mathfrak{g}_{3.5}^{0}$ \\[3pt]
    \hhline{-||----}
     & & & & \\[-8pt]
$-\eta_V\ \text{(4 fields)}$ & $-3.7926$ & 0.12141 & 1.3624, 1.7813 & [0.90691, 1.2253] \\[3pt]
    \hhline{-----}
    \end{tabular}
     \caption{Summary of the 10 new de Sitter solutions: Lie algebra underlying the 6d group manifold $\mmm$, and value of $-\eta_V$ computed with the four fields scalar potential \eqref{potential}.}\label{tab:sol}
  \end{center}
\end{table}

Let us first clarify notations of the Lie algebras. Useful references for those include \cite{Bock, Andriot:2010ju, Danielsson:2011au}, and more references can be found e.g.~in \cite{Andriot:2020lea}. The 6d group manifold $\mmm$ has been so far described through the structure constants $f^a{}_{bc}$, which actually correspond to an underlying Lie algebra. Identifying the latter is a first necessary step to know the geometry of the manifold $\mmm$, in particular and crucially, whether it is compact. This identification can however be difficult, since it typically requires to find an isomorphism, or change of basis, resulting in another set of structure constants for which we recognise the algebra; see \cite{Andriot:2020wpp} and the examples below for more details. The more structure constants we initially get, the less easy this task is. And indeed, for solution 19, we got all allowed structure constants non-zero, preventing us from successfully identifying the algebra (the same was true for many solutions of \cite{Andriot:2020wpp}). Once the algebra is identified, one should determine the compactness. The semi-simple algebras $\mathfrak{so}(2,1)$ and $\mathfrak{so}(3,1)$ are non-compact, and lead to a non-compact $\mmm$, while $\mathfrak{so}(3)$ (the real $\mathfrak{su}(2)$) is compact and so is $\mmm$. We will encounter three-dimensional solvable algebras, namely ${\rm Heis}_3$ (nilpotent), $\mathfrak{g}_{3.4}^{-1}$ (also known as $\mathfrak{iso}(1,1)$, with $f^2{}_{35} f^3{}_{25} > 0$) and $\mathfrak{g}_{3.5}^{0}$ (also known as $\mathfrak{iso}(2)$, with $f^2{}_{35} f^3{}_{25} < 0$). All admit lattices and can thus lead to a compact $\mmm$. In \cite{Andriot:2020wpp, Andriot:2020vlg} were identified 6 algebras among the 17 solutions, 4 of which gave a compact $\mmm$ and 2 non-compact. The 6d algebras indicated in Table \ref{tab:sol} did not appear there, hinting at new physics in our new solutions.

\begin{figure}[H]
\begin{center}
\begin{subfigure}[H]{0.45\textwidth}
\includegraphics[width=\textwidth]{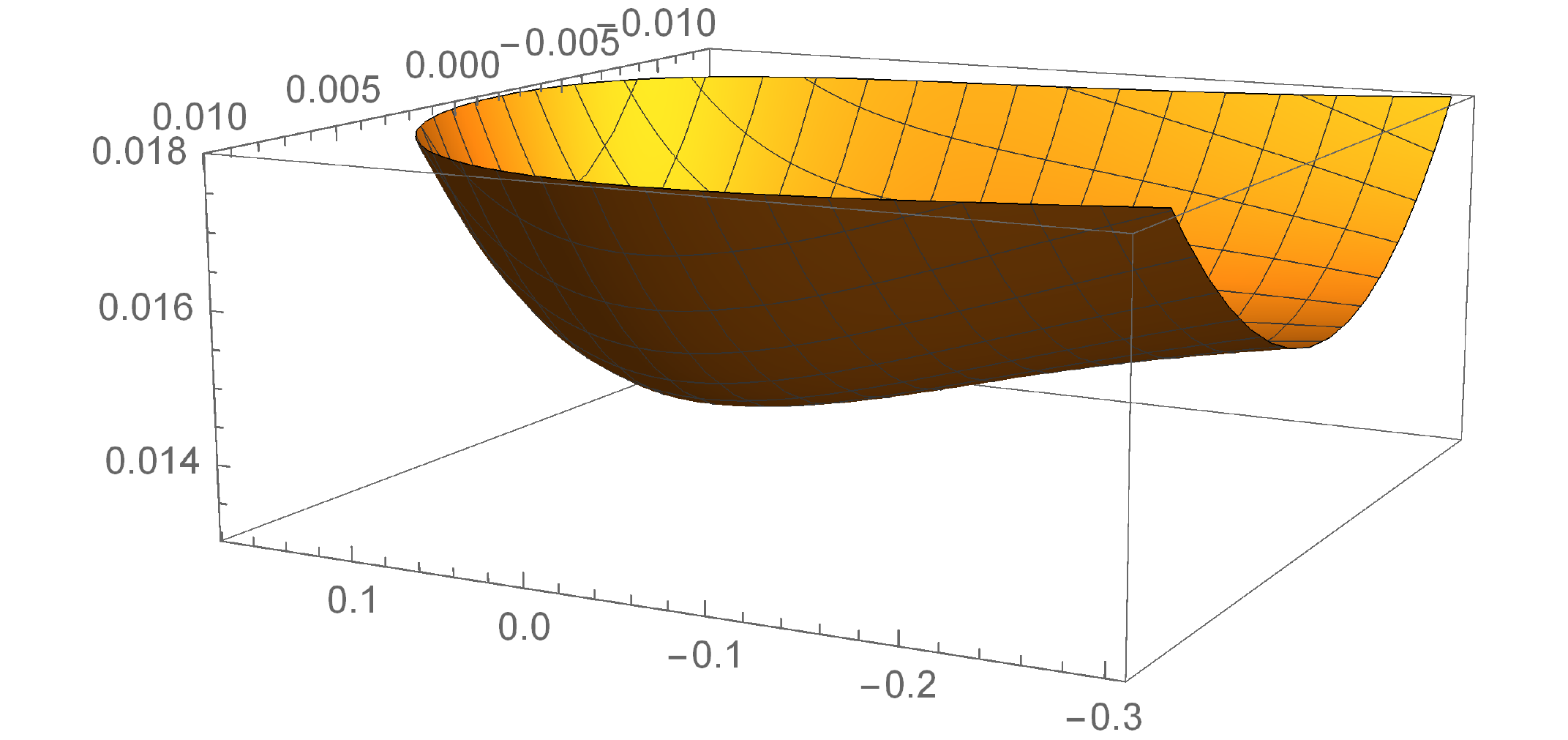}\caption{Solution 18}\label{fig:sol18}
\end{subfigure}
\qquad \ 
\begin{subfigure}[H]{0.40\textwidth}
\includegraphics[width=\textwidth]{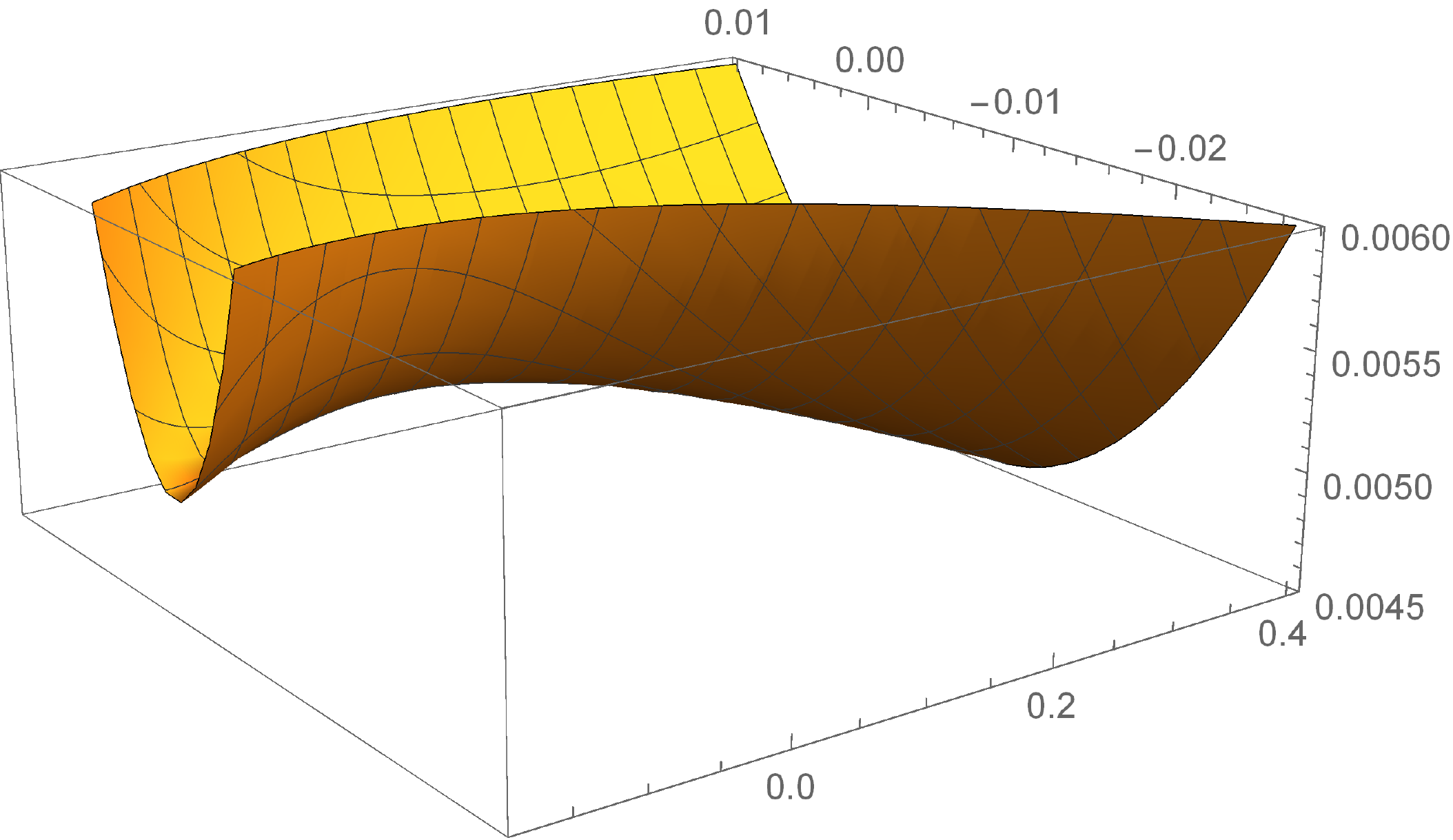}\caption{Solution 25}\label{fig:sol25}
\end{subfigure}
\caption{Potential $\frac{2}{M_p^2} V$ of \eqref{potential} around solutions 18 (Figure \ref{fig:sol18}) and 25 (Figure \ref{fig:sol25}). The two field fluctuations around the de Sitter extrema are one along the eigenvector $\vec{v}$ corresponding to the lowest eigenvalue of the mass matrix, and one along $\sigma_1$. We see the positive minimum for solution 18, and maximum in the tachyonic direction for solution 25.}\label{fig:pot}
\end{center}
\end{figure}

Regarding the values of $\eta_V$, all values obtained in \cite{Andriot:2020wpp} were such that $\eta_V \leq -2$, except for solution 14 which admitted the highest value $\eta_V \sim - 1.7067$. Violating the tachyon condition \eqref{C11} as does solution 14, we could hope to get higher $\eta_V$, and this is what we obtain in Table \ref{tab:sol}. These are actually the highest values obtained in known 10d classical supergravity de Sitter solutions. We come back below to the surprising values for solutions 18 and 19. Apart from those two, all values remain however close to 1, thus in agreement with the refined swampland de Sitter conjecture \cite{Garg:2018reu, Ooguri:2018wrx}. We illustrate two solutions and their stability in Figure \ref{fig:pot}.

We now give more details on each group of solutions
\begin{itemize}
  \item Solution 18: to our surprise, we obtained, for the first time among 10d de Sitter solutions, a positive value $\eta_V = 3.7926 $. This shows that proving the systematic presence of a tachyon, only based on equations to solve or on the 4d potential, cannot work. While this looks at first sight as a possible metastable de Sitter solution, this positive value is very likely linked to the non-compactness of the algebra and 6d manifold, which invalidates the solution. We identified its algebra to be the non-compact $\mathfrak{so}(2,1) \oplus {\rm Heis}_3$, by finding a change of basis bringing its structure constants to the non-zero following ones
      \beq
\mbox{Solution 18:}\quad  f^2{}_{35} = f^5{}_{23} = - f^3{}_{52} = 1 \ ,\quad  f^1{}_{46} = 1 \ .
\eeq
Alternatively, the analytic change of basis \eqref{chgbasis2021} also works and clearly maps to this algebra. The relation between non-compactness and metastability is not surprising from a gauged supergravity perspective.\footnote{A few (meta)stable de Sitter solutions are known in 4d gauged supergravities: an incomplete list includes \cite{Fre:2002pd, Ogetbil:2008tk, Roest:2009tt, Catino:2013syn, Cribiori:2020use} in ${\cal N}=2$ and \cite{deRoo:2002jf, deCarlos:2009qm} in ${\cal N}=4$ (see also \cite{DallAgata:2012plb} in ${\cal N}=8$). To the best of our knowledge, all such solutions involve either non-compact gaugings, de Roo-Wagemans angles or non-geometric fluxes. The last two are alike and make it difficult to have a higher dimensional origin from a standard geometric compactification. Furthermore, non-compact gaugings typically require non-compact extra dimensions as a higher dimensional origin, a situation therefore reminiscent of our solution 18 here. Constraints on gaugings for stability can be found in e.g.~\cite{Dibitetto:2010rg, Dibitetto:2011gm, Catino:2012dq, Danielsson:2012by}, while interesting criticisms on known stable de Sitter solutions with non-geometric fluxes were recently made in \cite{Plauschinn:2020ram}.} This situation however emphasizes that proving the presence of a systematic tachyon would require to take into account the compactness, which makes it even more difficult.
  \item Solution 19: its value $\eta_V = -0.12141 $ is unusually low in absolute value. As mentioned however, this solution admits all allowed structure constants non-zero, which makes the identification of its algebra difficult. The value of $\eta_V$ hints at the manifold being non-compact; if however it is not the case, this would be a promising example to get phenomenologically more interesting values of $\eta_V$ for e.g.~single-field inflation.
  \item Solutions 20-21: these solutions have several common features with solution 14. Their $\eta_V$ values are close: $\eta_V= -1.3624$ for solution 20, and $\eta_V= -1.7813$ for solution 21. They are also the only solutions among the new ones to have $T_{10}^2 <0$, as solution 14. We discuss those points further in section \ref{sec:under}. We identify their algebras to be $\mathfrak{so}(3) \oplus {\rm Heis}_3$, allowing for compact manifolds. We found a change of basis towards the following non-zero structure constants
\beq
\mbox{Solution 20-21:}\quad  f^2{}_{35} = f^3{}_{52} = f^5{}_{23} = 1 \ ,\quad  f^1{}_{46} = 1 \ . \\
\eeq
Alternatively, the following analytic change of basis
\bea
& {e^2}'=e^2 \ ,\ {e^4}'=e^4\ , \ {e^5}'=e^5 \ , \label{chgbasis2021}\\
& {e^1}'=e^1 - \frac{f^1{}_{35}}{f^2{}_{35}} e^2 \ ,\ {e^3}'=e^3 + \frac{f^2{}_{45}}{f^2{}_{35}} e^4 \ , \ {e^6}'=e^6 + \frac{f^2{}_{35} f^1{}_{45}- f^2{}_{45} f^1{}_{35} }{f^2{}_{35} f^1{}_{46}} e^5 \nn\ ,
\eea
takes the starting 10 non-zero structure constants $f^a{}_{bc}$ of these solutions to the only 4 non-zero $f^a{}_{bc}'$
\beq
f^2{}_{35} ,  f^3{}_{25} , f^5{}_{23} , f^1{}_{46} , f^1{}_{35} ,  f^1{}_{45} , f^2{}_{45} , f^5{}_{24} , f^6{}_{23} , f^6{}_{24}  \ \rightarrow \ f^2{}_{35}' ,  f^3{}_{25}' , f^5{}_{23}' , f^1{}_{46}' \ ,
\eeq
where $f^a{}_{bc}'= f^a{}_{bc}$, i.e.~the last ones are unchanged, while the other 6 are set to zero. Several vanish thanks to a Jacobi identity. This change of basis allows to preserve the volumes along the sources (see appendix C of \cite{Andriot:2020wpp}).
  \item Solutions 22-27: the algebra for those is identified to be $\mathfrak{g}_{3.4}^{-1} \oplus \mathfrak{g}_{3.5}^{0}$ (along respectively 235 $\oplus$ 146), allowing to have two compact three-dimensional solvmanifolds. We found a change of basis towards the following non-zero structure constants
\beq
\mbox{Solution 22-27:}\quad  f^2{}_{35} =  f^3{}_{25} = 1 \ ,\quad  f^1{}_{64} = - f^6{}_{14}= 1 \ ,
\eeq
or alternatively the following analytic change of basis
\bea
& {e^1}'=e^1 \ ,\ {e^4}'=e^4\ , \ {e^5}'=e^5 \ , \\
& {e^2}'=e^2 + \frac{f^3{}_{15}}{f^3{}_{25}} e^1 \ ,\ {e^3}'=e^3 + \frac{f^2{}_{45} f^1{}_{46}- f^2{}_{46} f^1{}_{45} }{f^2{}_{35} f^1{}_{46}} e^4 \ , \ {e^6}'=e^6 + \frac{f^1{}_{45}}{f^1{}_{46}} e^5 \nn\ ,
\eea
that takes the starting 8 non-zero structure constants $f^a{}_{bc}$ of these solutions to the only 4 non-zero $f^a{}_{bc}'$
\beq
f^2{}_{35} ,  f^3{}_{25} , f^1{}_{64} , f^6{}_{14} , f^1{}_{45} , f^1{}_{46} , f^2{}_{45} ,  f^2{}_{46} \ \rightarrow \ f^2{}_{35}' ,  f^3{}_{25}' , f^1{}_{64}' , f^6{}_{14}' \ ,
\eeq
where $f^a{}_{bc}'= f^a{}_{bc}$ remain unchanged while the other 4 are set to zero. This change of basis preserves once again the volumes along the sources. We also note that the $f^a{}_{bc}$ of these solutions give $R_5=0$. Finally, the values for these solutions are $\eta_V \in [-1.2253, -0.90691]$ (see appendix \ref{ap:sol}) which are interestingly high. We tried to go lower in $|\eta_V|$ without success.

\end{itemize}

We turn to the tachyons of these solutions (leaving solution 18 aside). A first point is that none of them admits a tachyon in the two fields $(\rho, \tau)$. In addition, by definition, they do not admit the four fields tachyon of \eqref{C11}. Finally, we test them upon the characterisation of the tachyon of solution 14 given by \eqref{C12}: among the new solutions, only 21 satisfies \eqref{C12} (but not 20). Therefore we now looked again for conditions characterising these new tachyons. The only interesting one is obtained using method 3 with the following combination
\bea
& b_{\sigma_1} =0 \ ,\ b_{\sigma_2} = -\frac{1}{6} + 4 c_{\sigma_2} (3 c_{\sigma_2} + 2 c_{\tau}) \ ,\ b_{\rho}=\frac{1}{3} + 8 (6 c_{\sigma_2}^2 + 4 c_{\sigma_2} c_{\tau} + c_{\tau}^2) \ , \\
& b_{\tau}=\frac{1}{2} + 4 c_{\tau}^2  \ ,\ c_{\sigma_1} =0 \ ,\ c_{\rho}= 4 c_{\sigma_2} + 2 c_{\tau}\nn\\
\Rightarrow\ & LC|_0 = {\cal P}(c_{\tau}, c_{\sigma_2}) + \frac{1}{2} {\cal R}_4\ ,\quad {\cal P}(c_{\tau}, c_{\sigma_2}) = c_{\tau}^2 A_0 + c_{\sigma_2}^2 B_0  -c_{\sigma_2} c_{\tau} C_0 \ ,\nn\\
& A_0 = 16 ( g_s^2|F_5|^2 + |H^{(0)_1}|^2 + |H^{(2)_1}|^2) \ ,\ C_0 = -24(g_s^2|F_1|^2 + g_s^2|F_5|^2 + |H^{(0)_1}|^2 +3 |H^{(2)_1}|^2) \ ,\nn\\
& B_0 = 6 (9 g_s^2|F_1|^2 - 3 g_s^2|F_5|^2 - 3 |H^{(0)_1}|^2 + 9 |H^{(2)_1}|^2 + 6 R_4 + 6 R_5 + g_s T_{10}^2) \ ,\label{A020}
\eea
where we used \eqref{eqR6} to replace $R_6$ and introduce ${\cal R}_4$. Following method 3, assuming $A_0> 0$, we get the following sufficient condition for a tachyon
\beq
\boxed{C13}:\quad A_0> 0\ , \quad 4A_0 B_0 - C_0^2 < 0 \ \mbox{with values of}\ \eqref{A020} \ . \label{C13}
\eeq
This condition, although not very simple, is satisfied by solutions 14, 20 and 21 (and surprisingly solution 15), but not by solutions 22-27. It gives an interesting characterisation and ``class'' for the tachyons of the former solutions, and confirms their common features already mentioned. For solutions 22-27, we do find any condition capturing all of their tachyons, using methods 2 or 3 (solutions 25 and 26 satisfy however surprisingly \eqref{C7}). While all other solutions (except 18 and 19) are captured by only two conditions \eqref{C11} and \eqref{C13}, those last solutions remain out of characterisation, keeping us away from a complete set of complementary conditions proving the existence of a systematic tachyon. We summarize these results in Table \ref{tab:sum}.

\subsection{Interpretation of the new solutions}\label{sec:under}

We discuss here some analytic and observed features of the new solutions, trying to understand them. The starting point is the requirement of violating condition \eqref{C11}. We first note the following inequality on the quantity entering \eqref{C11}
\beq
 -2  R_3 (g_s^2 |F_1|^2  - {\cal R}_4) - g_s^2 |F_1|^2 {\cal R}_4 \leq g_s^2 |F_1|^2 ( -2  R_3  -  {\cal R}_4)
\eeq
We deduce that if
\beq
 -2  R_3  -  {\cal R}_4 < 0 \ ,
\eeq
then we get a tachyon captured by \eqref{C11}. This simpler condition for a tachyon was also captured by \eqref{C10} with $a = 0$. On the contrary, avoiding now such a tachyon, we get the {\it necessary} condition ${\cal R}_4 + 2 R_3 \leq 0$. This quantity appears in equation \eqref{eqR4R3}, that we rewrite as follows
\bea
{\cal R}_4 + 2 R_3 & = 2 R_4 + \frac{g_s}{3}  T_{10}^2 +2 R_2 - |H^{(2)_1}|^2 - g_s^2 |F_5|^2 \label{exprR4R3}\\
& + 2R_1 -  (g_s^2 |F_1|^2 + g_s^2|F_3|^2 + |H^{(0)_1}|^2 ) + \frac{g_s}{3} T_{10}^1  \ ,\nn
\eea
where the second line appears in a Bianchi identity. Indeed, we recall from \cite{Andriot:2019wrs} that the sourced Bianchi identity can be rewritten as follows, for a set $I$ with $F_{8-p}^{(0)_I}=0$ and our ansatz
\bea
\varepsilon g_s \frac{T_{10}^I}{p+1} = & - \frac{1}{2} \left|*_{\bot_I}H^{(0)_I} + \varepsilon \varepsilon_p g_s F_{6-p}^{(0)_I} \right|^2  - \frac{1}{2} \sum_{a_{||_I}} \left| *_{\bot_I}( \d e^{a_{||_I}})|_{\bot_I} - \varepsilon \varepsilon_p g_s\, \iota_{a_{||_I}} F_{8-p}^{(1)_I} \right|^2 \label{BI} \\
& + \frac{1}{2}|H^{(0)_I}|^2 + \frac{1}{2} g_s^2 |F_{6-p}^{(0)_I}|^2 + \frac{1}{2} g_s^2 |F_{8-p}^{(1)_I}|^2 + \frac{1}{2} |f^{{}_{||_I}}{}_{{}_{\bot_I} {}_{\bot_I}}|^2 \ , \nn
\eea
where we include a random sign $\varepsilon$ for convenience. Denoting the big squares by $|\dots |^2$, we get here with $I=1$
\beq
\varepsilon  \frac{g_s}{6} T_{10}^1 =  - \frac{1}{2} \sum | \dots |^2  + \frac{1}{2} \left( |H^{(0)_1}|^2 +  g_s^2 |F_{1}|^2 + g_s^2 |F_{3}|^2 -2 R_1 \right) \ ,  \label{BI1}
\eeq
for any sign $\varepsilon$. We deduce
\beq
 |H^{(0)_1}|^2 +  g_s^2 |F_{1}|^2 + g_s^2 |F_{3}|^2 -2 R_1 - \frac{g_s}{3} T_{10}^1 \geq 0 \ .
\eeq
We conclude from \eqref{exprR4R3} that
\beq
{\cal R}_4 + 2 R_3 \leq 2 R_4 + \frac{g_s}{3}  T_{10}^2   \ . \label{ineq}
\eeq

While the {\it necessary} condition ${\cal R}_4 + 2 R_3 \leq 0$ has to be obeyed by the new solutions, the stronger condition $2 R_4 + \frac{g_s}{3}  T_{10}^2 \leq 0$ need not. The latter is thus neither necessary nor sufficient to avoid the tachyon \eqref{C11}, but can be compatible with its absence. We still mention this second, stronger condition, because we observe on all the new solutions (as well as on solution 14) the related feature
\beq
R_4 \times  T_{10}^2 <0 \ ,\label{R4T102}
\eeq
with $T_{10}^2 <0$ only for solutions 20, 21 (and 14). This observation agrees with the idea that $2 R_4 + \frac{g_s}{3}  T_{10}^2 \leq 0$, or if positive, cannot be intuitively too large given \eqref{exprR4R3} and ${\cal R}_4 + 2 R_3 \leq 0$. It remains unclear why \eqref{R4T102} should hold in general.\footnote{Let us recall for completeness the constraints obtained in section 4.4 of \cite{Andriot:2019wrs}, generalizing those of \cite{Andriot:2018ept}. In the case where for some $I$, $T_{10} - T_{10}^I = \sum_{J\neq I} T_{10}^J \leq 0$, then $|f^{{}_{||_I}}{}_{{}_{\bot_I} {}_{\bot_I}}|^2 \neq 0$ and $0 < \lambda_I < 1$. We apply this here for $I=1$ with \eqref{lambda1}, in the case where $T_{10}^2 + T_{10}^3 \leq 0$, which is true when $T_{10}^2 \leq 0$: we then deduce $\lambda_1 >0$. In other words,
\beq
T_{10}^2 \leq 0 \Rightarrow T_{10}^2 + T_{10}^3 \leq 0 \Rightarrow R_4 +R_6 >0 \ . \label{nogo44}
\eeq
This provides a further relation between $R_4$ and $T_{10}^2$, appearing again compatible with \eqref{R4T102}, but not directly related. We also note that all our new solutions verify $T_{10}^2 + T_{10}^3 \leq 0$, except solution 18.}

Let us then discuss schematically the sign of $R_4$, in relation with the matter of non-compactness mentioned in section \ref{sec:descr}. We first need to recall some algebra. We work here in a basis where all three indices of $f^a{}_{bc}$ are different. Therefore, given three different $a,b,c$, we consider sets of three structure constants built from placing these $a,b,c$ in different positions (we do not distinguish those related by antisymmetry of lower indices): for instance $f^2{}_{35}, f^3{}_{25}, f^5{}_{23}$. Let us consider only one such set for simplicity, and study the consequences on the sign of $R_4$ as well as on the geometry. This depends on how many structure constants in this set are non-zero, and their sign:
\begin{itemize}
  \item 3 structure constants: if the three $f^2{}_{35}, f^3{}_{25}, f^5{}_{23}$ are non-zero, then we have a semi-simple (sub)algebra. Then, either their sign is identical under cyclic permutations of indices, as for $\mathfrak{so}(3)$, or it is not as for $\mathfrak{so}(2,1)$. The former is compact, the latter is not. In addition, $R_4$ is a sum of terms of the form $-f^2{}_{35} f^3{}_{25}$. So $R_4<0$ means that one of these terms is negative, as for $\mathfrak{so}(2,1)$, i.e.~the non-compact case; the sign of the third non-zero structure constant does not matter.
  \item 2 structure constants: if only two are non-zero, then we get a solvable (sub)algebra. As mentioned in section \ref{sec:descr}, $f^2{}_{35} f^3{}_{25} < 0$ corresponds to $\mathfrak{g}_{3.5}^{0}$ and $f^2{}_{35} f^3{}_{25} > 0$ to $\mathfrak{g}_{3.4}^{-1}$, both allowing for compact group manifolds. But only the latter gives $R_4<0$.
  \item 1 structure constant: if only one is non-zero, then we get a nilpotent (sub)algebra ${\rm Heis}_3$, giving by itself a nilmanifold which is compact. This does not contribute to $R_4$.
\end{itemize}
We summarize this schematic reasoning as follows:
\begin{itemize}
  \item $R_4 < 0$: we can either face a non-compact situation, as in solution 18, or a compact solvmanifold as $\mathfrak{g}_{3.4}^{-1}$, as in solutions 22-27.
  \item $R_4>0$: the manifold may either be based (partly) on a semi-simple algebra $\mathfrak{so}(3)$, being then compact, as in solutions 20, 21, or a compact solvable one, as in solution 14 (which is on $\mathfrak{g}_{3.5}^{0} \oplus \mathfrak{g}_{3.5}^{0}$).
\end{itemize}
Interestingly, it seems we have encountered all possibilities with our search for solutions. We hope this brief understanding of our solutions will allow better characterisations in the future, especially regarding the tachyons.

\newpage

\section{Summary and outlook}\label{sec:ccl}

In this work we focused on de Sitter solutions of 10d type II supergravities with intersecting $O_p/D_p$, which are natural candidates for classical and perturbative de Sitter string backgrounds. We studied the stability of these de Sitter solutions, motivated by the phenomenological importance of (in)stability with quasi de Sitter space-times, but also by the various, sometimes contradictory, expectations of the swampland program on stability of de Sitter solutions. All known examples are actually observed to be (sharply) unstable, with $\eta_V < -1$ (an account of values can be found e.g.~in \cite{Andriot:2018mav, Andriot:2020wpp}). The ideal objective of this work was to prove such a systematic instability, and to additionally get an upper bound on $\eta_V$. To that end, we followed a well-verified proposal \cite{Danielsson:2012et}, stating that the systematic tachyon should be among a restricted set of scalar fields, $(\rho, \tau, \sigma_I)$.

\begin{figure}[H]
\begin{center}
\begin{subfigure}[H]{0.72\textwidth}
\includegraphics[width=\textwidth]{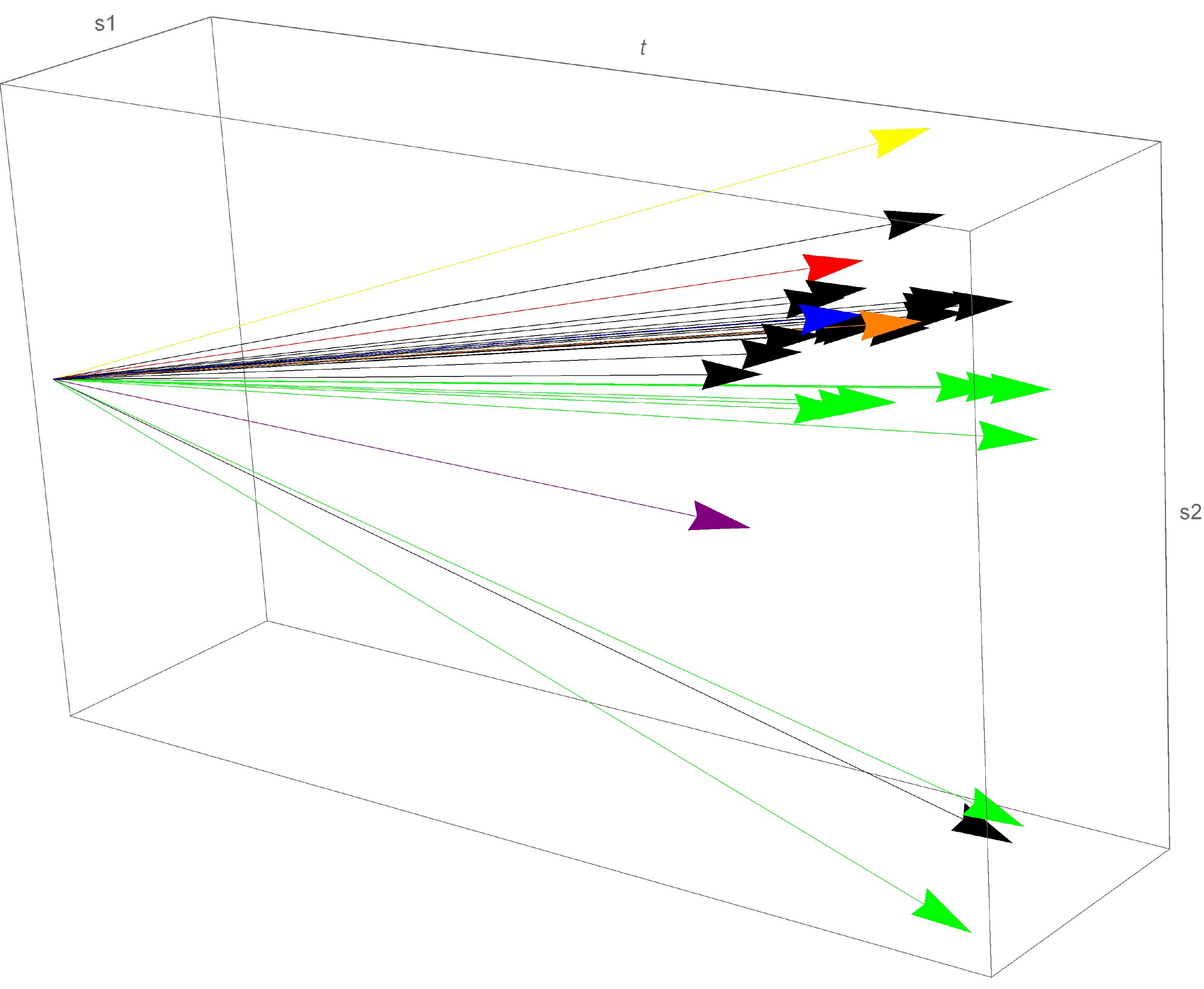}\caption{Tachyonic directions of solutions and sufficient conditions.}\label{fig:arrows3d}
\end{subfigure}
\quad
\begin{subfigure}[H]{0.55\textwidth}
\includegraphics[width=\textwidth]{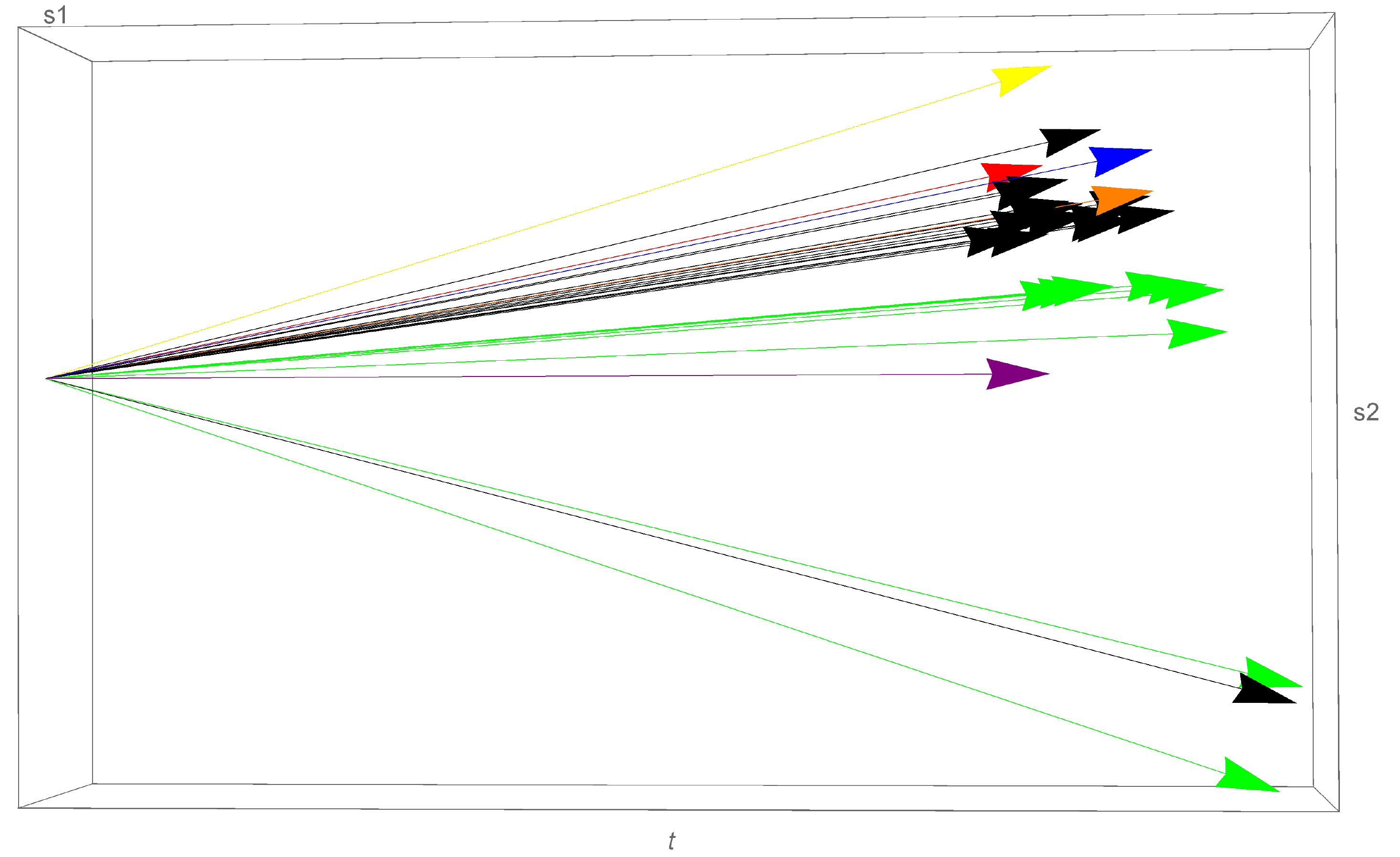}\caption{Figure \ref{fig:arrows3d} projected along $\sigma_1$.}\label{fig:arrows3ds1}
\end{subfigure}
\qquad \qquad \qquad \quad
\begin{subfigure}[H]{0.19\textwidth}
\includegraphics[width=\textwidth]{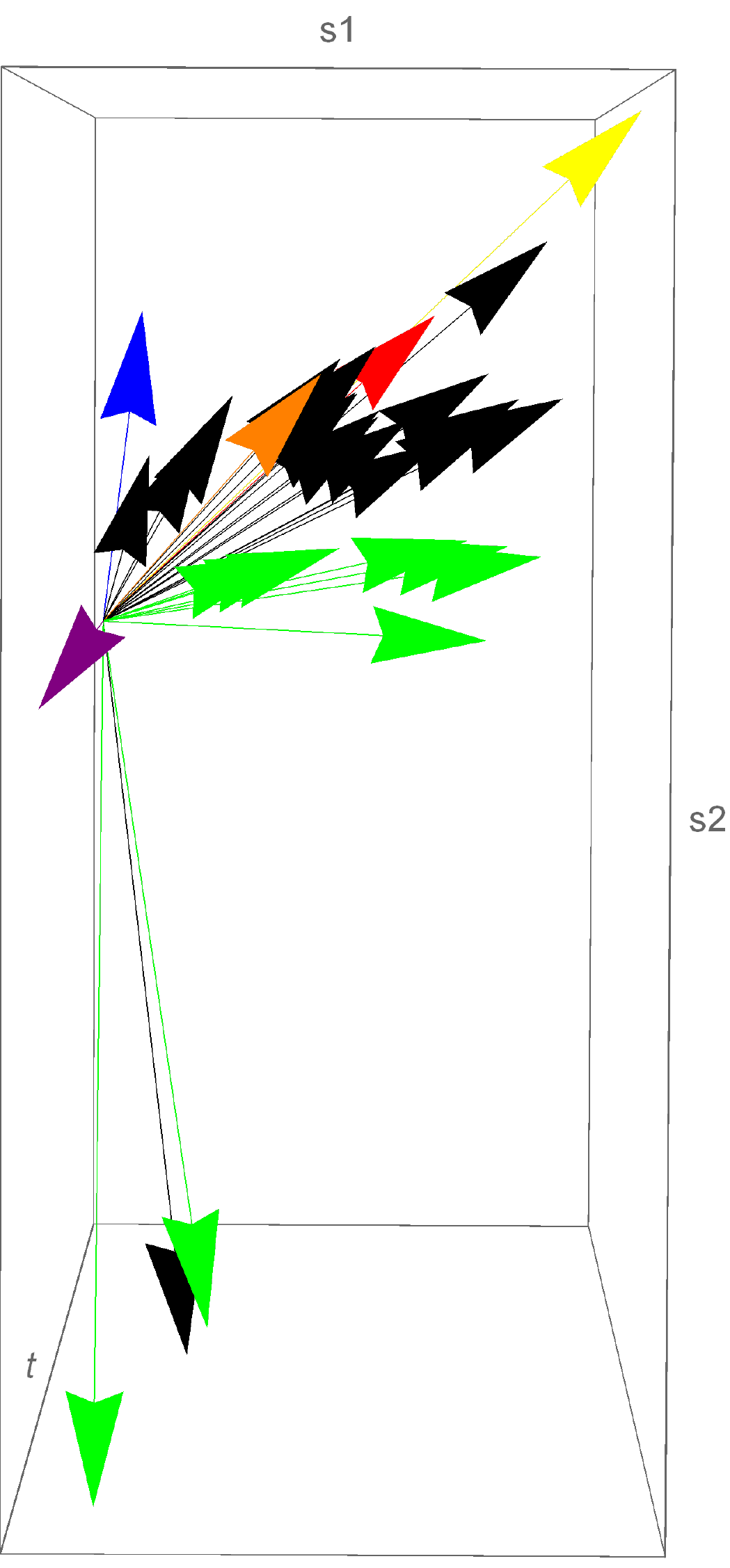}\caption{Idem along $\tau$.}\label{fig:arrows3dt}
\end{subfigure}
\caption{Tachyonic directions in field space along $(\tau, \sigma_1, \sigma_2)$. In black are displayed the tachyonic eigenvectors of solutions 1-17 of \cite{Andriot:2020wpp}, and in green those of the new solutions 19-27 found in this work (given by $\vec{v}$ in appendix \ref{ap:sol}). The blue vector is the eigenvector of the lowest eigenvalue of solution 18. The other colored vectors correspond to tachyonic directions in our sufficient conditions, given by $(c_{\rho}, c_{\tau}, c_{\sigma_1}, c_{\sigma_2})$: $C4$ (purple), $C7$ (red), $C8$ (orange), $C10$ (yellow). All 4-dimensional vectors are normalized to 1, so the different lengths displayed here encode the missing component along $\rho$; the latter is taken positive, fixing the overall sign.\\
We observe that all solutions of \cite{Andriot:2020wpp}, except 14, have a tachyon close to a direction captured by a sufficient condition. On the contrary, the tachyons of the new solutions are separated. Solution 14, as well as the new solutions 20 and 21, are in a distinct set at the ``bottom'', corresponding to a negative component along $\sigma_2$. Finally, solution 18 is aside from all others. These observations agree with our results.}\label{fig:arrows}
\end{center}
\end{figure}

We focused on type IIB with $O_5/D_5$, a setting where 17 de Sitter solutions were recently found \cite{Andriot:2020wpp, Andriot:2020vlg}. We give in section \ref{sec:setting} the 4d action with scalar potential for the four fields $(\rho, \tau, \sigma_1, \sigma_2)$, and discuss mass matrices in appendix \ref{ap:M}. From this concrete starting point, we develop in section \ref{sec:methods} and appendix \ref{ap:form} three methods to achieve the objective. We first introduce the methods in section \ref{sec:summethods}. Method 1 in section \ref{sec:method1} consists in studying eigenvalues of $2\times 2$ blocks of the mass matrix. Methods 2 and 3 use a formalism, presented in section \ref{sec:interlude} and appendix \ref{ap:form}, that identifies a tachyonic field direction, given by four coefficients $c_{\phi^i}$ along each scalar field $\phi^i$ direction, as e.g.~in \eqref{genineqintro}. Method 2 in section \ref{sec:method2} looks for ``universal'' tachyons with fixed constants $c_{\phi^i}$, while method 3 in section \ref{sec:method3} leaves some freedom among these coefficients, restricting them to a solution-dependent range. Thanks to these three methods, we obtain 13 sufficient conditions on solution data for having a tachyon, $C1$-$C13$, summarized in Table \ref{tab:sum}. These conditions can be viewed as characterising classes of solutions and their tachyons, and we could experience a diversity in those. Therefore, we did not manage to prove the systematic existence of a tachyon, which would have amounted to find a finite set of complementary sufficient conditions for a tachyon. As indicated in Table \ref{tab:sum}, we still identified general classes of tachyons, in the sense of capturing many tachyonic solutions with one condition. A general illustration of this situation is provided in Figure \ref{fig:arrows}.

The most successful condition in that respect is $C11$ in \eqref{C11}, that captured all but one solutions of \cite{Andriot:2020wpp}. We then ran a new search for de Sitter solutions of type IIB supergravity with $O_5/D_5$, asking to violate the condition \eqref{C11}. This resulted in finding 10 new de Sitter solutions, presented and discussed in section \ref{sec:sol} (see in particular Table \ref{tab:sol}, and Figure \ref{fig:pot}), and listed explicitly in appendix \ref{ap:sol}. As expected from this requirement, these solutions exhibit new physics w.r.t.~\cite{Andriot:2020wpp}: they are found on new 6d group manifolds, and admit higher values of $\eta_V$.\footnote{The values of $\eta_V$ of our new de Sitter solutions are higher than those of \cite{Andriot:2020wpp}. In both cases however, these values are computed restricting to our four fields only, with the corresponding potential \eqref{potential}. Considering more scalar fields would lower the value, as could have been the case of older 10d type IIA supergravity de Sitter solutions with $O_6/D_6$.} Leaving aside two solutions, $\eta_V$ goes up to $-0.90691$. The two remaining solutions are special: one is actually stable with $\eta_V= 3.7926$, but turns out to be on a non-compact manifold. The 6d geometry of the other one is not identified, but it admits the surprising value $\eta_V = -0.12141 $. Finally, we found some characterisation of tachyons of these new de Sitter solutions, but failed to find conditions for all of them. Overall, we can still claim to have identified 2 broad classes ($C11$ and $C13$) capturing 19 among 26 tachyonic de Sitter solutions. In addition, we note that the hypothesis of a systematic tachyon still holds (phrased as conjecture 2 in \cite{Andriot:2019wrs}), once the criterion of compactness is taken into account.

Method 2 allowed in addition to set a bound on $\eta_V$ for each condition found ($C3$-$C10$). Similarly to the variety of conditions found, the bounds are scattered, and range between $[-\frac{4}{3} , -\frac{25}{3422}] \approx [-1.33333 , -0.00730567]$ with $C10$ and $C8$. From this range, it is unfortunately difficult to conclude anything relevant to phenomenology, nor on a favored swampland statement on stability.\footnote{Interestingly, a similar situation occurred in compactifications with non-geometric fluxes \cite{CaboBizet:2020cse}, whose string origin remains however difficult to establish. Indeed, as illustrated there in Figures 1 and 8, a large number of de Sitter solutions have $\eta_V \sim -6$. However few de Sitter points lie outside this line, with the highest value being $\eta_V = -0.032873$ (private communication from N.~Cabo Bizet).} We note though that most de Sitter solutions found have an $\eta_V$ close to $-1$, remaining consistent with the refined de Sitter conjecture of \cite{Garg:2018reu, Ooguri:2018wrx}, while having possible interest for multi-field inflation or quintessence models of the kind \cite{Brown:2017osf, Garcia-Saenz:2018ifx, Achucarro:2018vey, Bjorkmo:2019aev, Bjorkmo:2019fls} and more recently \cite{Cicoli:2020noz, Akrami:2020zfz, Aragam:2020uqi, Farakos:2021mwt}. The validity of these 10d supergravity de Sitter solutions as classical perturbative string backgrounds remains to be checked following \cite{Andriot:2020vlg}, while we recall that no example of such a de Sitter background is known up-to-date.

\begin{table}[H]
  \begin{center}
  		\hspace*{-2.0cm}
    \begin{tabular}{|c||c|c|c|c|c|c|c|c|c|c|c|c|c|c|c|c|c|}
    \hline
  & & & & & & & & & & & & & & & & & \\[-8pt]
Solutions  & 1 & 2,7 & 3,5,6,11 & 4 & 8 & 9 & 10 & 12 & 13,16,17 & 14 & 15 & 18 & 19 & 20 & 21 & 22-24,27 & 25,26 \\[3pt]
    \hhline{==================}
   & & & & & & & & & & & & & & & & & \\[-8pt]
 $C1$ \eqref{C1} &  &  &  &  &  &  &  &  &  &  &  &  &  &  &  &  &  \\[3pt]
    \hhline{-||-----------------}
   & & & & & & & & & & & & & & & & & \\[-8pt]
 $C2$ \eqref{C2} &  &  &  &  &  &  &  &  &  &  &  &  &  &  &  &  &  \\[3pt]
    \hhline{-||-----------------}
   & & & & & & & & & & & & & & & & & \\[-8pt]
 $C3$ \eqref{C3} &  &  &  &  &  &  &  &  &  &  &  &  &  &  &  &  &  \\[3pt]
    \hhline{-||-----------------}
 & & & & & & & & & & & & & & & & & \\[-8pt]
 $C4$ \eqref{C4} & $\checkmark$ & $\checkmark$ & $\checkmark$ & $\checkmark$ & $\checkmark$ & $\checkmark$ &  &  &  &  &  &  &  &  &  &  &  \\[3pt]
    \hhline{-||-----------------}
   & & & & & & & & & & & & & & & & & \\[-8pt]
 $C5$ \eqref{C5} &  & $\checkmark$ & $\checkmark$ & $\checkmark$ & $\checkmark$ & $\checkmark$ &  &  &  &  &  &  &  &  &  &  &  \\[3pt]
    \hhline{-||-----------------}
   & & & & & & & & & & & & & & & & & \\[-8pt]
 $C6$ \eqref{C6} &  &  &  & $\checkmark$ &  &  &  &  &  &  & $\checkmark$ &  &  &  &  &  &  \\[3pt]
    \hhline{==================}
   & & & & & & & & & & & & & & & & & \\[-8pt]
 $C7$ \eqref{C7} & $\checkmark$ & $\checkmark$ & $\checkmark$ & $\checkmark$ & $\checkmark$ &  &  & $\checkmark$ & $\checkmark$ &  & $\checkmark$ &  &  &  &  &  & $\checkmark$ \\[3pt]
    \hhline{-||-----------------}
   & & & & & & & & & & & & & & & & & \\[-8pt]
 $C8$ \eqref{C8} & $\checkmark$ & $\checkmark$ & $\checkmark$ & $\checkmark$ & $\checkmark$ & $\checkmark$ & $\checkmark$ & $\checkmark$ &  &  &  &  &  &  &  &  &  \\[3pt]
    \hhline{-||-----------------}
   & & & & & & & & & & & & & & & & & \\[-8pt]
 $C9$ \eqref{C9} &  &  &  &  &  &  &  &  &  & $\checkmark$ &  &  &  &  & $\checkmark$ &  &  \\[3pt]
    \hhline{-||-----------------}
   & & & & & & & & & & & & & & & & & \\[-8pt]
 $C10$ \eqref{C10} & $\checkmark$ &  & $\checkmark$ & $\checkmark$ & $\checkmark$ &  & $\checkmark$ & $\checkmark$ & $\checkmark$ &  & $\checkmark$ &  &  &  &  &  &  \\[3pt]
    \hhline{-||-----------------}
   & & & & & & & & & & & & & & & & & \\[-8pt]
 $C11$ \eqref{C11} & $\checkmark$ & $\checkmark$ & $\checkmark$ & $\checkmark$ & $\checkmark$ & $\checkmark$ & $\checkmark$ & $\checkmark$ & $\checkmark$ &  & $\checkmark$ &  &  &  &  &  &  \\[3pt]
    \hhline{-||-----------------}
   & & & & & & & & & & & & & & & & & \\[-8pt]
 $C12$ \eqref{C12} &  & $\checkmark$ & $\checkmark$ & $\checkmark$ &  & $\checkmark$ &  &  &  & $\checkmark$ & $\checkmark$ &  &  &  & $\checkmark$ &  &  \\[3pt]
    \hhline{-||-----------------}
   & & & & & & & & & & & & & & & & & \\[-8pt]
 $C13$ \eqref{C13} &  &  &  &  &  &  &  &  &  & $\checkmark$ & $\checkmark$ &  &  & $\checkmark$ & $\checkmark$ &  &  \\[3pt]
    \hline
    \end{tabular}
     \caption{Sufficient conditions for a tachyon, $C1$-$C13$, found in this work, and whether they are satisfied ($\checkmark$) by solutions 1-17 of \cite{Andriot:2020wpp} and the new solutions 18-27 of this paper. Conditions $C1$-$C6$ characterise tachyons in the two fields $(\rho,\tau)$ only, and $C7$-$C13$ tachyons in the four fields $(\rho,\tau,\sigma_1,\sigma_2)$. The new solutions 18-27 were found by asking to violate $C11$, the most general tachyonic characterisation found for solutions of \cite{Andriot:2020wpp}. We did not manage to characterise tachyons of solutions 22,23,24,27, while solutions 18,19 are special.}\label{tab:sum}
  \end{center}
\end{table}

Making further progress on the characterisation of these instabilities remains crucial, and we believe that the formalism developed here should be helpful to future studies.\footnote{Note that we have not used the tachyonic eigenvector $\vec{v}$ given in appendix \ref{ap:sol}. Its coefficients could play the role of the $c_{\phi^i}$, so this extra data could convey interesting information. This is illustrated in Figure \ref{fig:arrows}.} Beyond the proof of a systematic tachyon, a characterisation of allowed values of $\eta_V$ is crucial for phenomenology. It remains a possibility, as suggested by the TCC \cite{Bedroya:2019snp}, that both metastable and unstable solutions exist. In that case, finding new examples of de Sitter solutions is important, using for instance the technique of violating tachyons characterisations as here.

Sharp instabilities of de Sitter solutions, meaning the condition $\eta_V <-1$, was connected in the swampland program \cite{Gautason:2018gln, Lust:2019zwm} to the absence of scale separation in anti-de Sitter solutions (see related discussions in \cite{Andriot:2020vlg}). A well-known (candidate) counter-example to the latter is the classical DGKT solution \cite{DeWolfe:2005uu, Acharya:2006ne}, for which it was shown that scale separation gets improved with string classicality. By analogy, it is legitimate to suggest that a mildly unstable de Sitter solution, of phenomenological interest, would be found together with a classical one (which has not yet been discovered). As argued in \cite{Andriot:2020vlg} though, a parametric control on the classicality, and on $\eta_V $, is then not expected. On this note, solution 19 with its comparatively low $|\eta_V|$ deserves a closer look.\\

Let us add a final word on the addition of anti-branes $\bar{D}_p$. It has been shown in \cite{Kallosh:2018nrk} for a type IIA de Sitter solution with $O_6/D_6$ that adding $\bar{D}_6$ could remove the tachyon present among the fields $(\rho, \tau, \sigma_I)$, by generating another critical point in the potential. Similar arguments were given in 3d in \cite{Farakos:2020idt}, adding on top $O_p/D_p$ of different size $p$. Note however that in both papers, no new complete 10d de Sitter solution was found once the $\bar{D}_p$ was added (in particular the Bianchi identities were not checked). It remains natural to ask whether we notice here a change in our conditions for tachyons, e.g.~a falsification, once $\bar{D}_p$ are added. In short, the answer is no, although conditions may then be slightly more difficult to verify.

In more details, adding $\bar{D}_5$ amounts in the potential to replace $T_{10}^I \rightarrow T_{10}^I + \bar{T}_{10}^I$, with $\bar{T}_{10}^I \leq 0$ \cite{Andriot:2020wpp}; a different replacement should be done in the sourced Bianchi identities, which however have not played any role here. One can verify that the reasonings to establish the signs of relevant quantities \eqref{entriessign} are unchanged. The signs used to established our conditions for tachyons are then also unchanged. So the conditions found for tachyons are formally unaltered by the introduction of $\bar{D}_5$. However, some conditions involve quantities that have to be negative, e.g.~$-T_{10}$ or $-T_{10}^1$. Upon the replacements made when introducing $\bar{D}_5$, these quantities would get a little more positive, therefore making it quantitatively more difficult to satisfy the condition for a tachyon. This remains to be seen in detail, since fluxes and other fields entering conditions would also change. Finding explicit new solutions when including $\bar{D}_5$ could help to check whether conditions are now violated for the reason just mentioned.  Finding different tachyon conditions, especially some requiring the sourced Bianchi identity, could also make the change more manifest. Removing this way a systematic tachyon from de Sitter solutions remains for sure an interesting option.

\vspace{0.1in}

\subsection*{Acknowledgements}

We thank N.~Cabo Bizet, N.~Cribiori, P.~Marconnet and T.~Wrase for useful exchanges on this project. We acknowledge support from the Austrian Science Fund (FWF): project number M2247-N27.

\newpage

\begin{appendix}

\section{Mass matrices}\label{ap:M}

In this appendix, we compute explicitly mass matrices introduced in section \ref{sec:setting} and discussed in appendix \ref{ap:t}, and verify some of their relations. Focusing on the setting with four scalar fields, with field metric $g_{ij}$ given in \eqref{gij}, we see that we can work separately on the two diagonal blocks $(\rho, \tau)$ and $(\sigma_1, \sigma_2)$. We start with $(\rho, \tau)$ and compute the corresponding $2\times 2$ mass matrix of coefficient $M^i{}_k= g^{ij} \nabla_j \del_k V= g^{ij} (\del_j \del_k V - \Gamma^l_{jk} \del_l V)$. The only non-zero Christoffel symbols are $\Gamma^{\rho}_{\rho \rho} = - \rho^{-1}$ and $\Gamma^{\tau}_{\tau \tau} = - \tau^{-1}$. From this we get
\beq
M_{(\rho, \tau)}= g^{-1} \nabla \del V = \frac{1}{M_p^2}
\left( \begin{array}{cc} \frac{2}{3} \rho \del_{\rho} ( \rho \del_{\rho} V) & \frac{2}{3} \rho^2 \del_{\rho} \del_{\tau} V \label{Mmatrix}\\
\frac{1}{2} \tau^2 \del_{\rho} \del_{\tau} V & \frac{1}{2} \tau \del_{\tau} ( \tau \del_{\tau} V) \end{array}\right) \ .
\eeq
From it, we can check
\beq
\hat{M} = \delta^{-1} \del_{\hat{\phi}} \del_{\hat{\phi}} V = \left(\frac{\del \hat{\phi}}{ \del \phi} \right) g^{-1}  \nabla  \del V  \left(\frac{\del \hat{\phi}}{ \del \phi} \right)^{-1} \ ,
\eeq
as introduced in section \ref{sec:setting}. Using the diffeomorphism and derivatives \eqref{relderrhotau} for $(\rho, \tau)$, we indeed verify explicitly
\beq
\hat{M}_{(\rho, \tau)} = \left( \begin{array}{cc} \del_{\hat{\rho}}^2  V & \del_{\hat{\rho}} \del_{\hat{\tau}} V \\
\del_{\hat{\rho}} \del_{\hat{\tau}} V & \del_{\hat{\tau}}^2  V \end{array}\right) = \left( \begin{array}{cc} \sqrt{\frac{3}{2}} \rho^{-1} & 0 \\
0 &  \sqrt{2} \tau^{-1} \end{array} \right) M_{(\rho, \tau)} \left( \begin{array}{cc} \sqrt{\frac{3}{2}} \rho^{-1} & 0 \\
0 &  \sqrt{2} \tau^{-1} \end{array} \right)^{-1} \ .
\eeq

We proceed similarly with the block $(\sigma_1 , \sigma_2)$. We compute the corresponding mass matrix: the only non-zero Christoffel symbols are $\Gamma^{1}_{11} = - \sigma_1^{-1}$ and $\Gamma^{2}_{22} = - \sigma_2^{-1}$, even though the metric has off-diagonal entries. From this we obtain
\bea
M_{(\sigma_1 , \sigma_2)} &= g^{-1} \nabla \del V \label{Ms1s2}\\
&= \frac{1}{M_p^2} \frac{1}{18}
\left( \begin{array}{cc} 2 \sigma_1 \del_{\sigma_1} ( \sigma_1 \del_{\sigma_1} V) + \sigma_1 \sigma_2 \del_{\sigma_1} \del_{\sigma_2} V  & \sigma_1 \del_{\sigma_2} ( \sigma_2 \del_{\sigma_2} V) + 2  \sigma_1^2 \del_{\sigma_1} \del_{\sigma_2} V \\
\sigma_2 \del_{\sigma_1} ( \sigma_1 \del_{\sigma_1} V) + 2  \sigma_2^2 \del_{\sigma_1} \del_{\sigma_2} V  & 2 \sigma_2 \del_{\sigma_2} ( \sigma_2 \del_{\sigma_2} V) + \sigma_1 \sigma_2 \del_{\sigma_1} \del_{\sigma_2} V  \end{array}\right) \ .\nn
\eea
We also compute, using $(\sigma_1 , \sigma_2)$ diffeomorphism \eqref{diffeosigma}
\bea
M &=  \left(\frac{\del \hat{\phi}}{ \del \phi} \right)^{-1} \hat{M} \left(\frac{\del \hat{\phi}}{ \del \phi} \right) \label{Ms1s22}\\
&= \frac{1}{2} \left( \begin{array}{cc} \del_{\hat{\sigma}_1}^2 V + \del_{\hat{\sigma}_2}^2 V + \frac{4}{\sqrt{3}} \del_{\hat{\sigma}_2} \del_{\hat{\sigma}_1} V &  \frac{\sigma_1}{\sigma_2} \left(\del_{\hat{\sigma}_1}^2 V - \del_{\hat{\sigma}_2}^2 V - \frac{2}{\sqrt{3}} \del_{\hat{\sigma}_2} \del_{\hat{\sigma}_1} V  \right) \\
\frac{\sigma_2}{\sigma_1} \left(\del_{\hat{\sigma}_1}^2 V - \del_{\hat{\sigma}_2}^2 V + \frac{2}{\sqrt{3}} \del_{\hat{\sigma}_2} \del_{\hat{\sigma}_1} V  \right) & \del_{\hat{\sigma}_1}^2 V + \del_{\hat{\sigma}_2}^2 V - \frac{4}{\sqrt{3}} \del_{\hat{\sigma}_2} \del_{\hat{\sigma}_1} V  \end{array}\right) \ . \nn
\eea
We then verify explicitly this relation between mass matrices, as done (reverse-wise) for $(\rho,\tau)$, by comparing \eqref{Ms1s22} to \eqref{Ms1s2}, using the derivatives \eqref{reldersigma} with $\del_{\hat{\sigma}_1}^2 V = \frac{1}{M_p^2} \frac{1}{12} \left( \sigma_1 \del_{\sigma_1} + \sigma_2 \del_{\sigma_2}  \right)^2 V$, etc.\\

Finally, let us illustrate the change of basis providing the tachyonic direction introduced in section \ref{sec:summethods} or appendix \ref{ap:t}, restricting to the $(\hat{\rho},\hat{\tau})$ block. To go from $\hat{M}$ to $\hat{M}'$, we perform an SO(2) transformation (block diagonal piece of the O(4) considered in appendix \ref{ap:t}, completed to have a determinant equal to 1), i.e.~a rotation $R$. We get explicitly
\bea
\hat{M}'& = R \hat{M} R^{-1} = \frac{1}{\hat{c}_{\hat{\rho}}^2 + \hat{c}_{\hat{\tau}}^2} \left( \begin{array}{cc} \hat{c}_{\hat{\rho}} & \hat{c}_{\hat{\tau}} \\ -\hat{c}_{\hat{\tau}} & \hat{c}_{\hat{\rho}} \end{array}\right) \hat{M}  \left( \begin{array}{cc} \hat{c}_{\hat{\rho}} & -\hat{c}_{\hat{\tau}} \\ \hat{c}_{\hat{\tau}} & \hat{c}_{\hat{\rho}} \end{array}\right) \\
& = \frac{1}{\hat{c}_{\hat{\rho}}^2 + \hat{c}_{\hat{\tau}}^2} \left( \begin{array}{cc} \hat{c}_{\hat{\rho}}^2 \del_{\hat{\rho}}^2  V + \hat{c}_{\hat{\tau}}^2 \del_{\hat{\tau}}^2  V + 2 \hat{c}_{\hat{\rho}} \hat{c}_{\hat{\tau}} \del_{\hat{\rho}} \del_{\hat{\tau}} V  & \hat{c}_{\hat{\rho}} \hat{c}_{\hat{\tau}} (\del_{\hat{\tau}}^2  V - \del_{\hat{\rho}}^2  V ) +  (\hat{c}_{\hat{\rho}}^2 - \hat{c}_{\hat{\tau}}^2)  \del_{\hat{\rho}} \del_{\hat{\tau}} V  \\
\hat{c}_{\hat{\rho}} \hat{c}_{\hat{\tau}} (\del_{\hat{\tau}}^2  V - \del_{\hat{\rho}}^2  V ) +  (\hat{c}_{\hat{\rho}}^2 - \hat{c}_{\hat{\tau}}^2)  \del_{\hat{\rho}} \del_{\hat{\tau}} V  &  \hat{c}_{\hat{\tau}}^2 \del_{\hat{\rho}}^2  V + \hat{c}_{\hat{\rho}}^2 \del_{\hat{\tau}}^2  V - 2 \hat{c}_{\hat{\rho}} \hat{c}_{\hat{\tau}} \del_{\hat{\rho}} \del_{\hat{\tau}} V \end{array}\right) \ . \nn
\eea
As in appendix \ref{ap:t}, we are interested in the top left coefficient:
\beq
\frac{1}{\hat{c}_{\hat{\rho}}^2 + \hat{c}_{\hat{\tau}}^2} \left(\hat{c}_{\hat{\rho}}^2 \del_{\hat{\rho}}^2  V + \hat{c}_{\hat{\tau}}^2 \del_{\hat{\tau}}^2  V + 2 \hat{c}_{\hat{\rho}} \hat{c}_{\hat{\tau}} \del_{\hat{\rho}} \del_{\hat{\tau}} V \right)  = \frac{1}{\hat{c}_{\hat{\rho}}^2 + \hat{c}_{\hat{\tau}}^2} \left(\hat{c}_{\hat{\rho}} \del_{\hat{\rho}} + \hat{c}_{\hat{\tau}} \del_{\hat{\tau}} \right)^2  V = \del_{\hat{t}_{c}}^2 V  \ .
\eeq
We verify explicitly that it can be recast into the derivative of a field $\hat{t}_{c}$, that eventually serves as the tachyon.

\section{Formalism for methods 2 and 3}\label{ap:form}

In this appendix, we present the formalism required by methods 2 and 3 of sections \ref{sec:method2} and \ref{sec:method3}. We prove some key formulas to be used there. The results obtained here are summarized in section \ref{sec:interlude}, and were briefly introduced in section \ref{sec:summethods}.

\subsection{The tachyonic direction and the bound on $\eta_V$}\label{ap:t}

We consider all four fields and the $4\times 4$ mass matrices $M$ and $\hat{M}$. We note though that the following reasoning does not depend on the number of fields, and can be generalized. By definition, $\hat{M}$ is real and symmetric. The same is true for any matrix $\hat{M}'$ obtained from $\hat{M}$, by any constant O(4) transformation $O$
\beq
\hat{M}' = O \hat{M} O^{-1} = O \hat{M} O^{T} \ .
\eeq
By virtue of the lemmas of section \ref{sec:summethods}, the minimal eigenvalue of $\hat{M}'$ is thus smaller than any of its diagonal entries, such as for instance its top left corner element $\hat{M}'{}^1{}_1$. Eigenvalues of $\hat{M}'$ are however the same as those of $\hat{M}$ or $M$. We deduce that for any (constant) O(4) transformation,
\beq
\mbox{minimal eigenvalue of $M$ or $\hat{M}$} \leq \hat{M}'{}^1{}_1 \ .
\eeq
With $\hat{M}^i{}_j = \delta^{ik} \del_{\hat{\phi}^k} \del_{\hat{\phi}^j} V$, we now rewrite this top left corner element as follows
\bea
\hat{M}'{}^1{}_1 & = O^1{}_i \delta^{ij} \del_{\hat{\phi}^j} \del_{\hat{\phi}^k} V \, O^{T}{}^k{}_1 = O^1{}_i \delta^{ij} \del_{\hat{\phi}^j} \del_{\hat{\phi}^k} V \, \delta_{1m} O^m{}_n \delta^{nk} = O^1{}_i \delta^{ij} \del_{\hat{\phi}^j} \, O^1{}_n \delta^{nk} \del_{\hat{\phi}^k} V  \ ,\nn\\
{\rm i.e.}\ \ \hat{M}'{}^1{}_1 & = \del_{\hat{t}_c}^2 V \ {\rm with}\ \del_{\hat{t}_c} = O^1{}_i \delta^{ij} \del_{\hat{\phi}^j} \ .
\eea
Since $O \in$ O(4), elements of its first line can be parameterized in full generality by 4 real parameters $\hat{c}_{\hat{\phi}^i}$ as
\beq
O^1{}_i = \frac{\hat{c}_{\hat{\phi}^i}}{\sqrt{\sum_i \hat{c}_{\hat{\phi}^i}^2} } \ .
\eeq
In other words, for any real $(\hat{c}_{\hat{\rho}} , \hat{c}_{\hat{\tau}} , \hat{c}_{\hat{\sigma}_1} , \hat{c}_{\hat{\sigma}_2})$, we can write
\beq
\hat{M}'{}^1{}_1 = \del_{\hat{t}_c}^2 V \quad {\rm with} \quad  \hat{c}_{\hat{\rho}} \del_{\hat{\rho}} + \hat{c}_{\hat{\tau}} \del_{\hat{\tau}} + \hat{c}_{\hat{\sigma}_1} \del_{\hat{\sigma}_1} + \hat{c}_{\hat{\sigma}_2} \del_{\hat{\sigma}_2} = \sqrt{\hat{c}_{\hat{\rho}}^2 + \hat{c}_{\hat{\tau}}^2 + \hat{c}_{\hat{\sigma}_1}^2 + \hat{c}_{\hat{\sigma}_2}^2}\ \del_{\hat{t}_c} \ ,
\eeq
and deduce
\beq
\mbox{minimal eigenvalue of $M$ or $\hat{M}$} \leq \del_{\hat{t}_c}^2 V \ . \label{mintc}
\eeq
Physically, we understand that this is simply a change of field basis. Even though not needed here, one can verify that $\hat{t}_c$ is a canonical field in this O(4)-rotated basis.\footnote{One can introduce the column matrices $\left(  \del_{\hat{\phi}} \right) $ and $\left(  \d \hat{\phi} \right)$ of $(i1)$-coefficients $\delta^{ij} \del_{\hat{\phi}^j}$ and $\d \hat{\phi}^i$. One has by definition $\left(  \d \hat{\phi} \right)^T  \left(  \del_{\hat{\phi}} \right) = 1\!\!1$. Thanks to orthonormality, one has  $\left( O \left(  \d \hat{\phi} \right) \right)^T \, O \left(  \del_{\hat{\phi}} \right) = 1\!\!1$. We can then introduce the dual one-form $\d \hat{t}_c = O^1{}_i \d \hat{\phi}^i$. From this one can consider the kinetic term and show that $\hat{t}_c$ is a canonical field; see also section 4.3 of \cite{Andriot:2020lea}.} This field will be understood as the tachyon, or tachyonic direction, since we will look for cases where $\del_{\hat{t}_c}^2 V < 0$. We have changed basis from $M$ to $\hat{M}$ to $\hat{M}'$ in order to isolate as a diagonal entry the (canonical) tachyonic direction in the mass matrix. Explicit illustration of this is given in appendix \ref{ap:M} where we compute the mass matrices.

We now further relate $\hat{t}_c$ to the initial, non-canonical fields. For this we need the explicit field space diffeomorphism to relate $\del_{\hat{\phi}^i}$ to $\del_{\phi^j}$. In our case, we can treat the fields two by two, $(\rho, \tau)$ and $(\sigma_1, \sigma_2)$, because they do not mix and the metric \eqref{gij} is block diagonal. We then use \eqref{relderrhotau} and \eqref{reldersigma} to rewrite the combination of derivatives in terms of the initial fields
\bea
& \hat{c}_{\hat{\rho}} \del_{\hat{\rho}} + \hat{c}_{\hat{\tau}} \del_{\hat{\tau}} + \hat{c}_{\hat{\sigma}_1} \del_{\hat{\sigma}_1} + \hat{c}_{\hat{\sigma}_2} \del_{\hat{\sigma}_2} \\
& =\frac{1}{M_p} \left( c_{\rho} \rho \del_{\rho} + c_{\tau} \tau \del_{\tau} +  \frac{c_{\sigma_1} + c_{\sigma_2}}{2} (\sigma_1 \del_{\sigma_1} + \sigma_2 \del_{\sigma_2}) + \frac{c_{\sigma_1} - c_{\sigma_2}}{2} (\sigma_1 \del_{\sigma_1} - \sigma_2 \del_{\sigma_2})  \right) \nn\\
& =\frac{1}{M_p} \left( c_{\rho} \rho \del_{\rho} + c_{\tau} \tau \del_{\tau} +  c_{\sigma_1} \sigma_1 \del_{\sigma_1}  + c_{\sigma_2} \sigma_2 \del_{\sigma_2} \right) \ , \nn
\eea
where we introduced the constants
\beq
\hat{c}_{\hat{\rho}} =  \sqrt{\frac{3}{2}}\, c_{\rho}  \ ,\ \hat{c}_{\hat{\tau}} = \sqrt{2}\, c_{\tau} \ , \ \hat{c}_{\hat{\sigma}_1} = \sqrt{3}\, (c_{\sigma_1} + c_{\sigma_2}) \ , \ \hat{c}_{\hat{\sigma}_2} = 3\, (c_{\sigma_1} - c_{\sigma_2}) \ . \label{constantsap}
\eeq
We recall that the above formalism is valid for any real constants $(\hat{c}_{\hat{\rho}} , \hat{c}_{\hat{\tau}} , \hat{c}_{\hat{\sigma}_1} , \hat{c}_{\hat{\sigma}_2})$, so the same is true for arbitrary real constants $(c_{\rho} , c_{\tau}, c_{\sigma_1} , c_{\sigma_2})$.

Before using the above to prove the existence of tachyon, let us introduce another field $\hat{t}_b$ for the first derivatives of the potential. We introduce similarly another O(4) transformation, parameterized similarly by constants $\hat{b}_{\hat{\phi}^i}$, defining $\hat{t}_b$ such that $\del_{\hat{t}_b} = O^1{}_i \delta^{ij} \del_{\hat{\phi}^j}$. Doing so, one has, with $(  \del_{\hat{\phi}} V ) $ the column matrix of $(i1)$-coefficients $\delta^{ij} \del_{\hat{\phi}^j} V$,
\bea
|\nabla V|^2 = g^{ij} \del_{i} V \del_{j} V = \delta^{ij} \del_{\hat{\phi}^i} V \del_{\hat{\phi}^j} V \geq & \sum_{i=1\dots 4} \delta^{ij} \del_{\hat{\phi}^i} V \del_{\hat{\phi}^j} V = ( \del_{\hat{\phi}} V )^T\,  (\del_{\hat{\phi}} V) \\
&\ = (O (\del_{\hat{\phi}} V) )^T\,  O (\del_{\hat{\phi}} V ) \geq (\del_{\hat{t}_b} V)^2 \ ,\nn
\eea
implying
\beq
|\nabla V| \geq |\del_{\hat{t}_b} V| \geq -\del_{\hat{t}_b} V \ . \label{mintb}
\eeq

We now make use of this formalism to prove the existence of a tachyon, and further get bounds on parameters $\eta_V$ and $\epsilon_V$. First, from \eqref{mintc}, we obtain that a tachyon exists if $\del_{\hat{t}_c}^2 V < 0$. We deduce the following sufficient condition for a tachyon
\begin{empheq}[innerbox=\fbox]{align}
\left( c_{\rho} \rho \del_{\rho} + c_{\tau} \tau \del_{\tau} +  c_{\sigma_1} \sigma_1 \del_{\sigma_1}  + c_{\sigma_2} \sigma_2 \del_{\sigma_2} \right)^2 V < 0  \label{conddel2ap}
\end{empheq}
for any real constants $(c_{\rho} , c_{\tau}, c_{\sigma_1} , c_{\sigma_2})$. Furthermore, from \eqref{mintc} and the above rewritings, we deduce
\beq
\left( \hat{c}_{\hat{\rho}}^2 + \hat{c}_{\hat{\tau}}^2 + \hat{c}_{\hat{\sigma}_1}^2 + \hat{c}_{\hat{\sigma}_2}^2 \right) \eta_V \leq \frac{1}{V} \left( c_{\rho} \rho \del_{\rho} + c_{\tau} \tau \del_{\tau} +  c_{\sigma_1} \sigma_1 \del_{\sigma_1}  + c_{\sigma_2} \sigma_2 \del_{\sigma_2} \right)^2 V \ .
\eeq
Similarly, we get from \eqref{mintb}
\beq
\sqrt{\hat{b}_{\hat{\rho}}^2 + \hat{b}_{\hat{\tau}}^2 + \hat{b}_{\hat{\sigma}_1}^2 + \hat{b}_{\hat{\sigma}_2}^2} \, \sqrt{2 \epsilon_V} \geq - \frac{\left( b_{\rho} \rho \del_{\rho} + b_{\tau} \tau \del_{\tau} +  b_{\sigma_1} \sigma_1 \del_{\sigma_1}  + b_{\sigma_2} \sigma_2 \del_{\sigma_2} \right) V}{V} \ ,
\eeq
where $\hat{b}_{\hat{\phi}^i}$ and $b_{\phi^i}$ are related as the $c$'s in \eqref{constantsap}, and are equally arbitrary. We conclude the following implication of inequalities
\begin{empheq}[innerbox=\fbox]{align}
& V + \left( b_{\rho} \rho \del_{\rho} + b_{\tau} \tau \del_{\tau} +  b_{\sigma_1} \sigma_1 \del_{\sigma_1}  + b_{\sigma_2} \sigma_2 \del_{\sigma_2} \right) V \nn\\
& \ \ + \left( c_{\rho} \rho \del_{\rho} + c_{\tau} \tau \del_{\tau} +  c_{\sigma_1} \sigma_1 \del_{\sigma_1}  + c_{\sigma_2} \sigma_2 \del_{\sigma_2} \right)^2 V \leq 0 \label{genineqap}\\
\Rightarrow & \, - \sqrt{\hat{b}_{\hat{\rho}}^2 + \hat{b}_{\hat{\tau}}^2 + \hat{b}_{\hat{\sigma}_1}^2 + \hat{b}_{\hat{\sigma}_2}^2} \, \sqrt{2 \epsilon_V} + \left( \hat{c}_{\hat{\rho}}^2 + \hat{c}_{\hat{\tau}}^2 + \hat{c}_{\hat{\sigma}_1}^2 + \hat{c}_{\hat{\sigma}_2}^2 \right) \eta_V \leq -1 \nn
\end{empheq}
for arbitrary constants $b$'s and $c$'s, related as in \eqref{constantsap}. In other words, if we find constants for which we can prove the first inequality, we deduce the second one, giving us bounds on the parameters. In particular, considering only the first derivatives, i.e.~restricting to $c_{\phi^i}=0$, we obtain
\bea
& V + \left( b_{\rho} \rho \del_{\rho} + b_{\tau} \tau \del_{\tau} +  b_{\sigma_1} \sigma_1 \del_{\sigma_1}  + b_{\sigma_2} \sigma_2 \del_{\sigma_2} \right) V \leq 0 \\
\Rightarrow \ &  \sqrt{2 \epsilon_V}  \geq \frac{1}{\sqrt{\hat{b}_{\hat{\rho}}^2 + \hat{b}_{\hat{\tau}}^2 + \hat{b}_{\hat{\sigma}_1}^2 + \hat{b}_{\hat{\sigma}_2}^2}} \ . \nn
\eea
This reproduces and generalizes the formula for the lower bound on $\epsilon_V$, proved differently in \cite{Hertzberg:2007wc, Andriot:2019wrs}, and used extensively in \cite{Andriot:2020lea} to compute the constant $c$ of the swampland de Sitter conjecture \cite{Obied:2018sgi}. Secondly, from \eqref{conddel2ap}, \eqref{genineqap} is sufficient to prove the existence of a tachyon (at a de Sitter extremum), giving in addition the following upper bound on $\eta_V$
\beq
\eta_V|_0 \leq -\frac{1}{\hat{c}_{\hat{\rho}}^2 + \hat{c}_{\hat{\tau}}^2 + \hat{c}_{\hat{\sigma}_1}^2 + \hat{c}_{\hat{\sigma}_2}^2} = -\frac{1}{\frac{3}{2} c_{\rho}^2 + 2 c_{\tau}^2 + 3 (c_{\sigma_1}+c_{\sigma_2})^2 + 9 (c_{\sigma_1}-c_{\sigma_2})^2} \ . \label{boundetaVap}
\eeq
Finally, the above can be rewritten as follows: with $a\in [0,1]$, starting with an inequality as
\bea
& V + \left( b_{\rho} \rho \del_{\rho} + b_{\tau} \tau \del_{\tau} +  b_{\sigma_1} \sigma_1 \del_{\sigma_1}  + b_{\sigma_2} \sigma_2 \del_{\sigma_2} \right) V \label{genineqaap}\\
& + \left( c_{\rho} \rho \del_{\rho} + c_{\tau} \tau \del_{\tau} +  c_{\sigma_1} \sigma_1 \del_{\sigma_1}  + c_{\sigma_2} \sigma_2 \del_{\sigma_2} \right)^2 V \leq (1-a) V \nn
\eea
one deduces the following bound on $\eta_V$
\beq
\eta_V|_0 \leq  -\frac{a}{\frac{3}{2} c_{\rho}^2 + 2 c_{\tau}^2 + 3 (c_{\sigma_1}+c_{\sigma_2})^2 + 9 (c_{\sigma_1}-c_{\sigma_2})^2} \ . \label{boundetaVaap}
\eeq
This is equivalent to the above, and can be understood as a rescaling of the various coefficients. This way of writing the sufficient condition for a tachyon and related bound may however be useful. In addition, for $a=0$, we note that we loose any interesting bound, and we are back to simply proving the existence of a tachyon. Indeed, the condition \eqref{genineqaap} is then the same as \eqref{conddel2ap} up to first derivatives of the potential. Those are useful to simplify the expressions obtained, as already in the first method.

Methods 2 and 3 will consists in finding constants $b$'s and $c$'s to get conditions of the type \eqref{conddel2ap}, \eqref{genineqap} or \eqref{genineqaap}, the latter allowing to get a bound on $\eta_V$. To find useful, i.e.~simple enough conditions, we need a further tool dubbed the linear combination, that we now turn to.

\subsection{The linear combination}\label{ap:LC}

To prove the existence of a tachyon, and possibly get a bound on $\eta_V$, we are interested in linear combinations of the potential, its first and second derivatives, as given in \eqref{genineqap}. We thus introduce in full generality this linear combination for the four field potential \eqref{potential}
\bea
LC= & \frac{2}{M_p^{2}} \Big( V + \left( b_{\rho} \rho \del_{\rho} + b_{\tau} \tau \del_{\tau} +  b_{\sigma_1} \sigma_1 \del_{\sigma_1}  + b_{\sigma_2} \sigma_2 \del_{\sigma_2} \right) V \label{LC4}\\
&\phantom{\frac{2}{M_p^{2}} \Big(} + \left( c_{\rho} \rho \del_{\rho} + c_{\tau} \tau \del_{\tau} +  c_{\sigma_1} \sigma_1 \del_{\sigma_1}  + c_{\sigma_2} \sigma_2 \del_{\sigma_2} \right)^2 V \Big)\nn\\
=\, & \frac{1}{2} g_s^2 |F_1|^2\, \tau^{-4}\rho^{2} \sigma_1^{-2} \sigma_2^{-2}\, \left(1 - 2 b_{\sigma_1} + b_{\sigma_2} - b_{\rho} + 2 b_{\tau}) + 4 (c_{\sigma_1} + c_{\sigma_2} - c_{\rho} + 2 c_{\tau})^2 \right) \nn\\
+\, & \frac{1}{2} g_s^2 |F_3|^2\, \tau^{-4}\, \left(1 - 4 b_{\tau} + 16 c_{\tau}^2\right) \nn\\
+\, & \frac{1}{2} |H^{(0)_1}|^2\, \tau^{-2}\rho^{-3} \sigma_1^{-6}\sigma_2^{6} \, \left(1 - 6 b_{\sigma_1} + 6 b_{\sigma_2} - 3 b_{\rho} - 2 b_{\tau} + (6 c_{\sigma_1} - 6 c_{\sigma_2} + 3 c_{\rho} + 2 c_{\tau})^2\right) \nn\\
+\, & \frac{1}{2} |H^{(2)_1}|^2\, \tau^{-2}\rho^{-3} \sigma_1^{6} \sigma_2^{-6} \, \left(1 + 6 b_{\sigma_1} - 6 b_{\sigma_2} - 3 b_{\rho} - 2 b_{\tau} + (-6 c_{\sigma_1} + 6 c_{\sigma_2} + 3 c_{\rho} + 2 c_{\tau})^2\right) \nn\\
-\, & R_1\, \tau^{-2}\rho^{-1} \sigma_1^{-8} \sigma_2^{4} \, \left(1 - 8 b_{\sigma_1} + 4 b_{\sigma_2} - b_{\rho} - 2 b_{\tau} + (8 c_{\sigma_1} - 4 c_{\sigma_2} + c_{\rho} + 2 c_{\tau})^2\right) \nn\\
-\, & R_2\, \tau^{-2}\rho^{-1} \sigma_1^{4} \sigma_2^{-8} \, \left(1 + 4 b_{\sigma_1} - 8 b_{\sigma_2} - b_{\rho} - 2 b_{\tau} + (-4 c_{\sigma_1} + 8 c_{\sigma_2} + c_{\rho} + 2 c_{\tau})^2\right) \nn\\
-\, & R_3\, \tau^{-2}\rho^{-1} \sigma_1^{4} \sigma_2^{4} \, \left(1 + 4 b_{\sigma_1} + 4 b_{\sigma_2} - b_{\rho} - 2 b_{\tau} + (-4 c_{\sigma_1} - 4 c_{\sigma_2} + c_{\rho} + 2 c_{\tau})^2\right) \nn\\
+\, & R_4\, \tau^{-2}\rho^{-1} \sigma_1^{-2} \sigma_2^{-2} \, \left(-1 + 2 b_{\sigma_1} + 2 b_{\sigma_2} + b_{\rho} + 2 b_{\tau} - (2 c_{\sigma_1} + 2 c_{\sigma_2} + c_{\rho} + 2 c_{\tau})^2\right) \nn\\
+\, & R_5\, \tau^{-2}\rho^{-1} \sigma_1^{4} \sigma_2^{-2} \, \left(-1 - 4 b_{\sigma_1} + 2 b_{\sigma_2} + b_{\rho} + 2 b_{\tau} - (-4 c_{\sigma_1} + 2 c_{\sigma_2} + c_{\rho} + 2 c_{\tau})^2\right) \nn\\
+\, & R_6\, \tau^{-2}\rho^{-1} \sigma_1^{-2} \sigma_2^{4} \, \left(-1 + 2 b_{\sigma_1} - 4 b_{\sigma_2} + b_{\rho} + 2 b_{\tau} - (2 c_{\sigma_1} - 4 c_{\sigma_2} + c_{\rho} + 2 c_{\tau})^2\right) \nn\\
+\, & g_s \frac{T_{10}^1}{24}\, \tau^{-3}\rho^{-\frac{1}{2}} \sigma_1^{-4}  \sigma_2^2 \, \left(-4 + 2 (8 b_{\sigma_1} - 4 b_{\sigma_2} + b_{\rho} + 6 b_{\tau}) - (8 c_{\sigma_1} - 4 c_{\sigma_2} + c_{\rho} + 6 c_{\tau})^2\right) \nn\\
+\, & g_s \frac{T_{10}^2}{24}\, \tau^{-3}\rho^{-\frac{1}{2}} \sigma_1^{2}  \sigma_2^{-4} \, \left(-4 + 2 (-4 b_{\sigma_1} + 8 b_{\sigma_2} + b_{\rho} + 6 b_{\tau}) - (-4 c_{\sigma_1} + 8 c_{\sigma_2} + c_{\rho} + 6 c_{\tau})^2\right) \nn\\
+\, & g_s \frac{T_{10}^3}{24}\, \tau^{-3}\rho^{-\frac{1}{2}} \sigma_1^2  \sigma_2^2 \, \left(-4  + 2 (-4 b_{\sigma_1} -4 b_{\sigma_2} + b_{\rho} + 6 b_{\tau}) - (-4 c_{\sigma_1} -4 c_{\sigma_2} + c_{\rho} + 6 c_{\tau} )^2 \right) \ . \nn
\eea
Ideally, we would like to show that this combination is negative, $LC \leq 0$, allowing through \eqref{genineqap} to prove the existence of a tachyon and get a bound on $\eta_V$. We will not manage to do this in full generality, but we will still get such results up to conditions on solutions data, i.e.~fluxes, sources or curvatures. To achieve this, one considers coefficient combinations on the right-hand side, in front of entries of definite sign, e.g.~$1 - 4 b_{\tau} + 16 c_{\tau}^2 $ in front of $|F_3|^2 \geq 0$, and finds values such this coefficient combination is negative. The aim is then to find as many negative combinations as possible.

Some entries do not have a definite sign: $R_4, R_5, R_6, T_{10}^2$ (see \eqref{entriessign} and definitions in section \ref{sec:setting}). There are various possibilities for those. First, one can replace $T_{10}^2$ by $T_{10}^2=T_{10} - T_{10}^1 - T_{10}^3$, each of those having a definite sign. In addition, one can replace the $R_{4,5,6}$ by the following $X_{4,5,6}$ of definite sign: indeed, the $X_i$ are perfect squares, as can be seen in definitions \eqref{coeff}, with
\bea
& X_4=R_4+R_1+R_2 \ ,\ X_5=R_5+R_2+R_3 \ ,\ X_6=R_6+R_1+R_3 \label{Xi}\\
& R_{1,2,3}\leq 0 \ ,\ X_{4,5,6} \leq 0 \ .
\eea
Another option is to replace the $R_{4,5,6}$ using another combination of first derivatives, and the relation of ${\cal R}_4$ to the potential \eqref{VR4}: for instance one has the following combination
\bea
& b_{\sigma_1}=b_{\sigma_2} = \frac{1}{6} \ ,\ b_{\rho}= \frac{1}{3} \ ,\ b_{\tau}= \frac{1}{2} \ ,\ c_{\sigma_1}= c_{\sigma_2}= c_{\rho}= c_{\tau}= 0 \label{eqR4R3}\\
\hspace{-0.5cm} \Rightarrow\ & \frac{1}{2} {\cal R}_4 = R_4+R_1+R_2 - R_3 - \frac{1}{2} (g_s^2|F_1|^2 +g_s^2 |F_3|^2+ g_s^2 |F_5|^2 + |H^{(0)_1}|^2 + |H^{(2)_1}|^2 ) + \frac{g_s}{6} (T_{10} -  T_{10}^3) \ .\nn
\eea
We recognise equation (3.19) of \cite{Andriot:2017jhf}, namely the trace of the internal Einstein equation along the set $I=3$, combined with the 4d Einstein equation. We can use \eqref{eqR4R3} to replace $R_4$ in terms of other quantities of definite sign. Similarly, we have
\bea
& b_{\sigma_1}=- \frac{1}{6}\ ,\ b_{\sigma_2} = 0 \ ,\ b_{\rho}= \frac{1}{3} \ ,\ b_{\tau}= \frac{1}{2} \ ,\ c_{\sigma_1}= c_{\sigma_2}= c_{\rho}= c_{\tau}= 0 \nn\\
\Rightarrow\ & \frac{1}{2} {\cal R}_4 = R_5+R_2+R_3 - R_1 - \frac{1}{2} (g_s^2|F_3|^2 + 2 g_s^2|F_5|^2 + 2 |H^{(2)_1}|^2 ) + \frac{g_s}{6} (T_{10} -  T_{10}^1) \ , \label{eqR5}
\eea
to replace $R_5$ and
\bea
& b_{\sigma_1}=0\ ,\ b_{\sigma_2} =- \frac{1}{6} \ ,\ b_{\rho}= \frac{1}{3} \ ,\ b_{\tau}= \frac{1}{2} \ ,\ c_{\sigma_1}= c_{\sigma_2}= c_{\rho}= c_{\tau}= 0 \nn\\
\Rightarrow\ & \frac{1}{2} {\cal R}_4 = R_6+R_1+R_3  - R_2 - \frac{1}{2} (g_s^2|F_3|^2  + 2 g_s^2|F_5|^2 + 2 |H^{(0)_1}|^2 ) + \frac{g_s}{6} (T_{10} -  T_{10}^2) \ . \label{eqR6}
\eea
to replace $R_6$. We will make use of those to get entries of definite sign.

This method and formalism has been introduced in \cite{Andriot:2018ept, Andriot:2019wrs}. There however, only $\del_{\phi^i}^2 V$ were considered for second derivatives, instead of the general tachyonic direction we have here. The idea was that of having a negative diagonal entry, as explained in section \ref{sec:summethods}, but we did not manage to get such a proof. As shown with solutions of \cite{Andriot:2020wpp}, we now know why: in all 17 solutions found there, we always had $\del_{\phi^i}^2 V > 0$ for all four fields, despite the existence of a tachyon. We could also formally prove here in section \ref{sec:method1} that some of these second derivatives are always positive on a de Sitter extremum with our potential. This highlights the fact the tachyon is due to off-diagonal terms, i.e.~mixed second derivatives $\del_{\phi^1} \del_{\phi^2} V$. Such terms are now present with our tachyonic direction $\left( c_{\rho} \rho \del_{\rho} + c_{\tau} \tau \del_{\tau} +  c_{\sigma_1} \sigma_1 \del_{\sigma_1}  + c_{\sigma_2} \sigma_2 \del_{\sigma_2} \right)^2 V$, so we hope to get better results.

As a warm-up, we will restrict ourselves to the two fields $(\rho, \tau)$: the linear combination for those is obtained directly from the potential \eqref{potrhotau}, or by setting in the above  $b_{\sigma_i}=c_{\sigma_i} =0$ and $\sigma_i=1$. We get
\bea
LC=\frac{2}{M_p^2} \Big( V +\, & b_{\rho} \rho \del_{\rho} V + b_{\tau} \tau \del_{\tau} V + (c_{\rho} \rho \del_{\rho} + c_{\tau} \tau \del_{\tau})^2\, V \Big) \label{LC}\\
=\, & \frac{1}{2} g_s^2 |F_1|^2\, \tau^{-4}\rho^{2} (1 + 2 b_{\rho} - 4 b_{\tau} + 4 c_{\rho}^2 - 16 c_{\rho} c_{\tau} + 16 c_{\tau}^2)\nn\\
+\, & \frac{1}{2} g_s^2 |F_3|^2\, \tau^{-4} (1 - 4 b_{\tau} + 16 c_{\tau}^2) \nn\\
+\, & \frac{1}{2} g_s^2 |F_5|^2\, \tau^{-4}\rho^{-2} (1 - 2 b_{\rho} - 4 b_{\tau} + 4 c_{\rho}^2 + 16 c_{\rho} c_{\tau} + 16 c_{\tau}^2) \nn\\
+\, & \frac{1}{2} |H|^2\, \tau^{-2}\rho^{-3} (1 - 3 b_{\rho} - 2 b_{\tau} + 9 c_{\rho}^2 + 12 c_{\rho} c_{\tau} + 4 c_{\tau}^2) \nn\\
-\, & \R_6\, \tau^{-2}\rho^{-1} (1 - b_{\rho} - 2 b_{\tau} + c_{\rho}^2 + 4 c_{\rho} c_{\tau} + 4 c_{\tau}^2) \nn\\
+\, & g_s \frac{T_{10}}{6}\, \tau^{-3}\rho^{-\frac{1}{2}} \frac{1}{4}  (-4 + 2 b_{\rho} + 12 b_{\tau} - c_{\rho}^2 - 12 c_{\rho} c_{\tau} - 36 c_{\tau}^2) \nn
\eea
Interestingly, all entries have here a definite sign, so we can directly get some results. We will make use of these tools in methods 2 and 3, in sections \ref{sec:method2} and \ref{sec:method3}.

\section{List of solutions}\label{ap:sol}

We give in this appendix the list and complete data of the 10 new de Sitter solutions of 10d type IIB supergravity with intersecting $O_5/D_5$, presented in section \ref{sec:sol}. We obtained them using the techniques of \cite{Andriot:2020wpp}.\footnote{As indicated in footnote [44] of \cite{Andriot:2020vlg}, the sign $\varepsilon_5$ in the right-hand side of the sourced Bianchi identity (conventions of \cite{Andriot:2016xvq, Andriot:2019wrs}) was wrongly computed in \cite{Andriot:2020wpp} to be $-1$ instead of $+1$. This is only a convention that can easily be absorbed by $F_q \rightarrow -F_q$ without any impact. We have not corrected this sign in the present paper.} In particular, all equations are solved with a maximal error of $\varepsilon < 5 \cdot 10^{-15}$, and solution data are known to 16 significant digits. We round them here to 5 significant digits for readability. The numerical values of physical variables are given in units of $2\pi l_s$ as in \cite{Andriot:2020wpp}, and we only give those which are non-zero. $T_{10}[I]$ stands for $g_s T_{10}^I$, $F_q[a_1,\dots, a_q]$ for $g_s F_{q\, a_1,\dots, a_q}$, $H[a,b,c]$ for $H_{abc}$, $f[a,b,c]$ for $f^a{}_{bc}$. We provide for each solution the values of ${\cal R}_4$, ${\cal R}_6$, $\eta_V$ for the four fields potential \eqref{potential}, the eigenvalues ${\rm masses}^2$ of the $4\times 4$ mass matrix $M$, and a normalised eigenvector $\vec{v}$ for the tachyonic, or lowest, eigenvalue.

\subsection*{Solution 18}
\label{sol18}
\begin{equation*}
\begin{aligned}
&T_{10}[1] \rightarrow 10, T_{10}[2] \rightarrow 5.0453, T_{10}[3] \rightarrow -1.7065,
   F_1[5] \rightarrow 0.14502, F_1[6] \rightarrow 1.1791,\\[6pt]
&F_3[1, 3, 5] \rightarrow -0.032396,
   F_3[1, 3, 6] \rightarrow -0.17114, F_3[1, 4, 5] \rightarrow -0.22672,
   F_3[1, 4, 6] \rightarrow 0.041315,\\[6pt]
&F_3[2, 3, 5] \rightarrow 0.55732,
   F_3[2, 3, 6] \rightarrow 0.038508, F_3[2, 4, 6] \rightarrow -0.62288,
   H[1, 2, 5] \rightarrow 0.69483,\\[6pt]
&H[3, 4, 5] \rightarrow 0.46484,
   H[3, 4, 6] \rightarrow -0.82698, f[1, 3, 5] \rightarrow -0.034063,
   f[1, 4, 5] \rightarrow 0.012660,\\[6pt]
&f[1, 4, 6] \rightarrow 0.15053,
   f[2, 3, 5] \rightarrow 1.5799, f[2, 4, 5] \rightarrow 0.27154, f[3, 2, 5] \rightarrow 0.12647,\\[6pt]
&f[5, 2, 3] \rightarrow 1.2374, f[5, 2, 4] \rightarrow 0.21267,
   f[6, 2, 3] \rightarrow -0.15219, f[6, 2, 4] \rightarrow -0.026158\,.
\end{aligned}
\end{equation*}
\begin{equation*}
{\cal R}_4 =0.028430 \,, \quad  {\cal R}_6 = -0.44861 \,, \quad \eta_V= 3.7926 \,,
\end{equation*}
\begin{equation*}
\text{masses}^2 = (4.4552, 3.6110, 0.29362, 0.026956) \,, \;  \vec{v} = (0.49697, 0.84917, 0.046867, 0.17244) \,.
\end{equation*}

\subsection*{Solution 19}
\label{sol19}
\begin{equation*}
\begin{aligned}
&T_{10}[1] \rightarrow 10, T_{10}[2] \rightarrow 0.12358, T_{10}[3] \rightarrow -0.39675,
   F_1[5] \rightarrow 0.032992, F_1[6] \rightarrow 0.16433,\\[6pt]
&F_3[1, 3, 5] \rightarrow -0.52888,
   F_3[1, 3, 6] \rightarrow 0.10898, F_3[1, 4, 5] \rightarrow 0.29728,
   F_3[1, 4, 6] \rightarrow -0.056703,\\[6pt]
&F_3[2, 3, 5] \rightarrow 0.46336,
   F_3[2, 3, 6] \rightarrow -0.47205, F_3[2, 4, 5] \rightarrow -0.069697,
   F_3[2, 4, 6] \rightarrow 0.87424,\\[6pt]
&H[1, 2, 5] \rightarrow 0.0018436,
   H[1, 2, 6] \rightarrow 0.010455, H[3, 4, 5] \rightarrow 0.15470,
   H[3, 4, 6] \rightarrow -0.029492,\\[6pt]
&f[1, 3, 5] \rightarrow 0.058151,
   f[1, 3, 6] \rightarrow 0.32171, f[1, 4, 5] \rightarrow 0.11066,
   f[1, 4, 6] \rightarrow 0.56519,\\[6pt]
&f[2, 3, 5] \rightarrow -0.92088,
   f[2, 3, 6] \rightarrow -0.059007, f[2, 4, 5] \rightarrow -0.48068,
   f[2, 4, 6] \rightarrow -0.41926,\\[6pt]
& f[3, 1, 5] \rightarrow 0.0050094,
   f[3, 1, 6] \rightarrow -0.028443, f[3, 2, 5] \rightarrow 0.011003,
   f[3, 2, 6] \rightarrow -0.0028411,\\[6pt]
& f[4, 1, 5] \rightarrow 0.0060442,
   f[4, 1, 6] \rightarrow 0.060379, f[4, 2, 5] \rightarrow -0.0063015,
   f[4, 2, 6] \rightarrow 0.0012974,\\[6pt]
& f[5, 1, 3] \rightarrow -0.061155,
   f[5, 1, 4] \rightarrow -0.026674, f[5, 2, 3] \rightarrow -0.066026,
   f[5, 2, 4] \rightarrow -0.028799,\\[6pt]
& f[6, 1, 3] \rightarrow 0.012278,
   f[6, 1, 4] \rightarrow 0.0053553, f[6, 2, 3] \rightarrow 0.013256,
   f[6, 2, 4] \rightarrow 0.0057819 \,.
\end{aligned}
\end{equation*}
\begin{equation*}
{\cal R}_4 = 0.0031779 \,, \quad  {\cal R}_6 = -0.44861 \,, \quad \eta_V= -0.12141 \,,
\end{equation*}
\begin{equation*}
\text{masses}^2 = (1.7237, 0.060109, 0.0058657, -0.000096456) \,, \;  \vec{v} = (0.35911, 0.91283, 0.19064, 0.037871) \,.
\end{equation*}

\subsection*{Solution 20}
\label{sol20}
\begin{equation*}
\begin{aligned}
&T_{10}[1] \rightarrow 10, T_{10}[2] \rightarrow -0.19097, T_{10}[3] \rightarrow -1.1799,
   F_1[5] \rightarrow 0.15119, F_1[6] \rightarrow 0.17554,\\[6pt]
&F_3[1, 3, 5] \rightarrow -0.23901,
   F_3[1, 3, 6] \rightarrow -0.64579, F_3[1, 4, 5] \rightarrow -0.017596,
   F_3[1, 4, 6] \rightarrow 0.19754,\\[6pt]
&F_3[2, 3, 5] \rightarrow -0.16767,
   F_3[2, 3, 6] \rightarrow 0.88008, F_3[2, 4, 6] \rightarrow -0.22116,
   H[1, 2, 5] \rightarrow -0.0097969,\\[6pt]
& f[1, 4, 6] \rightarrow 0.71835,
   f[2, 3, 5] \rightarrow 0.36455, f[2, 4, 5] \rightarrow 1.2126,
   f[3, 2, 5] \rightarrow -0.046623,\\[6pt]
&f[5, 2, 3] \rightarrow 0.15822,
   f[5, 2, 4] \rightarrow 0.52631, f[6, 2, 3] \rightarrow -0.13627,
   f[6, 2, 4] \rightarrow -0.45329 \,.
\end{aligned}
\end{equation*}
\begin{equation*}
{\cal R}_4 = 0.019450 \,, \quad  {\cal R}_6 = -0.72144 \,, \quad \eta_V= -1.3624 \,,
\end{equation*}
\begin{equation*}
\text{masses}^2 = (2.6116, 0.37159, 0.028935, -0.0066248) \,, \;  \vec{v} = (0.10243, 0.94233, 0.030698, -0.31715) \,.
\end{equation*}

\subsection*{Solution 21}
\label{sol21}
\begin{equation*}
\begin{aligned}
&T_{10}[1] \rightarrow 10, T_{10}[2] \rightarrow -0.13516, T_{10}[3] \rightarrow -0.67689,
   F_1[5] \rightarrow 0.25041, F_1[6] \rightarrow -0.18033,\\[6pt]
&F_3[1, 3, 5] \rightarrow -0.41712,
   F_3[1, 3, 6] \rightarrow 0.57791, F_3[1, 4, 5] \rightarrow 0.0044424,
   F_3[1, 4, 6] \rightarrow 0.30091,\\[6pt]
&F_3[2, 3, 5] \rightarrow 0.055123,
   F_3[2, 3, 6] \rightarrow 0.87210, F_3[2, 4, 6] \rightarrow 0.22542,
   H[1, 2, 5] \rightarrow 0.012410,\\[6pt]
& H[3, 4, 5] \rightarrow -0.14114,
   H[3, 4, 6] \rightarrow -0.22801, f[1, 3, 5] \rightarrow -0.37658,
   f[1, 4, 5] \rightarrow 0.43325,\\[6pt]
&f[1, 4, 6] \rightarrow 0.64505,
   f[2, 3, 5] \rightarrow -0.28801, f[2, 4, 5] \rightarrow 1.0164,
   f[3, 2, 5] \rightarrow 0.035108,\\[6pt]
&f[5, 2, 3] \rightarrow -0.063291,
   f[5, 2, 4] \rightarrow 0.22337, f[6, 2, 3] \rightarrow -0.087890,
   f[6, 2, 4] \rightarrow 0.31018 \,.
\end{aligned}
\end{equation*}
\begin{equation*}
{\cal R}_4 = 0.023161 \,, \quad  {\cal R}_6 = -0.75279 \,, \quad \eta_V= -1.7813 \,,
\end{equation*}
\begin{equation*}
\text{masses}^2 = (1.9567, 0.28341, 0.030287, -0.010314) \,, \;  \vec{v} = (0.092869, 0.96324, 0.077710, -0.23978) \,.
\end{equation*}

\subsection*{Solution 22}
\label{sol22}
\begin{equation*}
\begin{aligned}
&T_{10}[1] \rightarrow 10, T_{10}[2] \rightarrow 0.11698, T_{10}[3] \rightarrow -0.25961,
   F_1[5] \rightarrow 0.22288,\\[6pt]
&F_3[1, 3, 5] \rightarrow 0.64570, F_3[1, 3, 6] \rightarrow 0.16612,
   F_3[1, 4, 6] \rightarrow 0.023090, F_3[2, 3, 5] \rightarrow -0.70283,\\[6pt]
&F_3[2, 3, 6] \rightarrow 0.63805, F_3[2, 4, 6] \rightarrow -0.49421,
   H[1, 2, 5] \rightarrow -0.012009, H[3, 4, 6] \rightarrow -0.21608,\\[6pt]
& f[1, 4, 5] \rightarrow 0.18908, f[1, 4, 6] \rightarrow -0.66397,
   f[2, 3, 5] \rightarrow 0.48234, f[2, 4, 5] \rightarrow 0.64579,\\[6pt]
&f[2, 4, 6] \rightarrow 0.72271, f[3, 1, 5] \rightarrow 0.040164,
   f[3, 2, 5] \rightarrow 0.036900, f[6, 1, 4] \rightarrow -0.067815 \,.
\end{aligned}
\end{equation*}
\begin{equation*}
{\cal R}_4 = 0.0028407 \,, \quad  {\cal R}_6 = -0.80087 \,, \quad \eta_V= -1.0525 \,,
\end{equation*}
\begin{equation*}
\text{masses}^2 = (1.6969, 0.099699, 0.0037153, -0.00074744) \,, \;  \vec{v} = (0.34472, 0.91135, 0.21299, 0.072416) \,.
\end{equation*}

\subsection*{Solution 23}
\label{sol23}
\begin{equation*}
\begin{aligned}
&T_{10}[1] \rightarrow 10, T_{10}[2] \rightarrow 0.84108, T_{10}[3] \rightarrow -1.7713,
   F_1[5] \rightarrow 0.28275,\\[6pt]
&F_3[1, 3, 5] \rightarrow 0.76288, F_3[1, 3, 6] \rightarrow 0.34090,
   F_3[1, 4, 6] \rightarrow 0.18689, F_3[2, 3, 5] \rightarrow 0.23849,\\[6pt]
&F_3[2, 3, 6] \rightarrow 0.70671, F_3[2, 4, 6] \rightarrow 0.32182,
   H[1, 2, 5] \rightarrow 0.047753, H[3, 4, 6] \rightarrow -0.19749,\\[6pt]
& f[1, 4, 5] \rightarrow 0.15989, f[1, 4, 6] \rightarrow -1.0319,
   f[2, 3, 5] \rightarrow -0.35628, f[2, 4, 5] \rightarrow 0.81719,\\[6pt]
& f[2, 4, 6] \rightarrow -0.32259, f[3, 1, 5] \rightarrow 0.078001,
   f[3, 2, 5] \rightarrow -0.24951, f[6, 1, 4] \rightarrow -0.41774 \,.
\end{aligned}
\end{equation*}
\begin{equation*}
{\cal R}_4 = 0.038665 \,, \quad  {\cal R}_6 = -0.77384 \,, \quad \eta_V= -1.2253 \,,
\end{equation*}
\begin{equation*}
\text{masses}^2 = (2.1222, 0.21211, 0.072030, -0.011844) \,, \;  \vec{v} = (0.59221, 0.79571, 0.10757, 0.067508) \,.
\end{equation*}

\subsection*{Solution 24}
\label{sol24}
\begin{equation*}
\begin{aligned}
&T_{10}[1] \rightarrow 10, T_{10}[2] \rightarrow 0.19660, T_{10}[3] \rightarrow -0.44890,
   F_1[5] \rightarrow 0.23918,\\[6pt]
&F_3[1, 3, 5] \rightarrow 0.70008, F_3[1, 3, 6] \rightarrow -0.072124,
    F_3[1, 4, 6] \rightarrow 0.044829, F_3[2, 3, 5] \rightarrow 0.60939,\\[6pt]
&F_3[2, 3, 6] \rightarrow 0.68494, F_3[2, 4, 6] \rightarrow 0.47273,
   H[1, 2, 5] \rightarrow 0.019075, H[3, 4, 6] \rightarrow -0.22522,\\[6pt]
& f[1, 4, 5] \rightarrow -0.11312, f[1, 4, 6] \rightarrow -0.73817,
   f[2, 3, 5] \rightarrow -0.45730, f[2, 4, 5] \rightarrow 0.70098,\\[6pt]
& f[2, 4, 6] \rightarrow -0.64254, f[3, 1, 5] \rightarrow 0.054422,
   f[3, 2, 5] \rightarrow -0.062521, f[6, 1, 4] \rightarrow -0.10923 \,.
\end{aligned}
\end{equation*}
\begin{equation*}
{\cal R}_4 = 0.0061178 \,, \quad  {\cal R}_6 = -0.79288 \,, \quad \eta_V= -0.95955 \,,
\end{equation*}
\begin{equation*}
\text{masses}^2 = (1.7245, 0.11797, 0.0082620, -0.0014676) \,, \;  \vec{v} = (0.38671, 0.89718, 0.19939, 0.075995) \,.
\end{equation*}

\vspace{0.1in}

\subsection*{Solution 25}
\label{sol25}
\begin{equation*}
\begin{aligned}
&T_{10}[1] \rightarrow 10, T_{10}[2] \rightarrow 0.28255, T_{10}[3] \rightarrow -0.66015,
   F_1[5] \rightarrow 0.24822,\\[6pt]
&F_3[1, 3, 5] \rightarrow 0.73650, F_3[1, 3, 6] \rightarrow 0.0099987,
    F_3[1, 4, 6] \rightarrow 0.070509, F_3[2, 3, 5] \rightarrow 0.52642,\\[6pt]
&F_3[2, 3, 6] \rightarrow 0.71248, F_3[2, 4, 6] \rightarrow 0.44686,
   H[1, 2, 5] \rightarrow 0.025515, H[3, 4, 6] \rightarrow -0.22535,\\[6pt]
& f[1, 4, 5] \rightarrow -0.051498, f[1, 4, 6] \rightarrow -0.80098,
   f[2, 3, 5] \rightarrow -0.43118, f[2, 4, 5] \rightarrow 0.74004,\\[6pt]
&  f[2, 4, 6] \rightarrow -0.57251, f[3, 1, 5] \rightarrow 0.064822,
   f[3, 2, 5] \rightarrow -0.090691, f[6, 1, 4] \rightarrow -0.15442 \,.
\end{aligned}
\end{equation*}
\begin{equation*}
{\cal R}_4 = 0.010180 \,, \quad  {\cal R}_6 = -0.78633 \,, \quad \eta_V= -0.90691 \,,
\end{equation*}
\begin{equation*}
\text{masses}^2 = (1.7623, 0.13189, 0.014432, -0.0023081) \,, \;  \vec{v} = (0.43152, 0.87989, 0.18312, 0.077785) \,.
\end{equation*}

\subsection*{Solution 26}
\label{sol26}
\begin{equation*}
\begin{aligned}
&T_{10}[1] \rightarrow 10, T_{10}[2] \rightarrow 0.60749, T_{10}[3] \rightarrow -1.3921,
   F_1[5] \rightarrow 0.26812,\\[6pt]
&F_3[1, 3, 5] \rightarrow -0.77550, F_3[1, 3, 6] \rightarrow -0.23680,
    F_3[1, 4, 6] \rightarrow -0.15385, F_3[2, 3, 5] \rightarrow -0.31347,\\[6pt]
&F_3[2, 3, 6] \rightarrow -0.72645, F_3[2, 4, 6] \rightarrow -0.36053,
   H[1, 2, 5] \rightarrow 0.041294, H[3, 4, 6] \rightarrow -0.20803,\\[6pt]
&  f[1, 4, 5] \rightarrow -0.10180, f[1, 4, 6] \rightarrow 0.96159,
   f[2, 3, 5] \rightarrow 0.36993, f[2, 4, 5] \rightarrow -0.80647,\\[6pt]
&  f[2, 4, 6] \rightarrow 0.38869, f[3, 1, 5] \rightarrow -0.077154,
   f[3, 2, 5] \rightarrow 0.19087, f[6, 1, 4] \rightarrow 0.31938 \,.
\end{aligned}
\end{equation*}
\begin{equation*}
{\cal R}_4 = 0.026903 \,, \quad  {\cal R}_6 = -0.77236 \,, \quad \eta_V= -1.0438 \,,
\end{equation*}
\begin{equation*}
\text{masses}^2 = (1.9625, 0.17832, 0.045753, -0.0070202) \,, \;  \vec{v} = (0.55137, 0.82094, 0.12983, 0.072063) \,.
\end{equation*}

\subsection*{Solution 27}
\label{sol27}
\begin{equation*}
\begin{aligned}
&T_{10}[1] \rightarrow 10, T_{10}[2] \rightarrow 0.69681, T_{10}[3] \rightarrow -1.5499,
   F_1[5] \rightarrow -0.27379,\\[6pt]
&F_3[1, 3, 5] \rightarrow 0.77212, F_3[1, 3, 6] \rightarrow -0.28080,
    F_3[1, 4, 6] \rightarrow 0.16865, F_3[2, 3, 5] \rightarrow -0.27968,\\[6pt]
&F_3[2, 3, 6] \rightarrow 0.71969, F_3[2, 4, 6] \rightarrow -0.34410,
   H[1, 2, 5] \rightarrow 0.044110, H[3, 4, 6] \rightarrow 0.20379,\\[6pt]
& f[1, 4, 5] \rightarrow -0.12773, f[1, 4, 6] \rightarrow -0.99090,
   f[2, 3, 5] \rightarrow 0.36320, f[2, 4, 5] \rightarrow 0.81222,\\[6pt]
& f[2, 4, 6] \rightarrow 0.35892, f[3, 1, 5] \rightarrow 0.077686,
   f[3, 2, 5] \rightarrow 0.21447, f[6, 1, 4] \rightarrow -0.35893 \,.
\end{aligned}
\end{equation*}
\begin{equation*}
{\cal R}_4 = 0.031484 \,, \quad  {\cal R}_6 = -0.77199 \,, \quad \eta_V= -1.1172 \,,
\end{equation*}
\begin{equation*}
\text{masses}^2 = (2.0231, 0.19153, 0.055590, -0.0087933) \,, \;  \vec{v} = (0.56984, 0.80991, 0.12007, 0.070019) \,.
\end{equation*}

\end{appendix}

\newpage

\providecommand{\href}[2]{#2}\begingroup\raggedright
\endgroup

\end{document}